\documentclass[%
  aip,jcp,
 amsmath,amssymb,
preprint,
longbibliography,%
]{revtex4-2}

\usepackage{graphicx}
\usepackage[position=top]{subfig}

\usepackage{epstopdf}
\usepackage{dcolumn}
\usepackage{bm}
\usepackage{verbatim}
\usepackage{xcolor}
\usepackage{mathptmx}
\usepackage{caption}

\usepackage{soul}
\usepackage{float}
\usepackage{ragged2e}

\usepackage[utf8]{inputenc}
\usepackage[T1]{fontenc}
\usepackage{mathptmx}

\begin{document}

\title{Potential Distribution Theory of Alchemical Transfer}

\author{Solmaz Azimi}
\altaffiliation{Ph.D. Program in Biochemistry, The Graduate Center of the City University of New York, New York, NY}
\author{Emilio Gallicchio}
\altaffiliation{Ph.D. Program in Biochemistry, The Graduate Center of the City University of New York, New York, NY}
\altaffiliation{Ph.D. Program in Chemistry, The Graduate Center of the City University of New York, New York, NY}
\email[Corresponding author: ]{egallicchio@brooklyn.cuny.edu}
\affiliation{Department of Chemistry and Biochemistry, Brooklyn College of the City University of New York, New York, NY}


\begin{abstract}
  We present an analytical description of the Alchemical Transfer Method (ATM) for molecular binding using the Potential Distribution Theory (PDT) formalism. ATM models the binding free energy by mapping the bound and unbound states of the complex by translating the ligand coordinates. PDT relates the free energy and the probability densities of the perturbation energy along the alchemical path to the probability density at the initial state, which is the unbound state of the complex in the case of a binding process. Hence, the ATM probability density of the transfer energy at the unbound state is first related by a convolution operation of the probability densities for coupling the ligand to the solvent and coupling it to the solvated receptor--for which analytical descriptions are available--with parameters obtained from maximum likelihood analysis of data from double-decoupling alchemical calculations. PDT is then used to extend this analytical description along the alchemical transfer pathway. We tested the theory on the alchemical binding of five guests to the TEMOA host from the SAMPL8 benchmark set. In each case, the probability densities of the perturbation energy for transfer along the alchemical transfer pathway obtained from numerical calculations match those predicted from the theory and double-decoupling simulations. The work provides a solid theoretical foundation for alchemical transfer, offers physical insights on the form of the probability densities observed in alchemical transfer calculations, and confirms the conceptual and numerical equivalence between the alchemical transfer and double-decoupling processes.
\end{abstract}

\maketitle

\section{Introduction}

The modeling of free energies is critical to the characterization of materials, chemical processes, molecular recognition, and many other areas of investigation where molecular-level insights are pursued.\cite{Chipot:Pohorille:book:2007} Bridging the gap between theoretical models and experimental observations is perhaps the most important role of free energy calculations in modern chemical research. By comparing calculated free energies to measured values, scientists can refine their models and gain deeper insights into chemical composition, chemical interactions, and the dynamical behavior of complex molecular systems. Alchemical free energy models exploit the state function property of the free energy to compute free energy differences between physical states by transforming the system's Hamiltonian along a sequence of non-physical states that cannot be realized in the laboratory.\cite{tuckerman2023statistical} 

The study of protein-ligand binding in drug discovery is one of the most common applications of alchemical free energy models.\cite{abel2017advancing,schindler2020large,ganguly2022amber,procacci2022relative,baumann2023broadening,Allen2022.05.23.493001} In the popular double-decoupling approach, the binding free energy of a receptor-ligand complex in solution is obtained as the difference between the free energy of alchemically turning on the ligand-solvent interactions in solution and the free energy of turning off the interactions between the bound ligand and the solvated receptor, each obtained from separate calculations.\cite{lee2020alchemical,Mey2020Best} The term Double-Decoupling Method (DDM), which refers to this approach, stems from considering each step as ``decoupling'' the ligand from its environment (the solvent or the receptor in the solvent) to reach a vacuum state.\cite{Gilson:Given:Bush:McCammon:97} The same process can be equivalently described in terms of the reverse steps of ``coupling'' the ligand to either the solution or the solvated receptor from vacuum.  

Conceptually, each coupling step of DDM is closely related to the particle insertion method originally introduced to model solvation free energies.\cite{Widom:82,valleau-torrie1977,simonson1993free,stamatopoulou1998cavity} The particle insertion method obtains the solvation free energy by performing an exponential average of the solute-solvent interaction energies resulting from random insertions of the solute into an ensemble of pure solvent configurations. The Potential Distribution Theory (PDT)\cite{Widom:82,PDTbook:2006} applies to particle insertion when considering the solute-solvent energy as the perturbation energy $u$. The main statement of PDT is that the solvation free energy and the probability density distribution of the solute-solvent energy $p_1(u)$ in the coupled state are determined by the probability density distribution $p_0(u)$ of the perturbation energy in the uncoupled ensemble.\cite{PDTbook:2006} The PDT relationship extends to the sequence of alchemical intermediate states corresponding to the progressive introduction of the solute-solvent interaction by means of a progress parameter $\lambda$,\cite{swope1984molecular} implying that knowledge of $p_0(u)$ determines the perturbation energy distributions $p_\lambda(u)$ and the free energy profile $\Delta G(\lambda)$ along the entire alchemical coupling process.\cite{PDTbook:2006,Gallicchio2011adv} 

We used these PDT results to model alchemical binding processes from a statistical perspective,\cite{Gallicchio2010} and to optimize alchemical potential energy functions.\cite{pal2019perturbation} In particular, Kilburg and Gallicchio\cite{kilburg2018analytical} developed an analytical model of $p_0(u)$  with parameters learned from the distribution of perturbation energies collected from alchemical simulations of molecular coupling processes. Using the PDT formula, the model reproduced the perturbation energies and the free energy profile along the alchemical binding pathway of host-guest complexes with implicit solvation.\cite{kilburg2018analytical} The model was later employed to study alchemically-induced phase transitions, their role in the rate of convergence of free energy estimates, and in devising optimized alchemical potential energy functions applicable with explicit solvation models.\cite{pal2019perturbation,khuttan2021alchemical}

Recently, a direct alchemical transfer route to the free energy of molecular binding in explicit solution that bypasses the ligand's vacuum state has been developed. In the resulting Alchemical Transfer Method (ATM),\cite{wu2021alchemical} the alchemical transformation is encoded in a coordinate transformation that directly translates the ligand from a position in the solvent bulk into the receptor binding site. ATM has been extensively validated against host-guest systems\cite{wu2021alchemical,azimi2022application,khuttan2023taming} and relative binding free energy protein-ligand benchmarks,\cite{azimi2022relative,chen2023performance,sabanes2023validation} and is considered a viable alchemical approach in applied research, especially with advanced many-body potential models not yet supported by standard alchemical models.\cite{eastman2023openmm,sabanes2024enhancing}

This work extends the Potential Distribution Theory (PDT) formalism to describe alchemical transfer processes. Unlike coupling processes, where interatomic interactions are created from the uncoupled state, alchemical transfer involves the gain of ligand-receptor interactions accompanied by the simultaneous loss of ligand-solvent interactions. Following the PDT formalism, we approach the problem by seeking the probability density of the perturbation energy for alchemical transfer at the initial state, where the ligand is in solution. Because the perturbation energy for alchemical transfer is the difference between the receptor-ligand and the solvent-ligand interaction energies, we model its probability distribution as the convolution of the probability distributions of the two components that we obtain by double-decoupling alchemical simulations. We show that the PDT applied to the convolution function successfully reproduces the perturbation energy probability densities throughout alchemical transfer processes.

We illustrate the PDT theory developed here by applying it to the alchemical transfer binding of a series of guests to a molecular host. The coupling processes of the guest from vacum to the solution and the host are simulated and analyzed in terms of the analytical model of alchemical binding of Kilburg and Gallicchio.\cite{kilburg2018analytical} This procedure provides optimized parameters for the analytical expressions of the $p_0(u)$ functions for each coupling process. We show that the convolution of the analytical models learned from alchemical coupling simulations matches the probability densities of the alchemical transfer perturbation energy at the solvated state obtained from alchemical transfer simulations of binding. 

These results demonstrate that the PDT is applicable to alchemical transfer and that double-decoupling and transfer processes are statistically equivalent because the probability densities of the perturbation energy of the second can be determined from the first. More generally, we illustrate that the PDT, which is traditionally applied to coupling processes, is also suitable to describe more complex alchemical processes such as alchemical transfer. The work also illustrates the fundamental concept underlying the PDT that the alchemical pathways connecting the same endpoints are interrelated because they originate from the same probability density kernel $p_0(u)$. Hence, a $p_0(u)$ model learned from one pathway yields information about all other alchemical pathways.

\section{Theory}

\subsection{Alchemical Transfer and Double-Decoupling for Modeling Molecular Binding Equilibria}

Consider the standard free energy, $\Delta G^{\circ}_b$, of the non-covalent association equilibrium between receptor R and ligand L to form the receptor-ligand complex RL
\begin{equation}
    R_{(aq)} + L_{(aq)} \rightleftharpoons RL_{(aq)} \, ,
    \label{eq:binding}
\end{equation}
that is related to the binding constant $K_b$ through
\begin{equation}
\Delta G^{\circ}_b = -k_B T \ln K_b ,
\end{equation}
where $k_B$ is Boltzmann's constant and $T$ is the temperature. A statistical mechanics expression for $K_b$ is\cite{Gilson:Given:Bush:McCammon:97,Gilson2007,Gallicchio2011adv,gallicchio2021comppeptsci} 
\begin{equation}
    K_b = \frac{C^\circ V_{\rm site} }{8 \pi^2} \langle e^{- \beta u(x)} \rangle_0 
    \label{eq:Kb}
\end{equation}
where $\beta = 1/(k_B T)$,
\begin{equation}
u(x) = U_1(x) - U_0(x)
\label{eq:u-def}
\end{equation}
is the binding energy of the configuration $x$ of the complex defined as the potential energy of the bound complex, $U_1(x)$, relative to the potential energy of the unbound configuration, $U_0(x)$, when receptor and ligand are uncoupled at large separation, $\langle \ldots \rangle_0$ represents the ensemble average in the uncoupled state, $V_{\rm site}$ is the volume of the receptor binding site, and $C^\circ$ is the one molar standard concentration. The free energy $\Delta G^\circ_{\rm id} = -k_B T \ln C^\circ V_{\rm site}/( 8 \pi^2 )$, is the ideal component of the standard binding free energy (the value of the standard binding free energy that would be observed if the ligand did not interact with the receptor) while the term
\begin{equation}
    \Delta G_b = -k_B T \ln \langle e^{-\beta u} \rangle_0
    \label{eq:DGb_excess}
\end{equation}
is the excess component of the binding free energy.

The objective of alchemical computational binding free models is to estimate the quantity $\Delta G_b$ in Equation \ref{eq:DGb_excess} as accurately and rapidly as possible. This is  done through a series of non-physical potential energy functions $U_\lambda(x)$, where $0 \le \lambda \le 1$ is the alchemical progress parameter, that interpolate between the potential energy functions, $U_0(x)$ and $U_1(x)$, which describe the unbound and bound states of the complex.\cite{McCammon:Straatsma:92,Jorgensen2010, cournia2017relative,cournia2020rigorous,rizzi2020sampl6} $\Delta G_b$ for the association process in Eq.\ (\ref{eq:binding}) can be modeled alchemically in explicit solvent by either alchemical transfer\cite{wu2021alchemical} or double-decoupling.\cite{Gilson:Given:Bush:McCammon:97,Gilson2007,Deng2008a,Mey2020Best} In the alchemical transfer method (ATM), the receptor and ligand are simulated together in a solvent box using a $\lambda$-dependent hybrid potential of $U_0(x)$ and $U_1(x)$, where $U_1(x)$ is sampled by translating the ligand into the receptor binding site from an arbitrary position in the solvent. 

\begin{figure}
    \centering
    \includegraphics[scale=0.5]{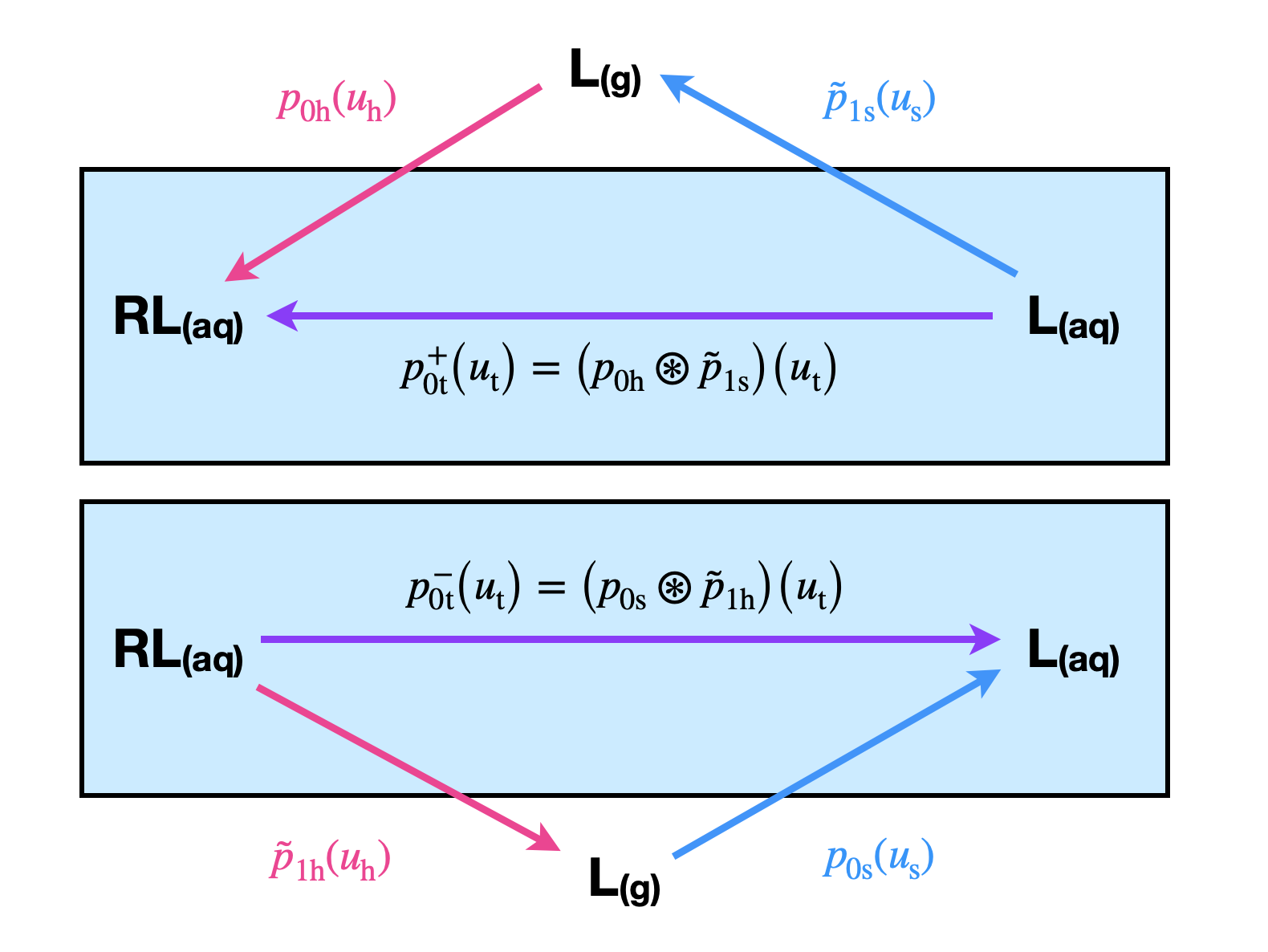}
    \captionsetup{justification=Justified}
    \caption{The binding free energy of ligand L to receptor R (top panel) in solution is estimated directly by alchemical transfer of the ligand into the receptor (horizontal arrow), or by double-decoupling, whereby the ligand is alchemically decoupled from the solvent to vacuum (blue arrow) followed by alchemical coupling of the ligand to the solvated receptor from vacuum (pink arrow). The probability densities of the ligand-receptor interaction energy, $u_{\rm h}$, along the coupling process are derived by the probability density, $p_{0{\rm h}}(u_{\rm h})$, at the initial decoupled state. Similarly, the decoupling process is described by the probability density $\tilde{p}_{1{\rm s}}(u_{\rm s})$ of the loss of ligand-solvent interaction energy at the solvent-coupled state. The probability density of the perturbation energy for alchemical transfer, $u_{\rm t}$, is given by the convolution of these two functions (horizontal purple arrow). The statistics of the alchemical dissociation process (bottom panel) is obtained analogously from the probability densities of the loss of ligand-receptor interactions and the gain of ligand-solvent interactions.}
    \label{fig:conv_scheme}
\end{figure}

The double-decoupling method (DDM), which is more widely used, models the binding process in two steps: first, L is alchemically transferred from vacuum to solution (the solvent coupling step), and then, in a separate simulation, L is alchemically transferred from vacuum to the binding site of R (the receptor coupling step). The excess binding free energy of the association between L and R is the difference between the free energies of the solvent and receptor coupling steps (Figure \ref{fig:conv_scheme}).

The two coupling steps of DDM can be modeled analytically (see below), taking advantage of the fact that the binding energy $u$ corresponds, in these cases, to the interaction energy between the ligand and the environment--the solvent or the receptor in the solvent.\cite{kilburg2018analytical} In alchemical transfer,\cite{wu2021alchemical,azimi2022relative} the perturbation energy is the sum of the loss of ligand-solvent interactions and gain of ligand-receptor interactions (Figure \ref{fig:conv_scheme}), and the corresponding perturbation energy distributions are obtained by the convolution of the distributions of the decoupling and coupling models (see below).

For either coupling or transfer, the ensemble average in Eq.\ (\ref{eq:DGb_excess}) can be expressed in the form\cite{Gallicchio2011adv}
\begin{equation}
    \langle e^{-\beta u} \rangle_0 = \int_{-\infty}^{+\infty} du \ e^{-\beta u} \ p_0(u)
    \label{eq:stat_average}
\end{equation}
 where $p_0(u)$ is the probability density of the binding energy at the initial state of the alchemical process, in which receptor and ligand are uncoupled, either when the ligand is placed in the solvent far away from the receptor as in ATM, or when the ligand is in vacuum as in the two coupling steps of DDM. As further discussed below, the function $p_0(u)$ plays a central role in this work.

The general expression of the $\lambda$-dependent alchemical potential energy function employed in this work is
\begin{equation}
U_{\lambda}(x) = U_{0}(x)+W_{\lambda}[u(x)],
\label{eq:pert_pot}
\end{equation}
where the perturbation energy $u(x)$ is defined by Eq.\ (\ref{eq:u-def}), $W_\lambda(u)$ is the alchemical perturbation energy function with the property that $W_0(u) = 0$, and $W_1(u) = u$ such that $U_\lambda(x)$ in Eq.\ (\ref{eq:pert_pot}) yields $U_0(x)$ and $U_1(x)$ at the endpoints of the alchemical transformation. The standard linear form $W_\lambda(u) = \lambda u$ satisfies this criteria, but non-linear versions can be more efficient in numerical applications.\cite{pal2019perturbation}  The specific expression and parameterization of $W_\lambda(u)$  employed in the calculations presented in this work are given in the Computational Details section. 

According to the Potential Distribution Theorem (PDT),\cite{PDTbook:2006} the probability density $p_0(u)$ of the perturbation energy at the initial state encodes all the information of the alchemical process, including the behavior of the intermediate $\lambda$ states. In particular, the PDT states that the probability distributions of the perturbation energy at the intermediate $\lambda$-states are given by\cite{Gallicchio2011adv,pal2019perturbation}
\begin{equation}
    p_\lambda(u) = \frac{e^{-\beta W_\lambda(u)} \ p_0(u)}{K(\lambda)} \, ,
    \label{eq:pdt}
\end{equation}
where
\begin{equation}
    K(\lambda) = \int_{-\infty}^{+\infty} e^{-\beta W_\lambda(u)} \ p_0(u) \, du
    \label{eq:Klambda}
\end{equation}
is the $\lambda$-dependent excess binding constant. In turn, the excess binding free energy profile is given by $\Delta G_b(\lambda) = -k_B T \ln K(\lambda)$, which, at $\lambda=1$, yields the excess binding free energy (Eq.\ (\ref{eq:DGb_excess})).\cite{Gallicchio2011adv,kilburg2018analytical,pal2019perturbation} Hence, knowledge of $p_0(u)$ determines the free energy profile and the perturbation energy distributions at all intermediate states along any alchemical pathway joining two given states. This work aims to construct a model for $p_0(u)$ applicable to direct alchemical transfer. 

Note that the PDT results summarized by Eqs.\ (\ref{eq:pdt}) and (\ref{eq:Klambda}) apply to alchemical models based on energy interpolation\cite{konig2020alternative} whose perturbation energy functions depend on only one or a few collective variables for which it is meaningful to consider probability densities as a function of $\lambda$. For example, it applies to alchemical transfer because its alchemical perturbation energy function $W_\lambda(u)$ in Eq.\ (\ref{eq:pert_pot}) depends on atomic coordinates only through the perturbation energy $u(x) = U_1(x) - U_0(x)$.\cite{wu2021alchemical} However, PDT does not apply to parameter interpolation alchemical models\cite{cournia2020rigorous,Mey2020Best,lee2020alchemical} or models based on $\lambda$-dependent soft-core pair potentials\cite{lee2020improved} whose potential energy functions depend directly on atomic coordinates in complex ways.

\subsection{Analytical Theory of Alchemical Coupling}

Consider $p_0(u)$, the probability density of the interaction energy, $u$, of two molecular entities, such as a solute with a solvent, in the decoupled state where the two molecular entities are not interacting. Here and elsewhere, the subscript "$0$" in $p_0(u)$ refers to the $\lambda = 0$ state of the alchemical transformation. For an alchemical coupling process, the initial state is the decoupled state of the system. The coupling energy of a configuration $x$ of the system is the perturbation energy [Eq.\ (\ref{eq:u-def})] where $U_1(x)$ is the potential energy of the system when the two molecular entities are interacting (\emph{coupled state}), and $U_0(x)$ is the potential energy when their interactions are turned off (\emph{decoupled state}). In this work, $U_1(x)$ is obtained from $U_0(x)$ by rigidly translating the solute into the solvent or the receptor from an arbitrary position in vacuum.\cite{Gallicchio2010}

Because it does not interact with the environment, a solute explores many positions and orientations in the decoupled state. Hence, multiple atomic collisions and overlaps will likely be found when a solute configuration generated in vacuum is transferred into the solvent. As a result of these collisions, perturbation energies $u$ corresponding to these configurations are likely to be large and positive.\cite{simonson1993free} In addition to these short-ranged repulsive interactions, solute-solvent interactions are characterized by long-ranged, slowly varying, and mostly favorable electrostatic and dispersion interactions.\cite{Chandler:Weeks:Andersen:83,Alper:Levy:90} 

Kilburg and Gallicchio\cite{kilburg2018analytical} exploited the distinct nature of collisional and long-range interactions to develop an analytical model for $p_0(u)$. They expressed the total interaction energy $u$ as the sum of (i) a collisional interaction energy ($u_c)$, representing unfavorable short-ranged, repulsive interactions, and (ii) a background interaction energy ($u_b$), representing mostly favorable, long-ranged, attractive interactions. They reasoned that the background interaction energy should follow linear response and central limit statistics because many individual interatomic interactions contribute to it.\cite{Alper:Levy:93, Levy1991LRT, Aqvist:Medina:Samuelsson:1994, Aqvist:Hansson:1996, Carlson:Jorgensen:1995, Levy:Gallicchio:98, Jones-Hertzog:Jorgensen:1997, simonson2002gaussian, Su:Gallicchio:Levy:2007}  Conversely, the collisional interaction energy is dominated by the closest, most repulsive pairwise atomic interaction and is thus expected to follow max statistics.\cite{GumbelBook} Starting with a Lennard-Jones pair-potential description of collisions,\cite{pal2019perturbation} Kilburg and Gallicchio developed an analytical statistical model of the collision energy and expressed the probability density in the decoupled ensemble, $p_0(u)$, of the total interaction energy, $u = u_c + u_b$,  as the convolution of the collisional and background statistical models. 

Specifically, $p_0(u)$ is written as\cite{kilburg2018analytical,pal2019perturbation}
\begin{equation}
    p_0(u) = b \mathcal{N}(u) + (1-b) C(u)
    \label{eq:basic_p0u}
\end{equation}
where
\begin{equation}
    \mathcal{N}(u_b) = \frac{1}{\sqrt{2 \pi \sigma^2}} \ e^{\frac{-(u_b - \bar u_0)^2}{2 \sigma^2}}.
    \label{eq:gaussian}
\end{equation}
is the normal distribution with mean $\bar u_0$ and standard deviation $\sigma$, $b$ is the probability that no collisions occur in the decoupled ensemble, and
\begin{equation}
    C(u) = \int_{-\infty}^{+\infty} du' \ \mathcal{N}(u')\mathcal{F}(u-u') = (\mathcal{N} \circledast \mathcal{F})(u).
    \label{eq:p0u_conv}
\end{equation}
is the convolution of the probability density, ${\mathcal F}(u_c)$, of the collisional interaction energy with the normal distribution of Eq.\ (\ref{eq:gaussian}) that represents the background interaction energy. Here and elsewhere in this work, the convolution operation ($\circledast$) arises whenever we interrogate the statistical behavior of a random energy variable $u$, which is the sum of two random variables whose statistics are known or assumed. 

The analytical expression of ${\mathcal F}(u_c)$ is\cite{pal2019perturbation}
\begin{equation}
  {\mathcal F}(u_c) = n_l \left[ 1 - \frac{(1+x_c)^{1/2}}{(1+x)^{1/2}} \right]^{n_l-1}
\frac {H(u)}{4 \epsilon} 
\frac {(1+x_c)^{1/2}}{x(1+x)^{3/2}}
    \label{eq:coll_func}
\end{equation}
in which, $x = \sqrt{1 + u_c/\epsilon + {\tilde u}/{\epsilon}}$, \ $x_c = \sqrt{1 + {\tilde u}/{\epsilon}}$, \ and
 $H(\cdot)$ is the Heaviside step function. The parameters of the collisional model have the following physical interpretations. The parameter $\tilde u$ represents the interaction energy above which the solute-solvent interaction energy follows max statistics.  $n_l$, which scales as the solute size, describes the number of statistically independent atom groups of the solute. Finally, $\epsilon$ is an effective Lennard-Jones potential energy prefactor that describes the rate of increase of the collisional energy as two atoms approach each other. The collisional parameters, together with the linear response parameters $\bar u_0$ and $\sigma$, specify the analytical model of alchemical coupling of Eq.\ (\ref{eq:basic_p0u}). As described below, the parameters of the analytical model for $p_0(u)$ are obtained by maximum likelihood analysis of simulation data.

To model the alchemical coupling of flexible polyatomic ligands that can adopt more than one conformation, in this work, we express $p_0(u)$ as the weighted average of modes described by the model above:\cite{khuttan2021alchemical}
\begin{equation}
   p_0(u) = \sum^{m}_{i} w_i p_{0,i}(u)
\label{eq:multimodal_p0u}
\end{equation}
where $p_{0,i}(u)$, with parameters $b_{i}$, $\bar u_{0,i}$, etc., is the $p_0(u)$ model specific for mode $i$, and the weight parameters $w_i$ represents the population of each mode in the decoupled ensemble. 

\subsection{Analytical Model of Alchemical Transfer}

In this section we derive an analytical model of alchemical transfer using the alchemical coupling formalism described above. Consider the thermodynamic scheme in Fig.\ \ref{fig:conv_scheme}, in which a guest is transferred from the solution to the binding site of a molecular host in order to alchemically estimate the binding free energy of the host-guest complex. Even though an analytical model of direct alchemical transfer is not available, the transfer process can be represented by sequential decoupling and coupling processes covered by the analytical formalism. The change in potential energy, $u_{\rm t}$, for transferring the ligand from the solvent to the receptor, is the sum of the loss of ligand-solvent interaction energy, $u_{\rm s}$, and the gain, $u_{\rm h}$, of the interaction energy between the ligand and the receptor (including the surrounding solvent). Hence, the probability density of $u_{\rm t}$ can be expressed as the convolution of the probability densities of $u_{\rm s}$ and  $u_{\rm h}$ collected in the initial ensemble where receptor and ligand are dissociated in solution.

Denoting $p_{1 \rm s}(u_{\rm s})$ as the probability density of the ligand-solvent interaction energy of the ligand in solution (the end-state of the solvation alchemical process), and $p_{0 \rm h}(u_{\rm h})$ as that of the ligand-receptor coupling energy in the initial state, the probability density of the perturbation energy for solvent to receptor transfer, $u_{\rm t} = u_{\rm h}-u_{\rm s}$, is given by the convolution
\begin{equation}
    p^{+}_{0\rm t}(u_{\rm t}) = ( p_{0 \rm h} \circledast {\tilde p}_{1 \rm s})(u_{\rm t})
    \label{eq:leg1_conv}
\end{equation}
where ${\tilde p}_{1 \rm s}(u_{\rm s}) = p_{1 \rm s}( -u_{\rm s} )$, is the probability density of the solute-solvent interaction energy loss in the solvated coupled ensemble (the initial state of the alchemical process, which is denoted by the blue arrow in the bottom panel of Fig.\ \ref{fig:conv_scheme}). (The meaning of the superscripted plus symbol in Eq.\ (\ref{eq:leg1_conv}) is specified below.) 

Because in the fully solvated state the ligand does not experience collisions with the solvent molecules, the solute-solvent interaction energy is expected to follow linear response and the corresponding probability density can be described by a normal distribution 
\begin{equation}
    \tilde p_{1 \rm s}(u_{\rm s})=\tilde{\mathcal{N}}_{1 \rm s}(u_{\rm s})
    \label{eq:p1h_nmodel}
\end{equation}
with mean $-\bar u_{\rm s}$ and standard deviation $\sigma_{\rm s}$, where $\bar u_{\rm s}$ and $\sigma_{\rm s}$ are the mean and standard deviation of the ligand-solvent interaction energy in the solvated coupled ensemble, respectively. Generally, as in Eq.\ (\ref{eq:multimodal_p0u}), $\tilde p_{1 \rm s}(u_{\rm s})$ is represented by a weighted sum of normal distributions. However, we assume one solvation mode for now to keep the notation simple. Furthermore, the assumption of linear response is expected to break down when, during the alchemical decoupling process, the interactions of the ligand with the solvent are weakened to the point that atomic clashes occur with significant probability. Nevertheless, following the Alchemical Transfer Method (ATM) protocol,\cite{wu2021alchemical} and as further discussed below, this model is applied only up to the alchemical intermediate state at $\lambda = 1/2$ before atomic clashes are observed.

The $p_{0 \rm h}(u_{\rm h})$ function of the second alchemical process, which corresponds to the gain of interactions between the solvated ligand and the receptor, is modeled in this work as that of the coupling processes of the guest in vacuum to the host (see Fig.\ \ref{fig:conv_scheme} top, pink arrow). This approximation is justified by the small influence of the solvent on the distribution of the internal degrees of freedom of the rigid guests considered here. Hence $p_{0 \rm h}(u_{\rm h})$ is represented by the analytical model of alchemical coupling described in the previous section. Here too, we generally consider multiple binding modes in practice, each of them described by the form in Eq.\ (\ref{eq:basic_p0u})
\begin{equation}
    p_{0 \rm h}(u_{\rm h}) = b \mathcal{N}_{0 \rm h}(u_{\rm h}) + (1-b)C_{0 \rm h}(u_{\rm h}) 
    \label{eq:p0b_model}
\end{equation}
where $\mathcal{N}_{0 \rm h}(u_{\rm h})$ is a normal distribution with mean $\bar u_{\rm h}$ and standard deviation $\sigma_{h}$, and
\begin{equation}
    C_{0 \rm h}(u_{\rm h}) = ( \mathcal{N}_{0 \rm h} \circledast {\mathcal F}_{\rm h}  ) (u_{\rm h})
    \label{eq:conv_C0b}
\end{equation}
is the convolution of the linear response and collisional models for coupling the guest to the host.

By inserting Eqs.\ (\ref{eq:p1h_nmodel}), (\ref{eq:p0b_model}), and (\ref{eq:conv_C0b}) in Eq.\ (\ref{eq:leg1_conv}) and using the linearity of the convolution operator, we finally obtain the following model for the probability density of alchemical transfer
\begin{equation}
    p^+_{0 \rm t}(u_{\rm t}) = b \mathcal{N}^+_{0 \rm t}(u_{\rm t})  + (1 - b) C^+_{0 \rm t }(u_{\rm t})
\label{eq:p0t_model}
\end{equation}
where
\begin{equation}
 \mathcal{N}^+_{0 \rm t}(u_{\rm t}) = ( \tilde{\mathcal{N}}_{1 \rm s} \circledast \mathcal{N}_{0 \rm h} ) (u_{\rm t})
\label{eq:n0t_normal}
\end{equation}
is the convolution of the normal distributions for desolvation and receptor coupling, which is itself a normal distribution with mean $\bar u_{0 \rm t} = \bar u_{0 \rm h} - \bar u_{1 \rm s}$ and standard deviation $\sigma_{\rm t} = \sqrt{\sigma^2_{\rm h} + \sigma^2_{\rm s}  } $, and
\begin{equation}
    C^+_{0 \rm t}(u_{\rm t}) = ( \mathcal{N}^+_{0 \rm t} \circledast {\mathcal F}_{\rm h} )  (u_{\rm t}) 
    \label{eq:c0t_model}
\end{equation}
is the convolution of the collisional distribution for the coupling to the host with the normal distribution in Eq.\ (\ref{eq:n0t_normal}).

Eqs.\ (\ref{eq:p0t_model})--(\ref{eq:c0t_model}) establish that, under the present assumptions, the $p_{0 \rm t}(u_{\rm t})$ distribution for the transfer process has the same form as that of a coupling process with parameters determined by specific combinations of those of the coupling process.  In particular, the mean perturbation energy linear response parameter of the transfer model ($\bar u_{0 \rm t}$) is the difference between the corresponding parameter of the receptor coupling model and the average solute-solvent interaction energy in the fully solvated state. The variance linear response parameter ($\sigma^2_{\rm t}$) is the sum of the variances of the desolvation and receptor coupling processes. The parameters of the collisional model ($b$, $\epsilon$, $\tilde u$, and $n_l$) are inherited directly from the receptor coupling collisional model. 

The expressions above have been derived for the simplest case of one solvent coupling mode and one receptor coupling mode. In general, each possible pair of modes with weights $w_{i\rm h}$ and $w_{j\rm s}$ of $m_{\rm h}$ receptor coupling modes and $m_{\rm s}$ solvent coupling modes, respectively, combine in the manner above to yield $m_{\rm h} m_{\rm s}$transfer modes each with weight $w_{ij\rm t} = w_{i\rm h} w_{j\rm s}$. 

The formalism so far describes the transfer process in the binding direction illustrated by the upper panel of Fig.\ (\ref{fig:conv_scheme}) and denoted by a `+' superscript in the expressions above. A similar prescription applies to the transfer unbinding process, where the bound ligand decouples from the receptor and couples to the solution. The corresponding analytical model $p^-_{0 \rm t}(u_{\rm t})$ for the unbinding distribution has the same form as the transfer model for binding (Eq. \ref{eq:p0t_model}) but with collisional parameters obtained from the model of coupling to the solvent and linear response parameters obtained from combining those of decoupling from the receptor and coupling to the solvent.    

\section{Methods}

\subsection{The Alchemical Transfer Method}

The double-decoupling and alchemical transfer binding free energy calculations reported in this work have been conducted using the Alchemical Transfer Method (ATM).\cite{wu2021alchemical, azimi2022application, azimi2022relative, khuttan2023taming, chen2023performance, sabanes2023validation, sabanes2024enhancing} Unlike alchemical approaches that modify the parameters of the energy function,\cite{cournia2020rigorous, ganguly2022amber} ATM relates the potential energy function of the final state ($U_1(x)$) to that of the initial state ($U_0(x)$) by a coordinate transformation.

Specifically, for the case of the solvation process of a ligand L from vacuum, denoted by $U_0(x_{\rm S}, x_{\rm L})$ the potential energy of the system when the ligand's coordinates $x_{\rm L}$ are such that the ligand is placed in vacuum far away from the solvent, whose molecules have coordinates $x_{\rm S}$. The potential energy of the system when the ligand is placed in the solvent is expressed in terms of $U_0(x_{\rm S}, \ x_{\rm L})$ as $U_1(x_{\rm S}, \ x_{\rm L}) = U_0(x_{\rm S}, \ x_{\rm L} + h) $, where $h$ is a displacement vector that brings the ligand from its position in vacuum to the corresponding position in the solvent. The binding process between a receptor R and a ligand L is described in a similar way by using a displacement vector that transfers the ligand from a position into the solvent to the  binding site of the  receptor. This formalism represents the unbound and bound states of the system by a single  set of degrees of freedom and to define the perturbation energy as the difference in the system's potential energy before and after the application of the  ligand displacement. For example, for the binding process we define the perturbation energy as
\begin{equation}
    u^{+} (x_{\rm R}, \ x_{\rm S}, \ x_{\rm L}) =  U_0 ( x_{\rm R}, \ x_{\rm S}, \ x_{\rm L} + h) - U_0 ( x_{\rm R}, \ x_{\rm S}, \ x_{\rm L})
    \label{eq:ATM-perte-leg1}
\end{equation}
and the corresponding alchemical potential energy function as
\begin{equation}
    U^+_\lambda (x_{\rm R}, \ x_{\rm S}, \ x_{\rm L}) = U_0 ( x_{\rm R}, \ x_{\rm S}, \ x_{\rm L}) + W_\lambda [ u^{+} (x_{\rm R}, \ x_{\rm S}, \ x_{\rm L}) ] .
    \label{eq:ATM-pert-pot-leg1}
\end{equation}

The alchemical potential energy function (\ref{eq:ATM-pert-pot-leg1}) cannot cover the entire alchemical binding pathway when the solvent is represented explicitly.\cite{wu2021alchemical} Instead, the alchemical process is decomposed into two legs. In the first leg, the system is taken from the unbound state to an alchemical intermediate state (at $\lambda = 1/2$, typically) using the potential  (\ref{eq:ATM-pert-pot-leg1}). The second leg proceeds in the unbinding direction starting from the bound state at $\lambda=1$ until it reaches the same alchemical intermediate, using the alchemical potential energy function
\begin{equation}
    U^-_{1-\lambda} (x_{\rm R}, \ x_{\rm S}, \ x_{\rm L}) = U_0 ( x_{\rm R}, \ x_{\rm S}, \ x_{\rm L} + h) + W_\lambda [ u^{-} (x_{\rm R}, \ x_{\rm S}, \ x_{\rm L}) ]
    \label{eq:ATM-pert-pot-leg2}
\end{equation}
where
\begin{equation}
    u^{-} (x_{\rm R}, \ x_{\rm S}, \ x_{\rm L}) =  U_0 ( x_{\rm R}, \ x_{\rm S}, \ x_{\rm L}) - U_0 ( x_{\rm R}, \ x_{\rm S}, \ x_{\rm L} + h)
    \label{eq:ATM-perte-leg2}
\end{equation}
is the perturbation energy corresponding to the second leg. 

To compute the binding free energy, the reversible work values  ($\Delta G^+$ and $\Delta G^-$) along each alchemical leg is calculated by thermodynamic reweighting\cite{Tan2012, Shirts2008a} and the excess binding free energy is estimated from their difference: 
\begin{equation}
    \Delta G_b = \Delta G^{+} - \Delta G^{-} \, .
\end{equation}

\section{Computational Details}

\subsection{Benchmark Systems}

\begin{figure}[h!]
\subfloat{\includegraphics[width=5in]{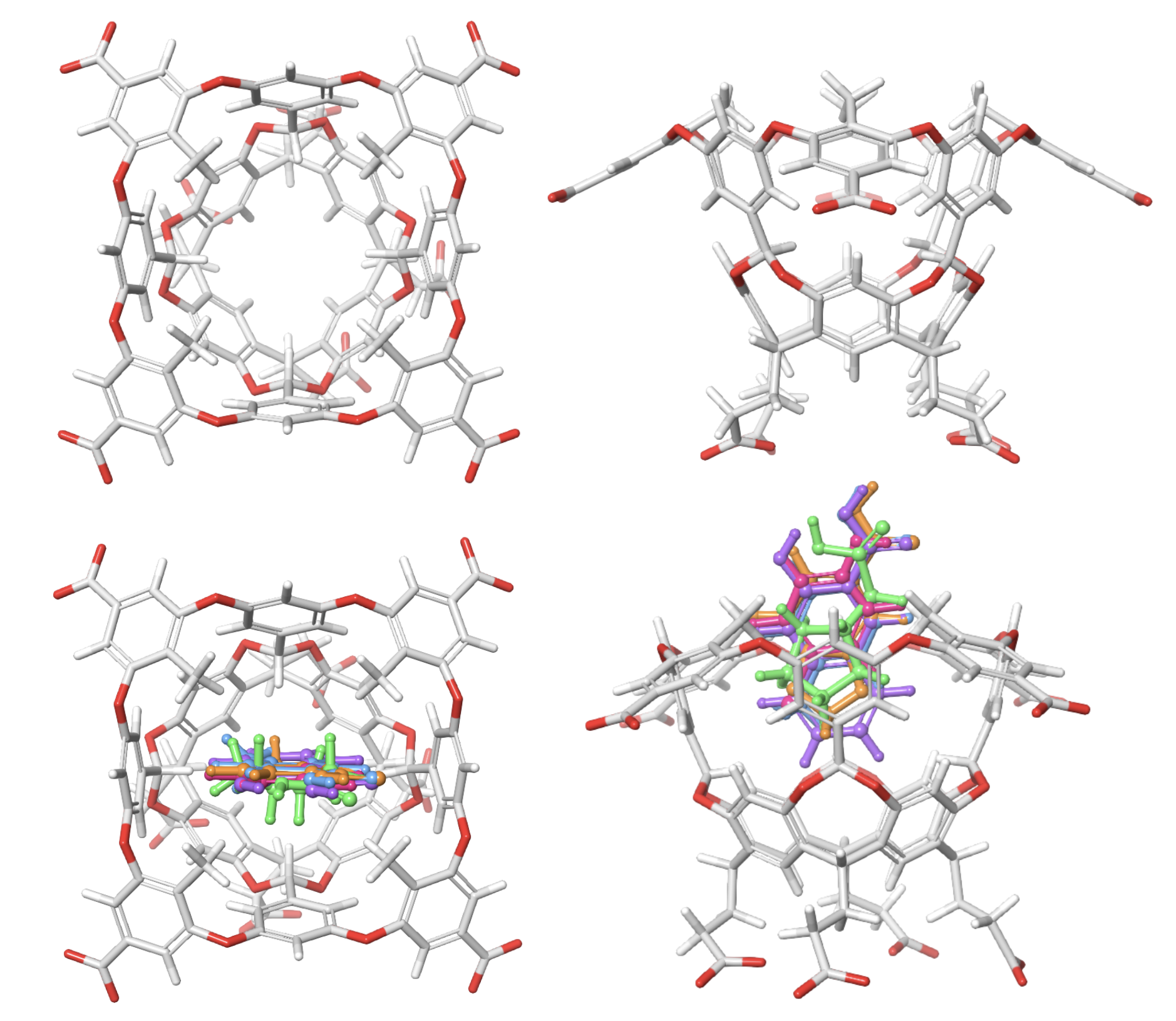}} \\
\subfloat{\includegraphics[width=5in]{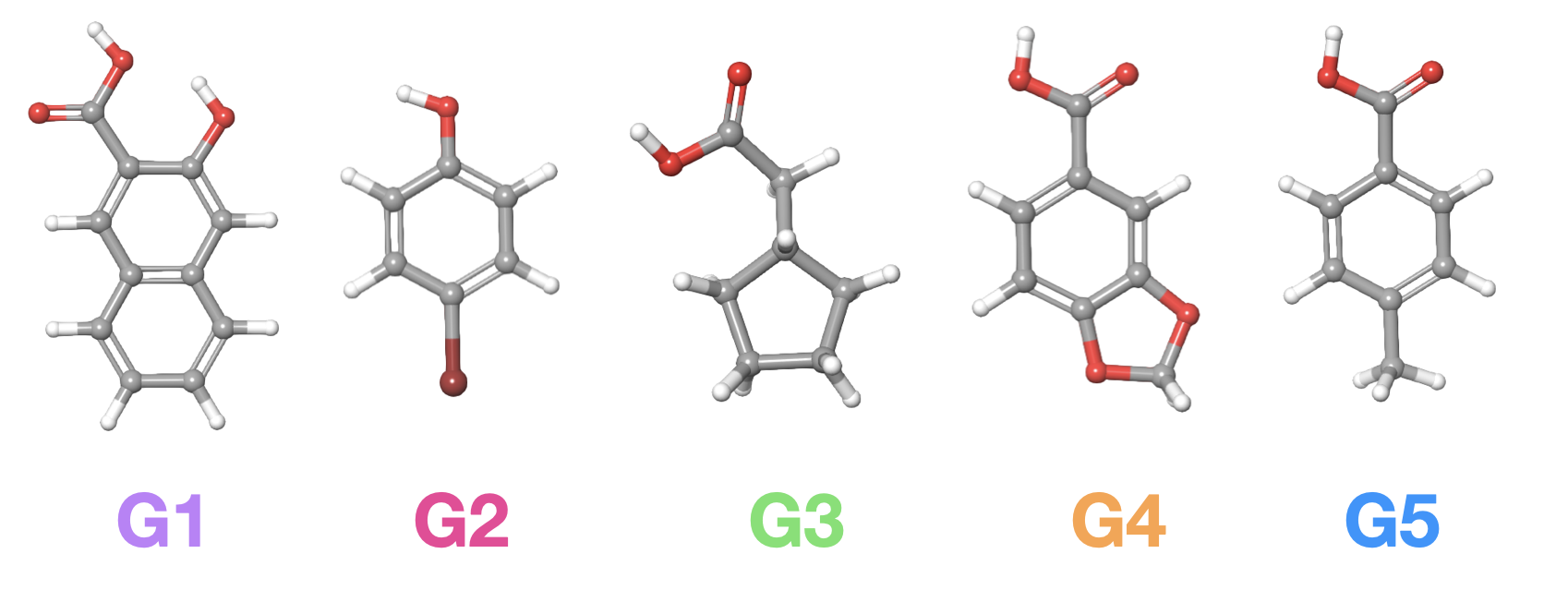}} \\
\captionsetup{justification=Justified}
    \caption{The SAMPL8 benchmark set considered in this work. Top: tetramethyl octa acid host (TEMOA) shown in licorice representation. Gray corresponds to carbon atoms, red to oxygen, and white to hydrogen. Middle: the five guests of the SAMPL8 benchmark set bound to TEMOA. Bottom: the chemical structures of the five guests, G1 to G5, shown in ball-and-stick (CPK) representation. Gray corresponds to carbon atoms, red to oxygen, brown to bromine, and white to hydrogen. The color of the labels corresponds to the color of the guests in the middle panel.}
\label{fig:chem_systems}
\end{figure}

We tested the proposed PDT theory for alchemical transfer on a subset of the SAMPL8 host-guest benchmark (Figure \ref{fig:chem_systems}). Specifically, we considered the binding of the five small guest compounds, named G1 to G5, plus water to the tetramethyl octa-acid (TEMOA) host.\cite{suating2020proximal} Experimentally, the SAMPL8 measurements at a pH where, with the exception of G2, the guests are expected to be deprotonated in solution and when bound to the host.\cite{azimi2022application} However, due to the difficulties of applying the double-decoupling process to ionized species,\cite{hahn2022bestpractices} in this work, we opted to carry out the numerical test of our theory on the neutral forms of the guests. The protocol was also tested on the transfer of one water molecule in water.

\subsection{System Setup and Simulation Settings}

The input files for the molecular simulations presented in this work are available in the GitHub repository listed in the Software and Data Availability section.  

In Eq.\ (\ref{eq:pert_pot}), we employ the softplus alchemical potential energy function\cite{pal2019perturbation,khuttan2021alchemical}
\begin{equation}
    W_\lambda[u_{\rm sc}(u)] = \frac{\lambda_2 - \lambda_1}{\alpha} \ln \left\{ 1 + e^{-\alpha [ u_{\rm sc}(u) -u_0 ])} \right\} + \lambda_2 \ u_{\rm sc}(u)
    \label{eq:softplus_pot}
\end{equation}
where the soft-core perturbation energy function $u_{\rm sc}(u)$ is defined as:
\begin{equation}
  u_{\rm sc}(u)=
\begin{cases}
u & u \le u_c \\
(u_{\rm max} - u_c ) f_{\rm sc}\left[\frac{u-u_c}{u_{\rm max}-u_c}\right] + u_c & u > u_c
\end{cases}
\label{eq:soft-core-general}
\end{equation}
with
\begin{equation}
f_\text{sc}(y) = \frac{z(y)^{a}-1}{z(y)^{a}+1} \label{eq:rat-sc} \, ,
\end{equation}
and
\begin{equation}
    z(y)=1+2 y/a + 2 (y/a)^2 .
\end{equation}

The soft-core perturbation energy function $u_{\rm sc}(u)$ is a monotonically increasing function of $u(x)=U_1(x)-U_0(x)$, designed to smoothly cap large values of the perturbation energy encountered along the alchemical transformation to the maximum value without affecting the end states.\cite{khuttan2021alchemical,wu2021alchemical,azimi2022relative}  In this work, we set  $u_c = 0$ kcal/mol and $u_{\rm max} = 50$ kcal/mol for the coupling and decoupling calculations. The transfer calculations for water-in-water and TEMOA-H$_2$O employed the parameters $u_c = 0$ kcal/mol and $u_{\rm max} = 50$ kcal/mol, whereas the transfer calculations for the guests to the host employed  $u_c = 100$ kcal/mol and $u_{\rm max} = 200$ kcal/mol. The $a$ parameter of the soft-core function was set to $0.0625$ in all cases. The softplus alchemical potential energy function above has been shown to eliminate or reduce alchemically-induced pseudo phase transitions that slow down the convergence of the free energy estimate.\cite{pal2019perturbation,khuttan2021alchemical,lu2013order} 

The parameters ($\lambda_1$, $\lambda_2$, $\alpha$, and $u_0$) of the softplus alchemical perturbation energy function [Eq.\ (\ref{eq:softplus_pot})] are functions of $\lambda$ and vary along the alchemical transformation according to a set schedule (see the Appendix). The softplus alchemical perturbation function reduces to the standard linear form $\lambda u_{\rm sc}$ when the schedule is such that $\lambda_1 = \lambda_2 = \lambda$.  The TEMOA-H$_2$O coupling and decoupling calculations employed a linear alchemical schedule. All other alchemical transformations employed non-linear schedules to accelerate conformational mixing (see Software and Data Availability section). The analytical models' parameters for $p_0(u)$ are independent of the alchemical schedules.

The host-guest systems were prepared from the original MOL2 files provided by the SAMPL8 organizers at  {\tt https://github.com\-/samplchallenges\-/SAMPL8\-/tree\-/master\-/host\_guest\-/GDCC}. All five guests were protonated and manually placed into the inner cavity of the TEMOA host with their polar ends directing out of the cavity (Figure \ref{fig:chem_systems}) using Maestro (Schr\"{o}dinger, Inc.). Force-field parameter assignments with the GAFF1.8/AM1-BCC force field and TIP3P solvation of the systems were performed using AmberTools 19 and the LEaP program.

\begin{figure}
    \centering
    \includegraphics[scale=0.42]{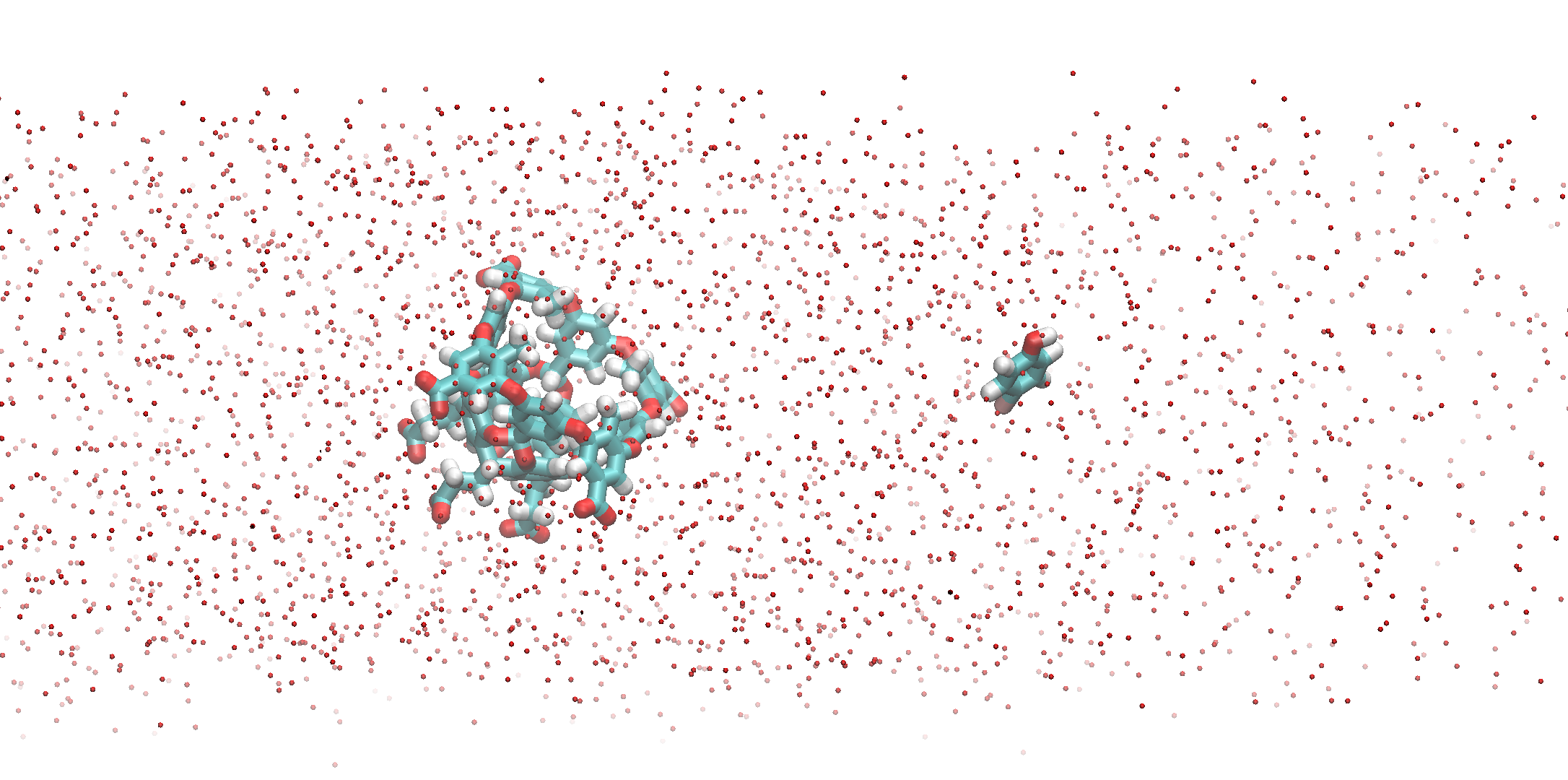}
    \caption{The water slab system that all calculations were conducted in. The system shown here is TEMOA and G2. Structures are styled in licorice, in which cyan are carbon atoms, red oxygen, white hydrogen, and light pink bromine. }
    \label{fig:slab}
\end{figure}

The simulations were conducted in a water slab (Figure \ref{fig:slab}) of approximate dimensions $40 \times 60 \times 42$ \AA $^3$ embedded in a $40 \times 60 \times 142$ \AA $^3$ periodic simulation box. The resulting system contains layers of water slabs of 42 \AA\ thickness separated by 100 \AA-thick vacuum regions along the $z$ direction. The evaporation of water molecules from the slab was prevented by imposing a flat-bottom harmonic restraint to the oxygen atoms of the water molecules along the $z$-direction with a force constant of 1.9 kcal/mol \AA$^2$ and a tolerance of 21 \AA\ from the center of the slab. The water solvent in the slab was minimized and thermalized at 300 K. In the alchemical decoupling calculations, the guest was transferred from the water slab or the host binding site to a position in the vacuum region displaced by 70 \AA\ along the $z$-direction. The alchemical transfer calculations employed a 30 \AA\ displacement vector along the $x$-direction parallel to the slab to bring the guest from the solvent to the binding site of the host. 

The heavy atoms of the lower cup of the TEMOA host, which were the first 40 atoms of the host as listed in the SAMPL8-provided files, were restrained by a flat-bottom harmonic potential with a force constant of 25  kcal/mol \AA$^2$ and a tolerance of 0.5 \AA.\cite{azimi2022application} A flat-bottom harmonic potential between the centers of mass of the host and the guest with a force constant of 25 kcal/mol \AA$^2$ was applied for a distance greater than $4.5$ \AA\ to define the binding site region ($V_{\rm site}$).  The corresponding ideal binding free energy term\cite{Gallicchio2011adv}
\begin{equation}
    \Delta G^\circ_{\rm id} = -k_B T \ln C^\circ V_{\rm site}
\end{equation}
is equal to $0.87$ kcal/mol in this case.

Alchemical transfer calculations started from a configuration at the alchemical intermediate obtained by slow annealing to $\lambda = 1/2$ in 250 ps starting from the initial state at $\lambda =0$ prepared by conventional energy minimization, thermalization, and relaxation steps. The annealing step establishes a suitable initial configuration of the system at $\lambda = 1/2$ without severe unfavorable repulsive interactions that would otherwise hamper molecular dynamics simulations. 

Asynchronous Hamiltonian molecular dynamics replica exchange\cite{Gallicchio2008, gallicchio2015asynchronous} with a 2 fs time-step and 22 replicas in $\lambda$-space was employed for conformational sampling using the AToM-OpenMM software package.\cite{AToM-OpenMM} Replicas were cycled in and out of the GPU devices every 40 ps. Perturbation energy samples and trajectory frames were saved with the same frequency.
Each replica was simulated for 60 ns.  Free energies and their corresponding uncertainties were estimated using UWHAM thermodynamic reweighting\cite{UWHAM} after discarding 1/3 of the initial trajectory.

\subsection{Parameter Optimization Protocol}

We obtained optimized parameters of the analytical model for $p_0(u)$ by means of the maximum-likelihood analysis of the distributions of perturbation energy samples of all $\lambda$-states from the corresponding alchemical molecular dynamics calculations.\cite{kilburg2018analytical} Specifically, because the calculations provide samples of the soft-core perturbation energies $u_{\rm sc}$, we consider the probability density of the soft-core perturbation energy given by
\begin{equation}
    p_{0}(u_{\rm sc})=p_{0}(u)/ u_{\rm sc}'(u)
    \label{fig:p0sc}
\end{equation}
where $u = u(u_{\rm sc})$ is the value of the inverse of the soft-core function at $u_{\rm sc}$ [Eq.\ (\ref{eq:soft-core-general})]  and $u_{\rm sc}'(u) \ge 0$ is the derivative of the soft-core function. Even though they are different functions, to simplify the notation, here we use the same symbol, $p_0$, for the probability density functions of the original, $p_0(u)$ and soft-core, $p_0(u_{\rm sc})$, perturbation energies and use their arguments to distinguish them.

The cost function is then expressed in terms of the likelihood function $\mathcal{L} (\theta)$  as
\begin{equation}
-\log \mathcal{L} (\theta) = - \sum_i \log p_{\lambda_i}(u_{{\rm sc},i} | \theta)
\label{eq:likelihood-func}
\end{equation}
where $\theta$ represents the collection of the parameters of the model that we seek to optimize and, $p_{\lambda_i}(u_{{\rm sc}} | \theta )$ is the analytical expression of the probability density of $u_{\rm sc}$ at the alchemical state at $\lambda = \lambda_i$. The latter is obtained from Eq.\ (\ref{eq:pdt}) using the perturbation energy function $W_{\lambda}[u_{\rm sc}(u)]$. The sum in Eq.\ (\ref{eq:likelihood-func}) runs over the samples from the molecular simulations at all $\lambda$-states, where $u_{{\rm sc},i}$ denotes the soft-core perturbation energy of the sample and $\lambda_i$ the value of $\lambda$ of the alchemical states from where the sample was collected.\cite{kilburg2018analytical}

\begin{figure}[h!]
    \subfloat[\centering Coupling]{\includegraphics[width=5.5in]{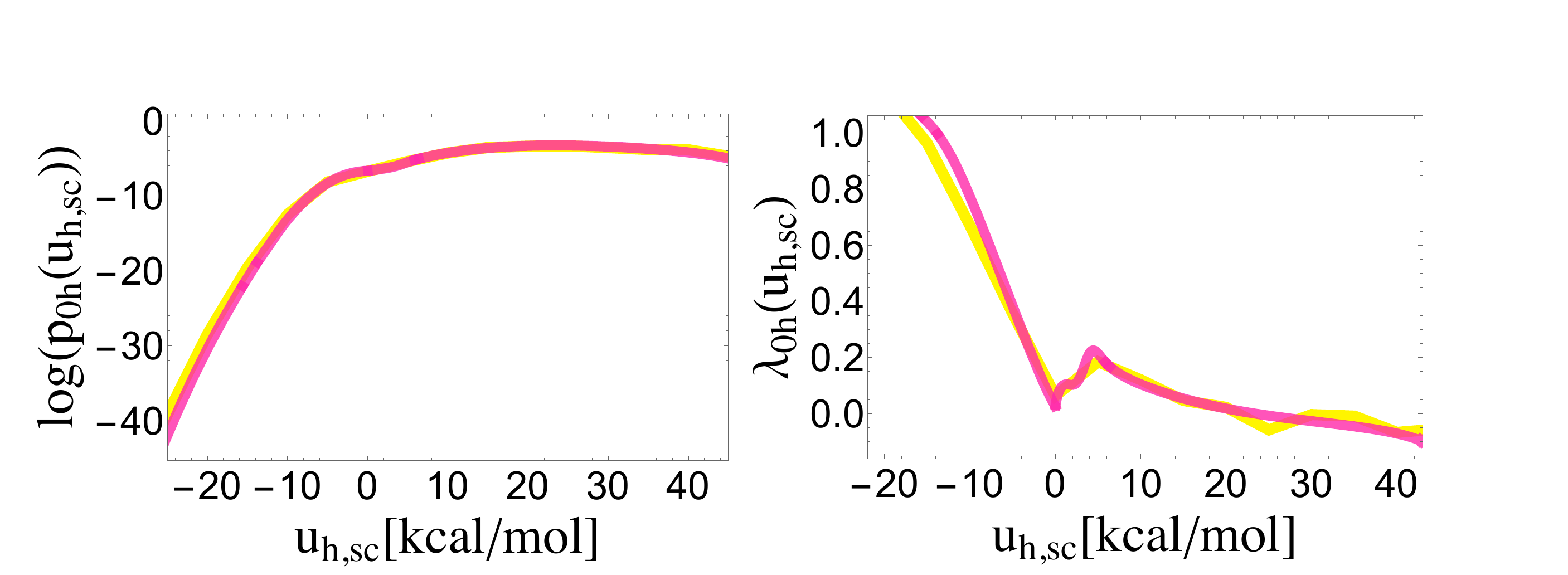}} \\
    \subfloat[\centering Solvation]{\includegraphics[width=5.5in]{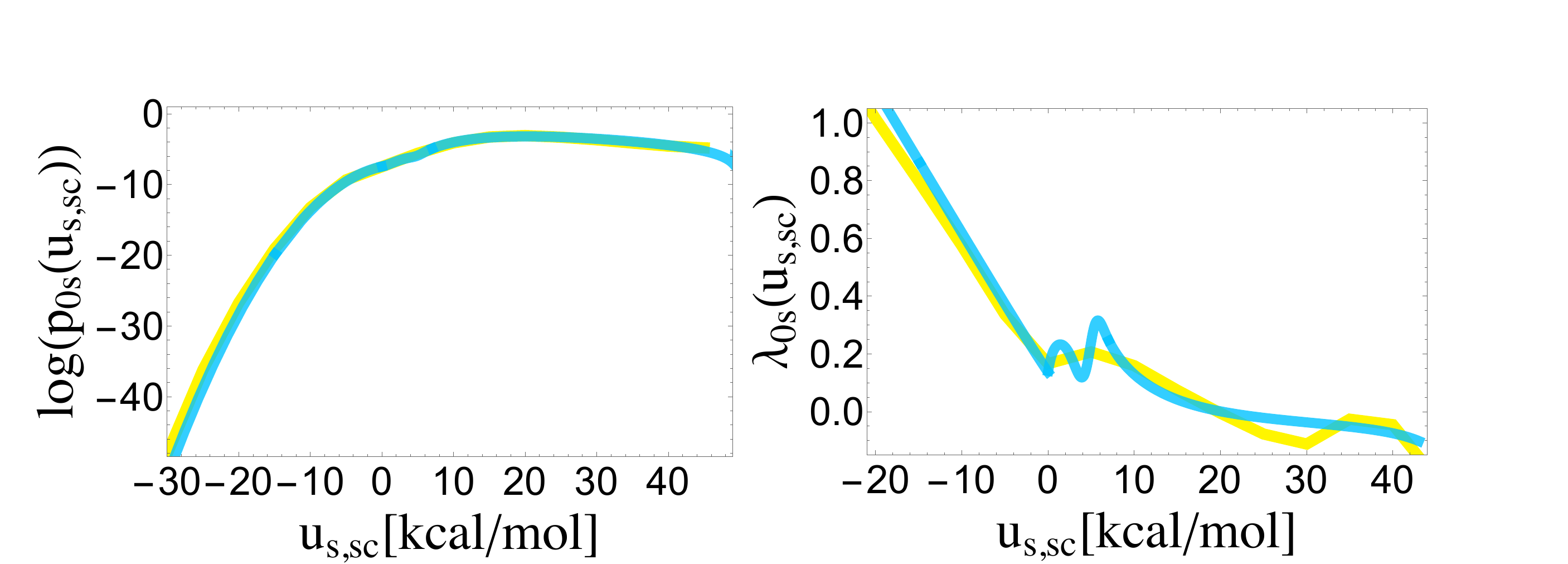}} \\
    \captionsetup{justification=Justified}
    \caption{The $\log p_{0{\rm h}}(u_{\rm h,sc})$ and  $\log p_{0{\rm s}}(u_{\rm s,sc})$ functions (left) and the  $\lambda_{0{\rm h}}(u_{\rm h,sc})$ and  $\lambda_{0{\rm s}}(u_{\rm s,sc})$ $\lambda$-functions (right) of the soft-core perturbation energies for (a) the coupling of water to TEMOA and (b) the solvation of water in water. The yellow curves are Gaussian kernel estimates from the samples of the alchemical molecular simulations. The red and blue curves are from the corresponding optimized analytical models; red is for host coupling and blue for solvent coupling.
   }
    \label{fig:logP0_lambdf}
\end{figure}

Initial guesses for the parameters of the analytical model for $p_0(u)$ were derived from the shapes of kernel density estimates of the $\log p_0(u_{\rm sc})$ function and of the corresponding $\lambda$-function\cite{pal2019perturbation,khuttan2021alchemical}
\begin{equation}
    \lambda_0(u_{\rm sc}) = k_B T \frac{d \log p_0(u_{\rm sc})}{du_{\rm sc}}
\end{equation}
(see Figure \ref{fig:logP0_lambdf}). Briefly, the UWHAM statistical inference analysis provides a statistical weight $W_{0,i}$ to each sample $i$ that represents the probability of observing it at $\lambda=0$, even though it might have been collected at some other $\lambda$-state. We obtained $p_0(u_{\rm sc})$ using the weighted Gaussian kernel estimate  
\begin{equation}
p_{0}(u_{\rm sc}) = \sum_i W_{0,i} \mathcal{N}(u_{\rm sc}| u_{{\rm sc},i}, \sigma)
\end{equation}
where $ \mathcal{N}(u | {\bar u}, \sigma)$ is the normal distribution with mean  ${\bar u}$ and standard deviation $\sigma$. The kernel estimate for the $\lambda$-function was obtained similarly using the derivative of the normal distribution as a kernel function. In this work, we performed Gaussian kernel estimates with  $\sigma = 1$ kcal/mol. 

The $\log p_0(u_{\rm sc})$ and $\lambda_0(u_{\rm sc})$ functions have characteristic shapes (see Figure \ref{fig:logP0_lambdf}) that provide information about the number of modes and their parameters. For example, following linear response, $\log p_0(u_{\rm sc})$ tends to vary quadratically at low energies when the system is nearly coupled. Similarly, the $\lambda$-function often varies linearly in this regime, and the onset and slope of the curve provide estimates for the analytical model's mean and standard deviation parameters. Conversely, deviations from quadratic and linear behaviors can be ascribed to contributions from multiple modes and can provide information about their parameters and relative statistical weight. The relationship between the collisional parameters ($b$, $\epsilon$, $\tilde u$, and $n_l$) and the shapes of the  $\log p_0(u_{\rm sc})$ and $\lambda_0(u_{\rm sc})$ is less obvious. Nevertheless, it was helpful to find initial guesses by studying the effect of varying the parameters on the agreement between the calculated and analytical curves. We used the Mathematica program (Wolfram, Inc.) for this purpose. The $\log p_0(u_{\rm sc})$ and $\lambda_0(u_{\rm sc})$ functions for the other systems in this work are in the Supplementary Information, section A. 

After finding initial guesses for the number of modes and their parameters, the parameters of the analytical model were refined by minimizing the cost function (Eq. \ref{eq:likelihood-func}) using a protocol implemented in TensorFlow available at {\tt https://github.com\-/Gallicchio-Lab\-/femodel-tf-optimizer}.\cite{kilburg2018analytical} The procedure involves the numerical integration steps to evaluate, for example, Eq.\ (\ref{eq:p0u_conv}). These were performed by Gauss-Hermite quadrature using 19 nodes. Parameter optimization was performed on the same set of perturbation energy samples used for the estimation of free energies.

\section{Results}

Here, we present a series of results to illustrate that the statistical behavior of alchemical transfer processes can be predicted by analyzing double-decoupling processes. The results are arranged as follows. First, a simple proof-of-principle case study is demonstrated for the transfer of a water molecule from one location in the solvent to another. We then present the results of double-decoupling alchemical calculations for the SAMPL8 series of TEMOA host-guest complexes and the analytical models of the alchemical probability densities. In each case, we show that the free energies and analytical models for the corresponding alchemical transfer processes obtained by the convolution of the decoupling processes agree with the alchemical transfer simulations' results.  

\subsection{Transfer of Water in Water}

The transfer free energy of a water molecule can be computed by first transferring the water molecule from solution to vacuum and then coupling it back to solvent (double-decoupling process). The transfer free energy can also be obtained by directly moving the water molecule from one location in the solvent to another (alchemical transfer process). Because the solvent is uniform, the initial and final states of the transfer process are thermodynamically equivalent, and the corresponding free energy is zero. Due to this symmetry, the decoupling legs from the initial and final states and the alchemical transfer's forward and reverse legs are also equivalent. Hence, we will present only one example of each.

This case study serves as a proof-of-principle test for investigating the hypothesis posed in this work. The alchemical coupling simulation yields the parameters of the analytical model for $p_{0{\rm s}}(u_{\rm s}$), the probability density of the interaction energy between the distinguished water molecule and the rest of the solvent. The optimized parameters of the analytical model for water coupling are listed in Table \ref{tab:water-water-models} under the ``H$_2$O coupling'' header. They indicate that the coupling of water to water is well described by one mode, that the mean and standard deviations of the background interaction energy in the decoupled ensemble are $2.41$ and $3.46$ kcal/mol, respectively (the  $\bar{u}_{0}$ and $\sigma$ parameters), that the probability of finding a configuration free of clashes is $5.77 \times 10^{-3}$ (the $b$ parameter), that the effective Lennard-Jones $\epsilon$ parameter for collisions is $3.9$ kcal/mol, that the minimum collision energy is $3.9$ kcal/mol (the $\tilde{u}$ parameter), and that the effective average number of colliding atoms is $2.50$ (the $n_l$ parameter).

When  $p_{0{\rm s}}(u_{\rm s}) $ is transformed according to the Potential Distribution Theorem prescription [Eq.\ (\ref{eq:pdt})], the  model yields the probability densities $p_{\lambda{\rm s}}(u_{\rm s}$) of the water-solvent interaction energy $u_s$ as a function of the coupling parameter $\lambda$ (Figure \ref{fig:coupl-solv-upto-g2}b). The weakly coupled states near $\lambda = 0$ are characterized by wide distributions with long tails at high interaction energies characteristic of frequent and severe atomic clashes. The interaction energies plotted in Figure \ref{fig:coupl-solv-upto-g2}b and elsewhere are damped down by the soft-core function [Eq.\ (\ref{eq:soft-core-general})]. The raw interaction energies of these states can be orders of magnitude greater and the corresponding distributions stretch towards large values.  As the coupling increases, the distributions shift to lower interaction energies. The features that arise at interaction energies just greater than zero are artifacts due to the soft-core function that terminates there. For $\lambda$ states above approximately $1/2$, the probability densities assume a Gaussian shape and shift towards lower energies proportionally to $\lambda$ at constant width, as expected from linear response,\cite{simonson2002gaussian, Alper:Levy:93} until they reach the fully coupled state at $\lambda=1$ described by $p_{1{\rm s}}(u_{\rm s}$).

\begin{figure}[htbp]
\includegraphics[scale=0.25]{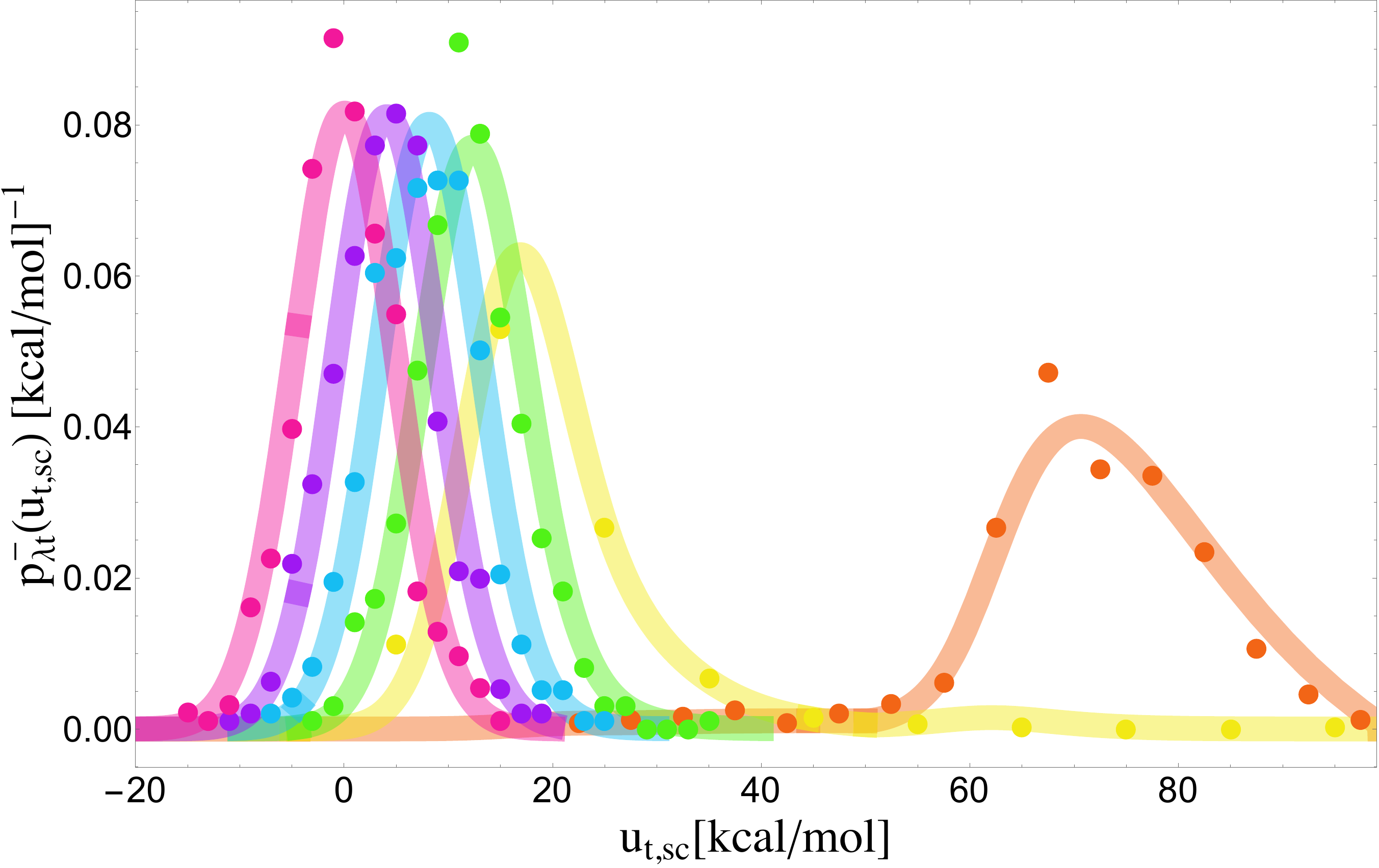}
\captionsetup{justification=Justified}
    \caption{The probability densities $p_{\lambda{\rm t}}(u_{\rm t,sc})$ of the soft-core perturbation energy of alchemical transfer collected from simulations of water transfer in water (dots) compared with the analytical probability densities (lines) predicted from the model of alchemical solvent coupling $p_{0{\rm s}}(u_{\rm s})$. $\lambda=0.1$ (yellow), $\lambda=0.2$ (green), $\lambda=0.3$ (blue), $\lambda=0.4$ (purple), and $\lambda=0.5$ (pink).}
    \label{fig:water-water-fit}
\end{figure}

Next, we take the convolution of $\tilde{p}_{{1}\rm s}(u_{\rm s}) = p_{1{\rm s}}(-u_{\rm s})$ and $p_{0{\rm s}}(u_{\rm s})$ that yields, according to Eq.\ (\ref{eq:p0t_model}), a model for the probability density $p_{0{\rm t}}(u_{\rm t})$ for the alchemical transfer of the water molecule from one place in the solvent to another. The parameters of the analytical model of $p_{0{\rm t}}(u_{\rm t})$ are listed in Table \ref{tab:water-water-models} under the ``H$_2$O transfer'' header. Furthermore, application of Eq.\ (\ref{eq:pdt}) to $p_{0{\rm t}}(u_{\rm t})$ yields analytical predictions of the probability densities of the transfer perturbation energies at all $\lambda$ values along the alchemical path. As shown in Figure \ref{fig:water-water-fit}, there is an excellent agreement between the analytical predictions of the $p_{\lambda{\rm t}}(u_{\rm t})$ alchemical transfer probability densities and the results of molecular simulation.

\begin{table*}[h!]
    \centering
    \captionsetup{justification=Justified}
  \caption{\label{tab:water_parameters}The parameters of the analytical model of alchemical coupling and alchemical transfer for a water molecule in explicit solvent.}
\begin{ruledtabular}
\begin{tabular}{lccccccc}
       & $w_i$               & $b$ & $\bar{u}_{0}$\footnote{kcal/mol} & $\sigma^{\rm a} $ & $\epsilon^{\rm a}$ & $\tilde{u}^{\rm a}$ & $n_l$ \\ \hline 
\multicolumn{8}{c}{H$_2$O coupling\footnote{from Table \ref{tab:solv_parameters}} } \tabularnewline
mode 1 & $1.0$ & $5.77 \times 10^{-3}$ & $2.41$             & $3.46$              & $3.9$           &   $3.9$          & $2.50$
       \tabularnewline
\multicolumn{8}{c}{H$_2$O transfer\footnote{Leg 1 and Leg 2}} \tabularnewline
mode 1 & $1.0$ & $5.82 \times 10^{-3}$ & $20.1$             & $4.89$              & $3.9$           &   $3.9$          & $2.50$
\end{tabular}
\end{ruledtabular}
\label{tab:water-water-models}
\end{table*}

\subsection{Binding Free Energy Estimates of the Host-Guest Complexes}

The results of the double-decoupling (DDM) and alchemical transfer (ATM) binding free energy calculations for the complexes of TEMOA with water and the five SAMPL8 guests are presented in Table \ref{tab:FE_estimates}. In general, the free energy estimates between the two methods are within statistical uncertainty. In particular, the two methods clearly agree for TEMOA-H$_2$O, TEMOA-G2, TEMOA-G4, and TEMOA-G5. The difference in the estimates for TEMOA-G1 and TEMOA-G3 is about 1 kcal/mol, which is just outside the level of confidence and could indicate slow convergence or a small bias. The statistical uncertainties of the DDM and ATM estimates are similar, considering that the ATM calculations are only half as computationally expensive as the combinations of the two coupling steps of the DDM calculations. 

G2 and G4 are the most favorable binders of the five guests, a rank that is consistently predicted by both DDM and ATM. The least favorable binder in the DDM calculations is G3, whereas ATM designates G1 as the weakest binder. Of the five guests, G1 is the bulkiest, containing 14 heavy atoms, and G3 is the only non-planar and non-aromatic guest. Evidently, binding either of these molecules to the cavity of TEMOA is a more challenging transformation than binding the G2, G4, and G5 guests, all of which contain a single aromatic ring. The double-decoupling free energies indicate that the higher affinities of the G2 and G4 guests are due to their stronger interactions with the host (the ``Host Coupling'' free energies in Table \ref{tab:FE_estimates}) relative to the other guests, while the variation of the hydration free energies (the ``Solvent Coupling'' free energies in Table \ref{tab:FE_estimates}), which oppose binding, are comparatively smaller. The computed binding free energies of the protonated guests presented here cannot be compared directly with the modeled and experimental binding affinities of the ionized forms of the guests reported earlier.\cite{azimi2022application,amezcua2022overview,suating2022nature} However, G2 and G4 are the strong binders in both cases, probably reflecting small variations of the guests' pKa's and the resulting ionization penalties.\cite{khuttan2023taming}  

Water's excess binding free energy, $\Delta G^\ast_{b,w} = -k_B T \ln \rho/\rho_0$, to the TEMOA cavity measures the average water density within the binding site volume relative to bulk.\cite{Widom1982} Taking into account the volume of the spherical binding site region (see Computational Details), the transfer free energy calculated of approximately $0.8$ kcal/mol for water (Table \ref{tab:FE_estimates}) is consistent with about three water molecules present within TEMOA's cavity displaced upon binding of the guest molecule. This number is consistent with direct observations.\cite{azimi2022application}

\begin{table*}[h!]
    \captionsetup{justification=Justified}
    \caption{Excess free energy estimates of the case studies in this work obtained by two methods: coupling/decoupling (DDM) and direct transfer (ATM). The difference between the host coupling and the solvent coupling free energies yields the DDM binding free energy of the designated ligand to host TEMOA. Uncertainties reported are twice the standard deviation.}
\begin{ruledtabular}
    \centering
    \begin{tabular}{lcccc}
   & Host Coupling\footnote{kcal/mol} & Solvent Coupling$^{\rm a}$ & DDM$^{\rm a}$ & ATM$^{\rm a}$  \\
   \hline

TEMOA-H$_2$O &
$-3.87 \pm 0.07$ & $-4.50 \pm 0.1$ & $0.628 \pm 0.1$ & $0.845 \pm 0.3$  

\tabularnewline
TEMOA-G$1$ &
$-14.2 \pm 0.1$ & $-2.51 \pm 0.1$ & $-11.7 \pm 0.2$ & $-10.5 \pm 0.2$ 

\tabularnewline
TEMOA-G$2$ &
$-17.0 \pm 0.09$ & $-4.32 \pm 0.1$ & $-12.7 \pm 0.1$ & $-12.7 \pm 0.2$

\tabularnewline
TEMOA-G$3$ &
$-14.3 \pm 0.1$ & $-3.50 \pm 0.1$ & $-10.8 \pm 0.2$ & $-11.6 \pm 0.3$

\tabularnewline
TEMOA-G$4$ &
$-17.6 \pm 0.1$ & $-3.54 \pm 0.1$ & $-14.1 \pm 0.2$ & $-13.9 \pm 0.3$

\tabularnewline
TEMOA-G$5$ &
$-15.1 \pm 0.1$ & $-3.19 \pm 0.1$ & $-11.9 \pm 0.2$ & $-11.7 \pm 0.3$ 
    
    \end{tabular}

    \label{tab:FE_estimates}
\end{ruledtabular}
\end{table*}

\subsection{Analytical Models for Coupling to the Host}

The binding of one water molecule to TEMOA from vacuum can be described analytically by a probability density function $p_{0h}(u_h)$ with two modes (Table \ref{tab:water_parameters}) of nearly equal statistical weight. The second mode corresponds to configurations that are more likely to clash with the host's atoms or water molecules in the cavity (the smaller $b$ parameter in Table \ref{tab:coupl_parameters}). The two modes probably reflect the position of the uncoupled water molecule either at the center or at the rim of the binding site volume--where it is more likely to find the host's atoms--or configurations that, by chance, have fewer or more water molecules bound to the apo form of the host. 

\begin{figure}
    \subfloat[\centering TEMOA-H$_2$O coupling]{{\includegraphics[width=3.2in]{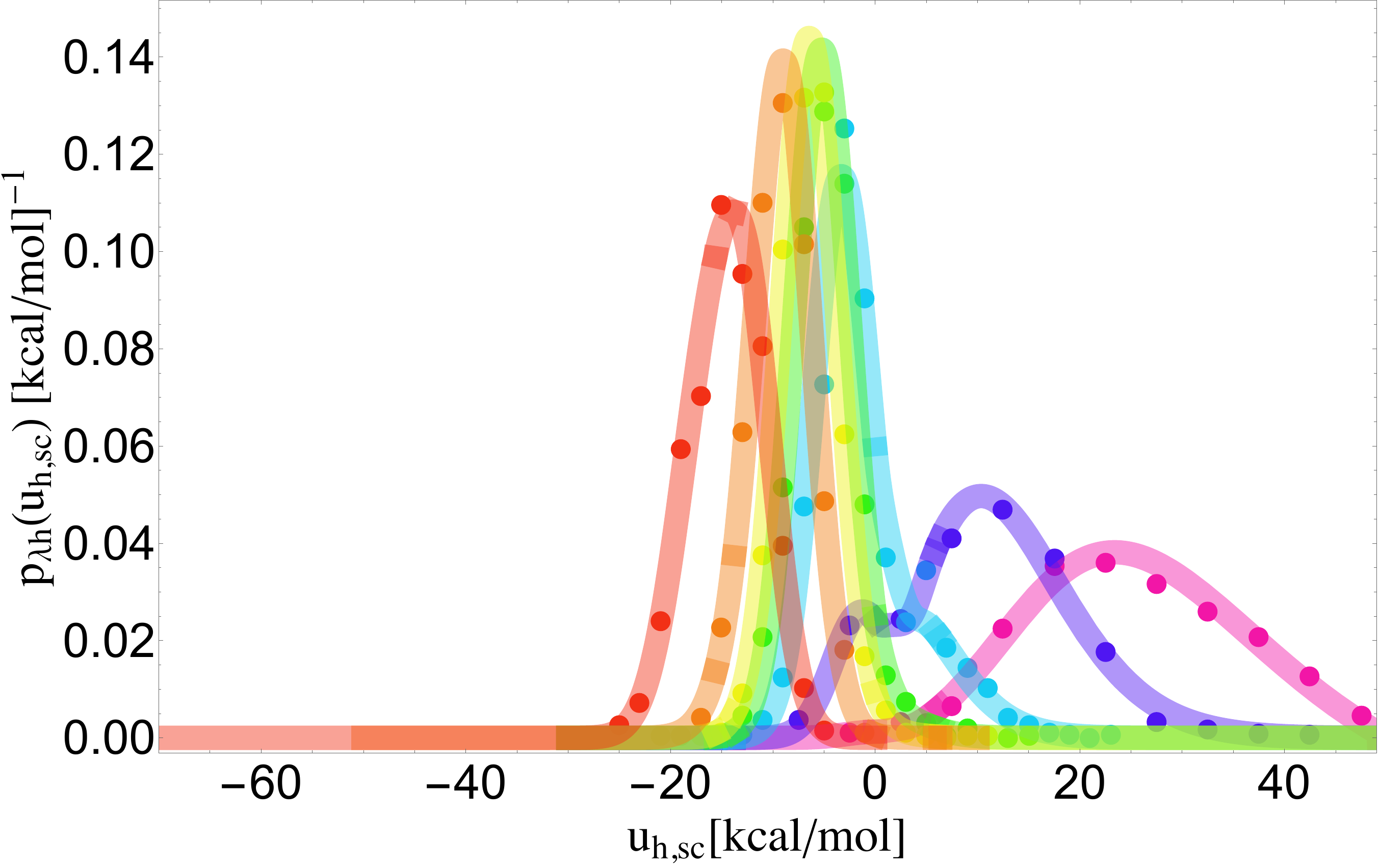} }} 
    \subfloat[\centering H$_2$O hydration]{{\includegraphics[width=3.2in]{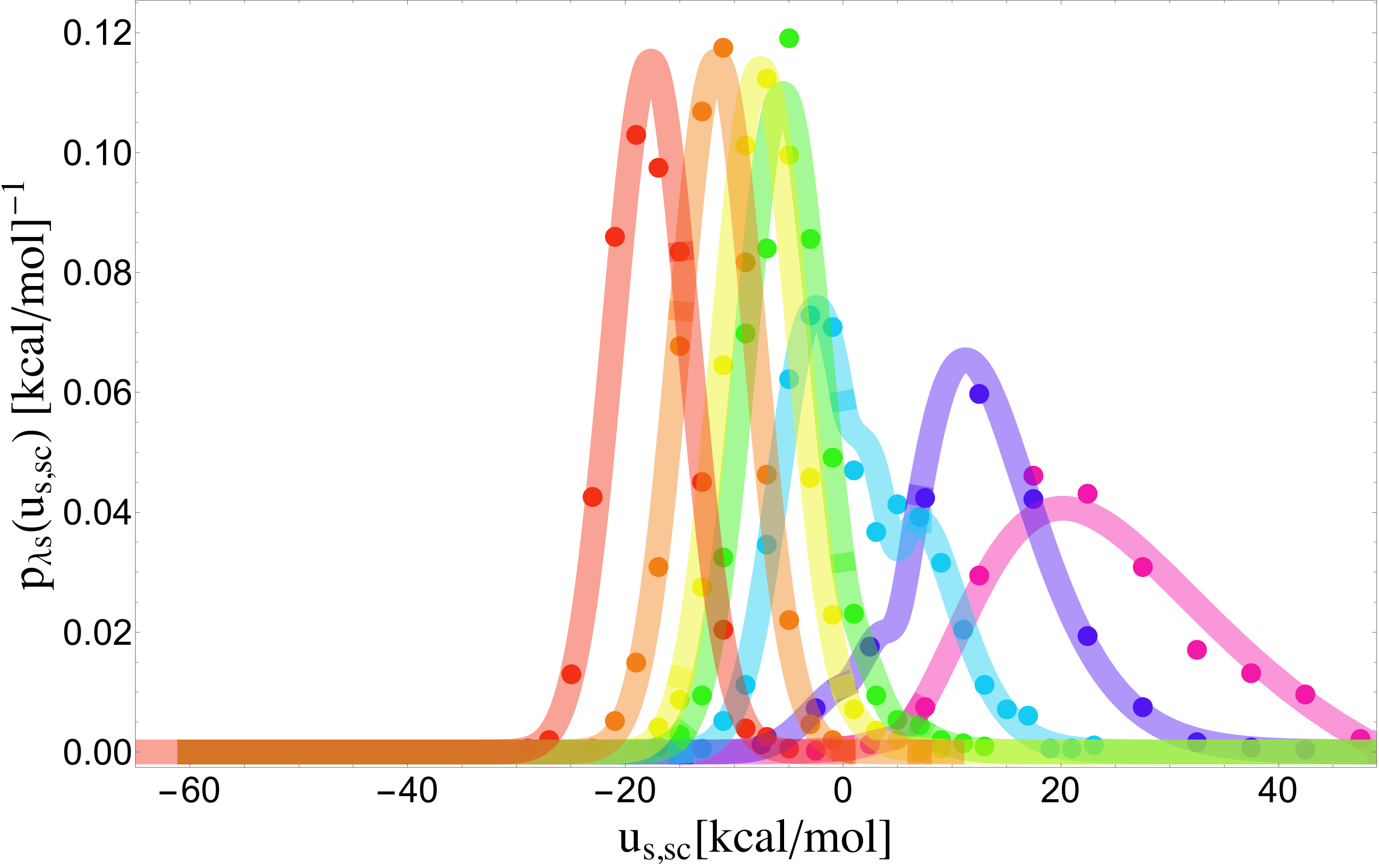} }} \\
    \subfloat[\centering TEMOA-G1 coupling]{{\includegraphics[width=3.2in]{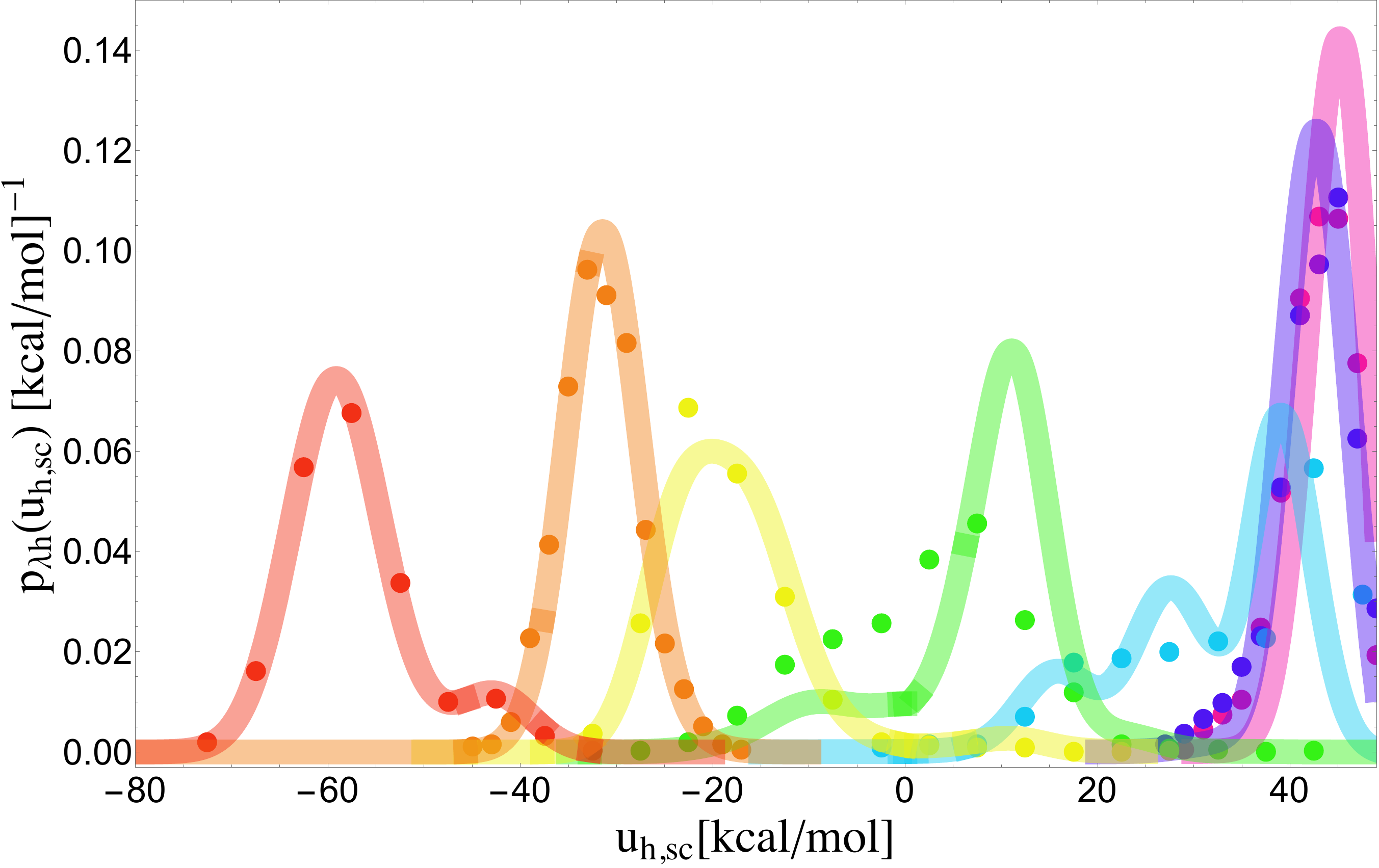} }} 
    \subfloat[\centering G1 hydration]{{\includegraphics[width=3.2in]{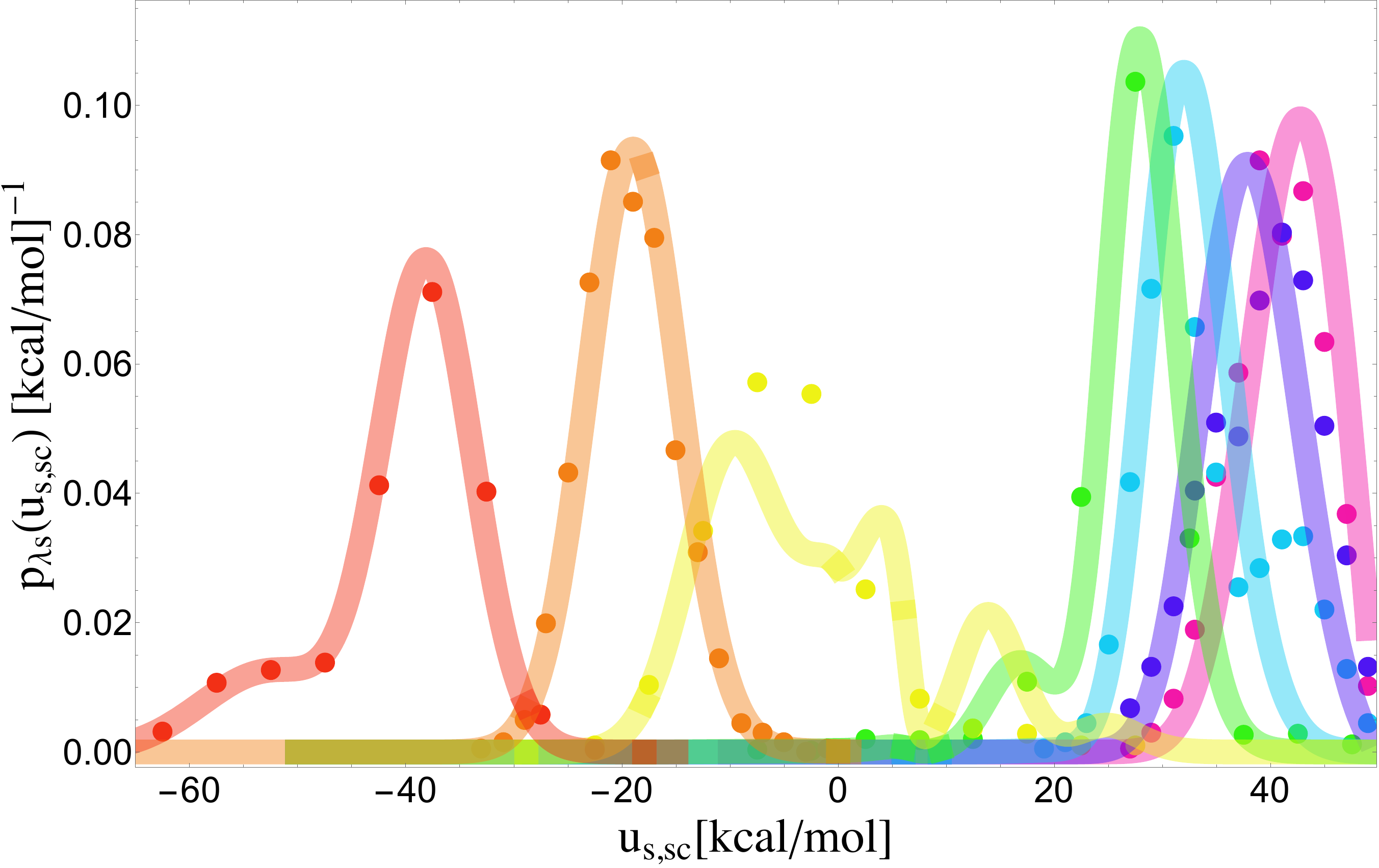} }} \\
    \subfloat[\centering TEMOA-G2 coupling]{{\includegraphics[width=3.2in]{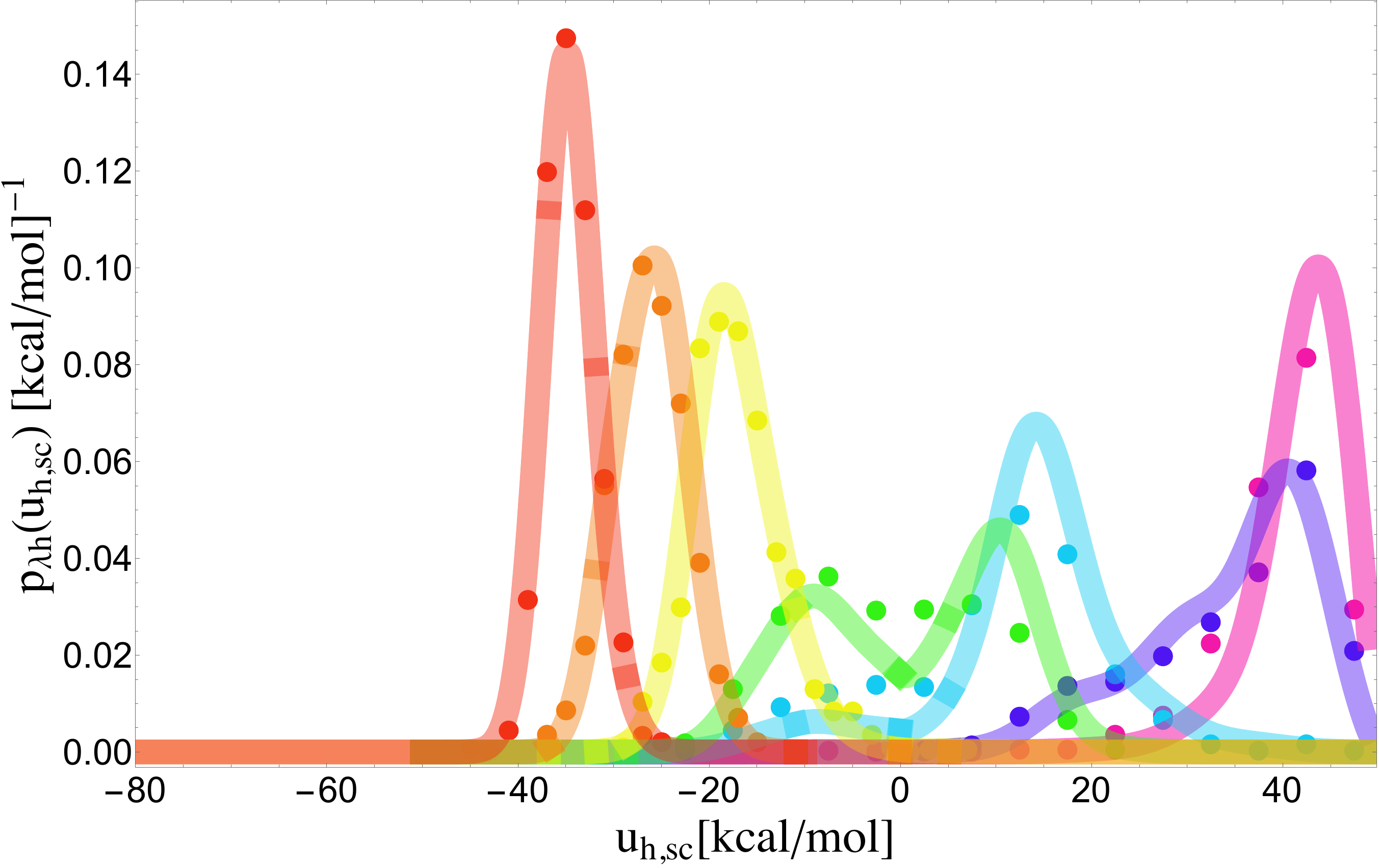} }} 
    \subfloat[\centering G2 hydration]{{\includegraphics[width=3.2in]{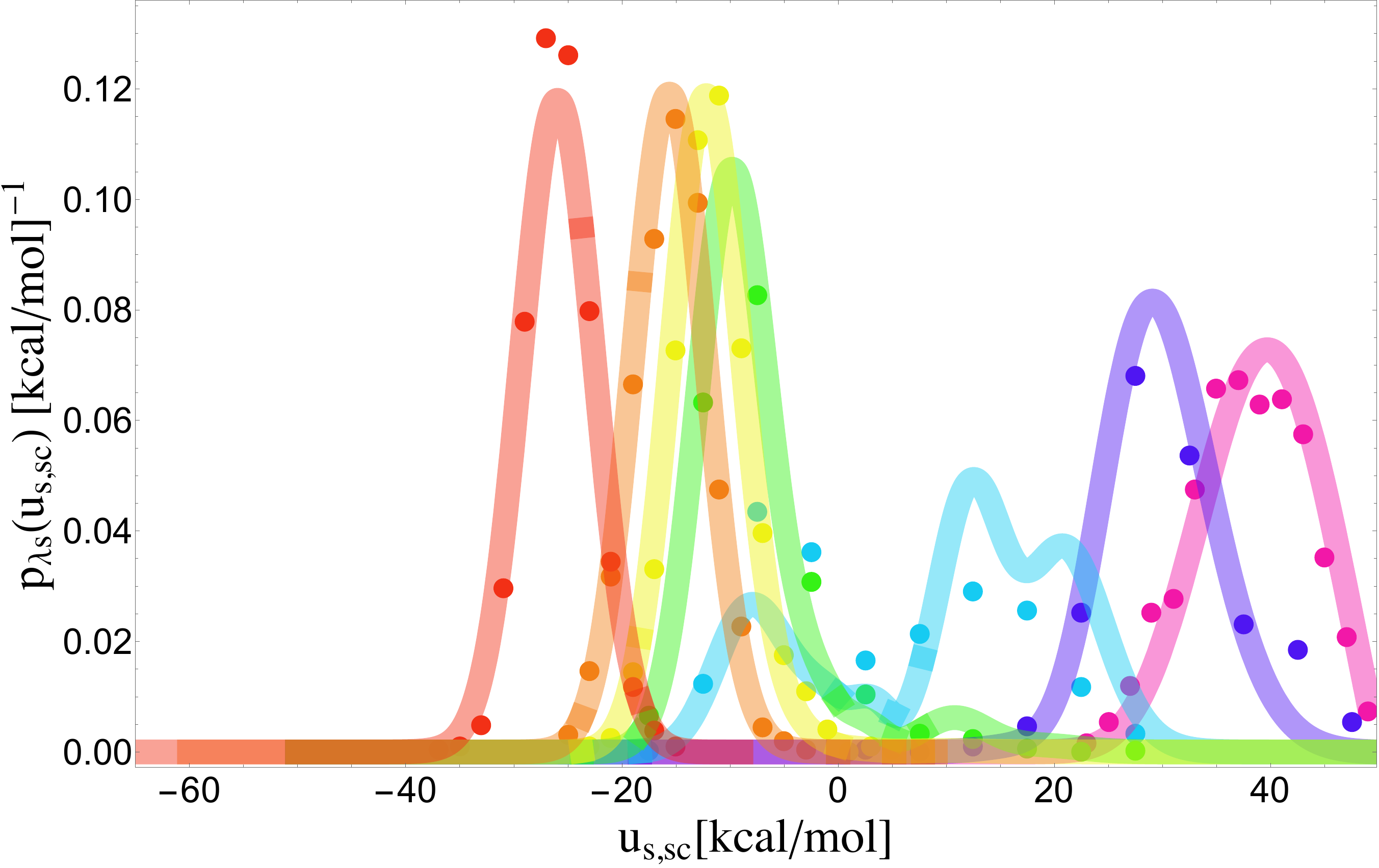} }}
\captionsetup{justification=Justified}
\caption{Probability densities  $p_{\lambda{\rm h}} (u_{\rm h,sc})$ and   $p_{\lambda{\rm s}} (u_{\rm s,sc})$ of the soft-core interaction energies for the coupling of a guest to TEMOA (left) and the coupling of a guest to water (right) collected from simulations (dots) and predicted from the analytical model's descriptions of the probability densities at the decoupled state  $p_{0{\rm h}} (u_{\rm h})$ and   $p_{0{\rm s}} (u_{\rm s})$ (lines). Each color corresponds to an alchemical state: pink $\lambda = 0$, purple $\lambda = 0.1$, blue $\lambda = 0.25$, green $\lambda = 0.4$, yellow $\lambda = 0.5$, orange $\lambda = 0.7$, and red $\lambda = 1$.}
\label{fig:coupl-solv-upto-g2}
\end{figure}

\begin{figure}
    \subfloat[\centering TEMOA-G3 coupling]{{\includegraphics[width=3.2in]{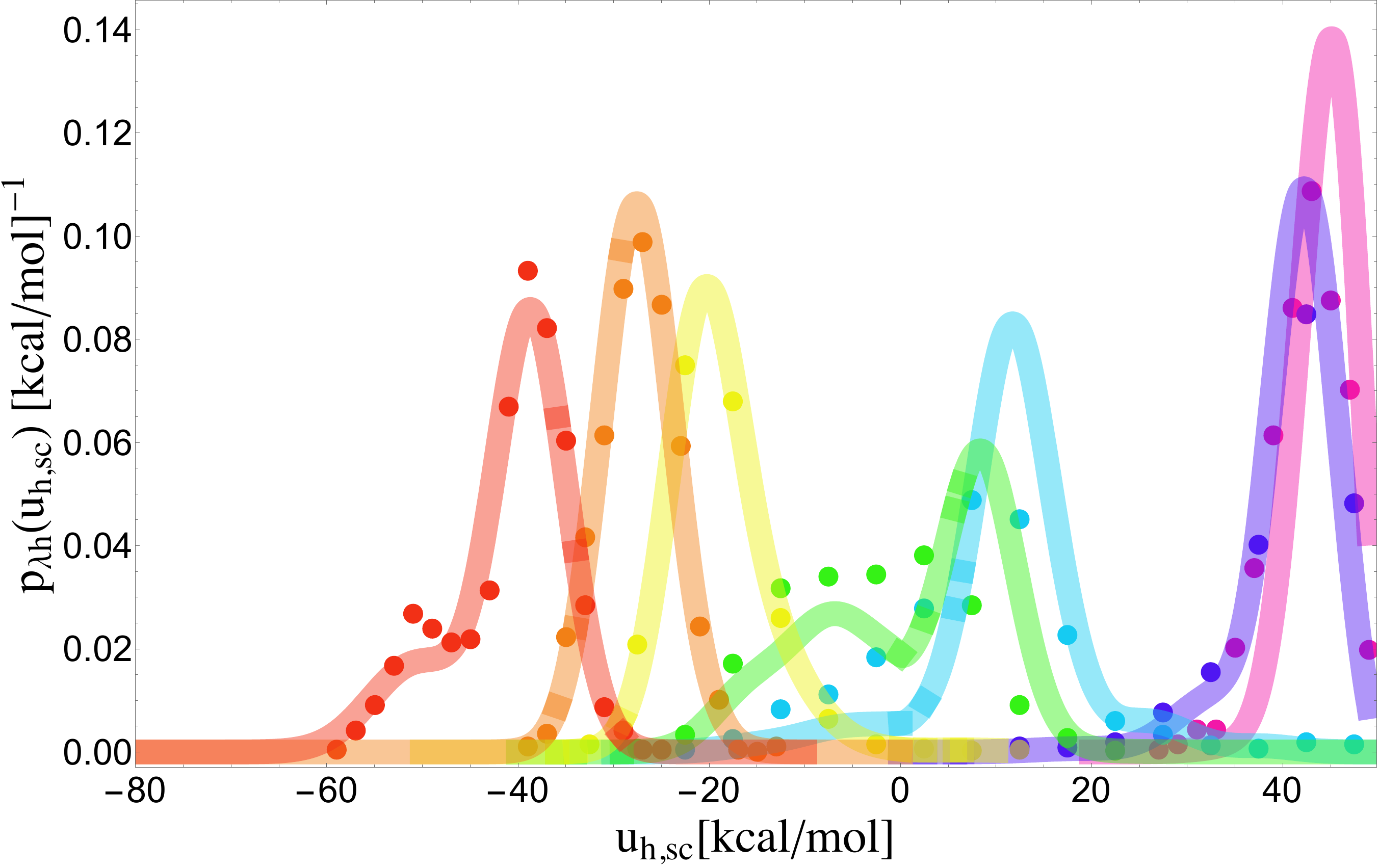} }} 
    \subfloat[\centering G3 hydration]{{\includegraphics[width=3.2in]{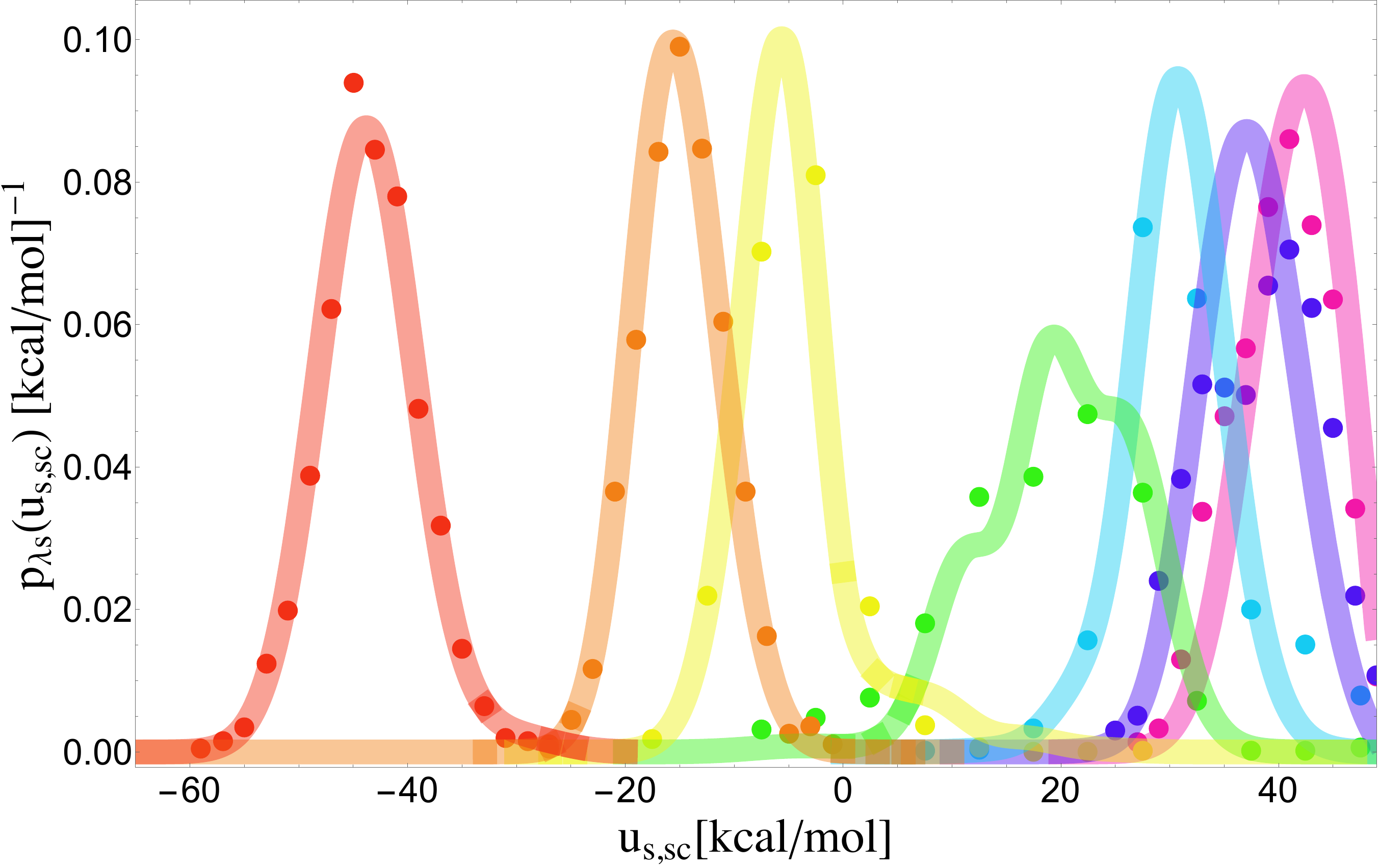} }}   \\
    \subfloat[\centering TEMOA-G4 coupling]{{\includegraphics[width=3.2in]{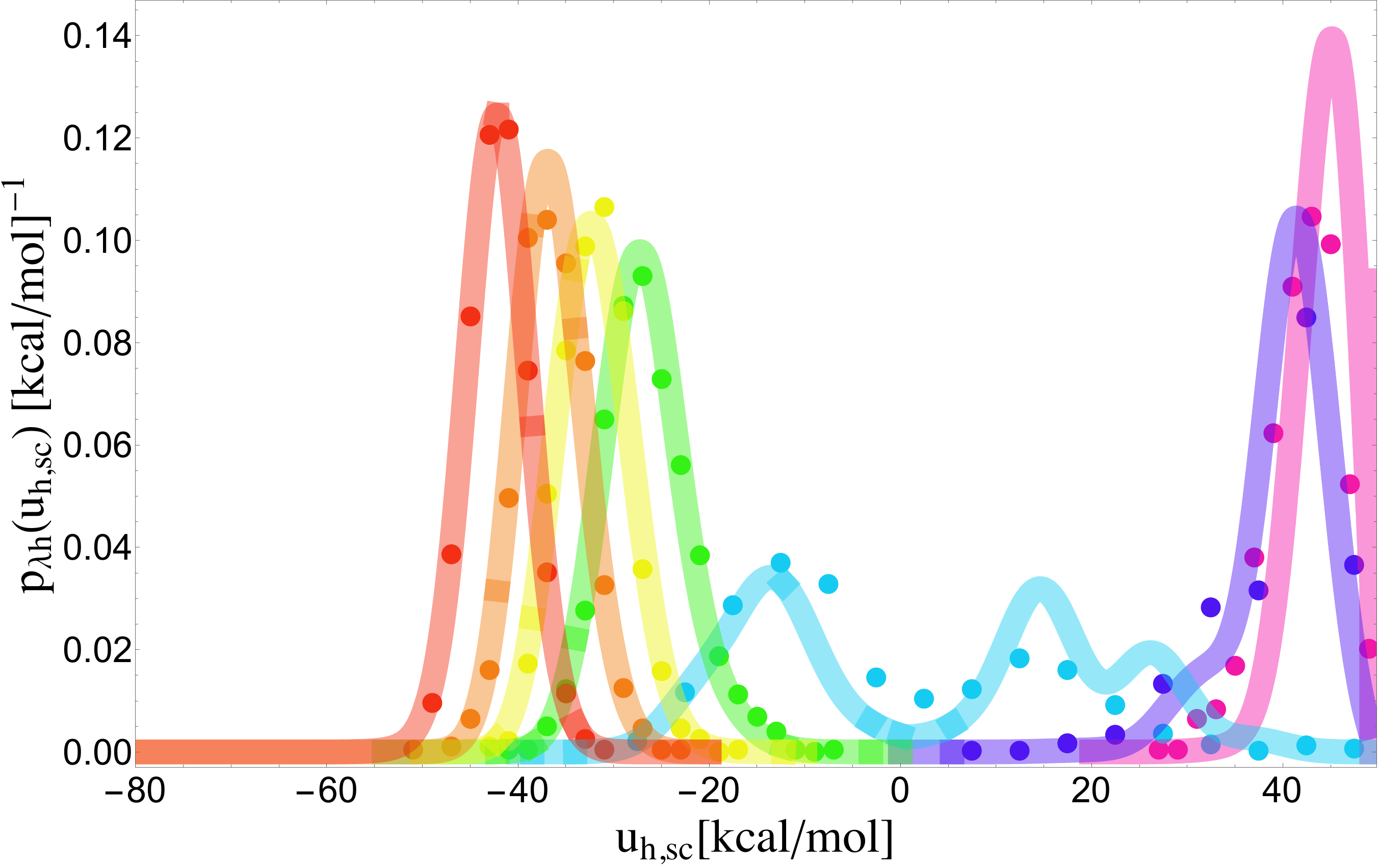} }} 
    \subfloat[\centering G4 hydration]{{\includegraphics[width=3.2in]{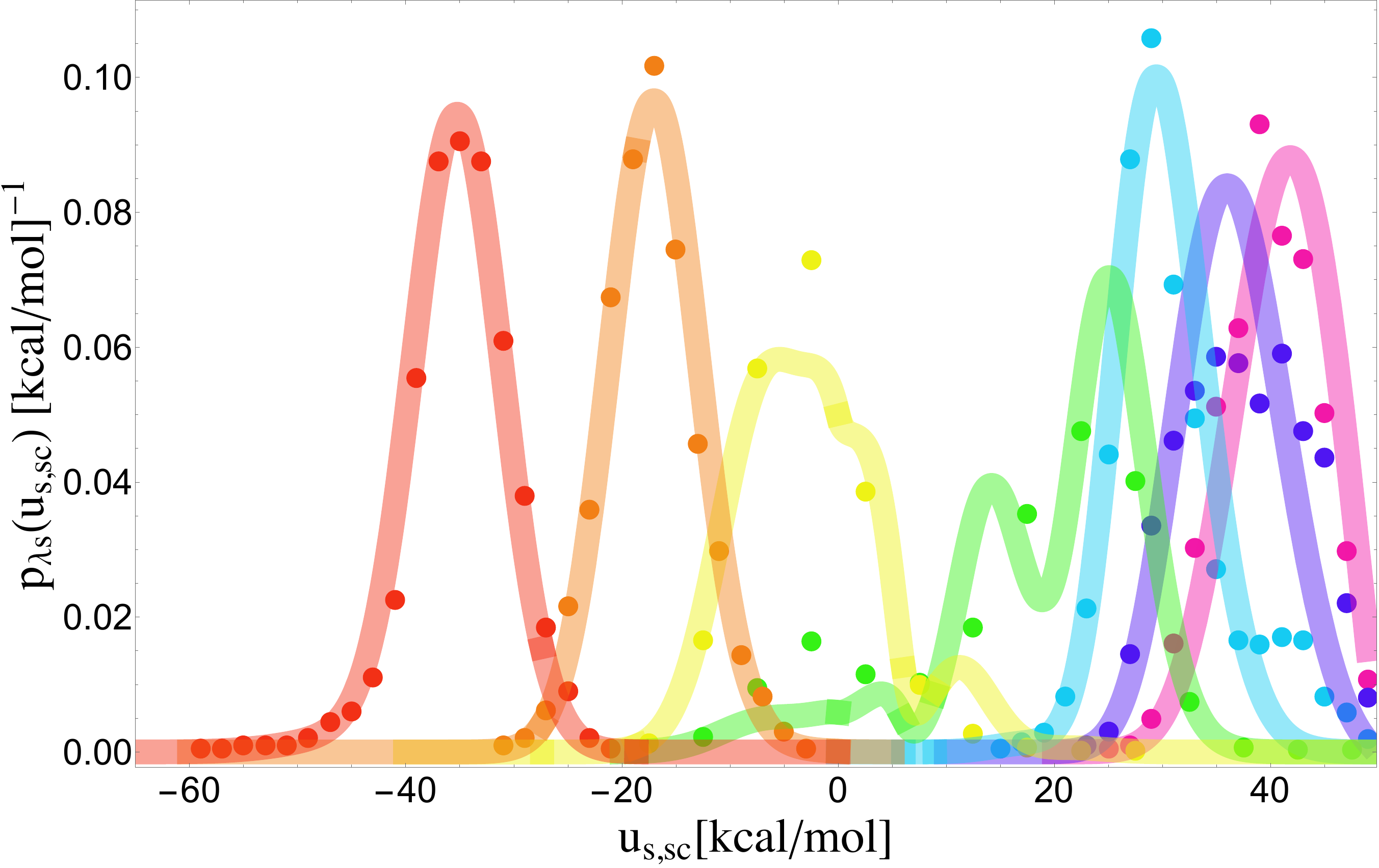} }} \\
    \subfloat[\centering TEMOA-G5 coupling]{{\includegraphics[width=3.2in]{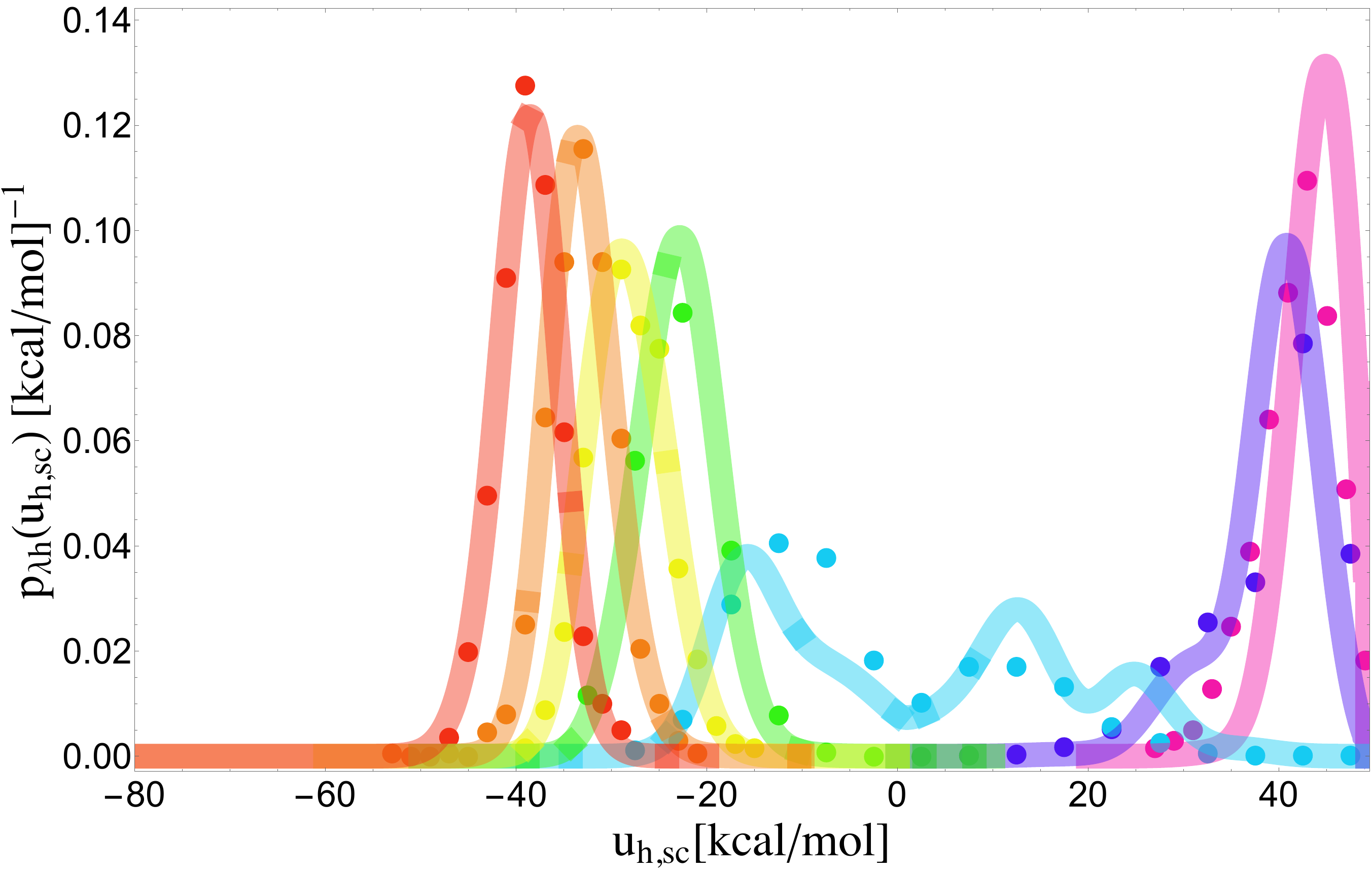} }} 
    \subfloat[\centering G5 hydration]{{\includegraphics[width=3.2in]{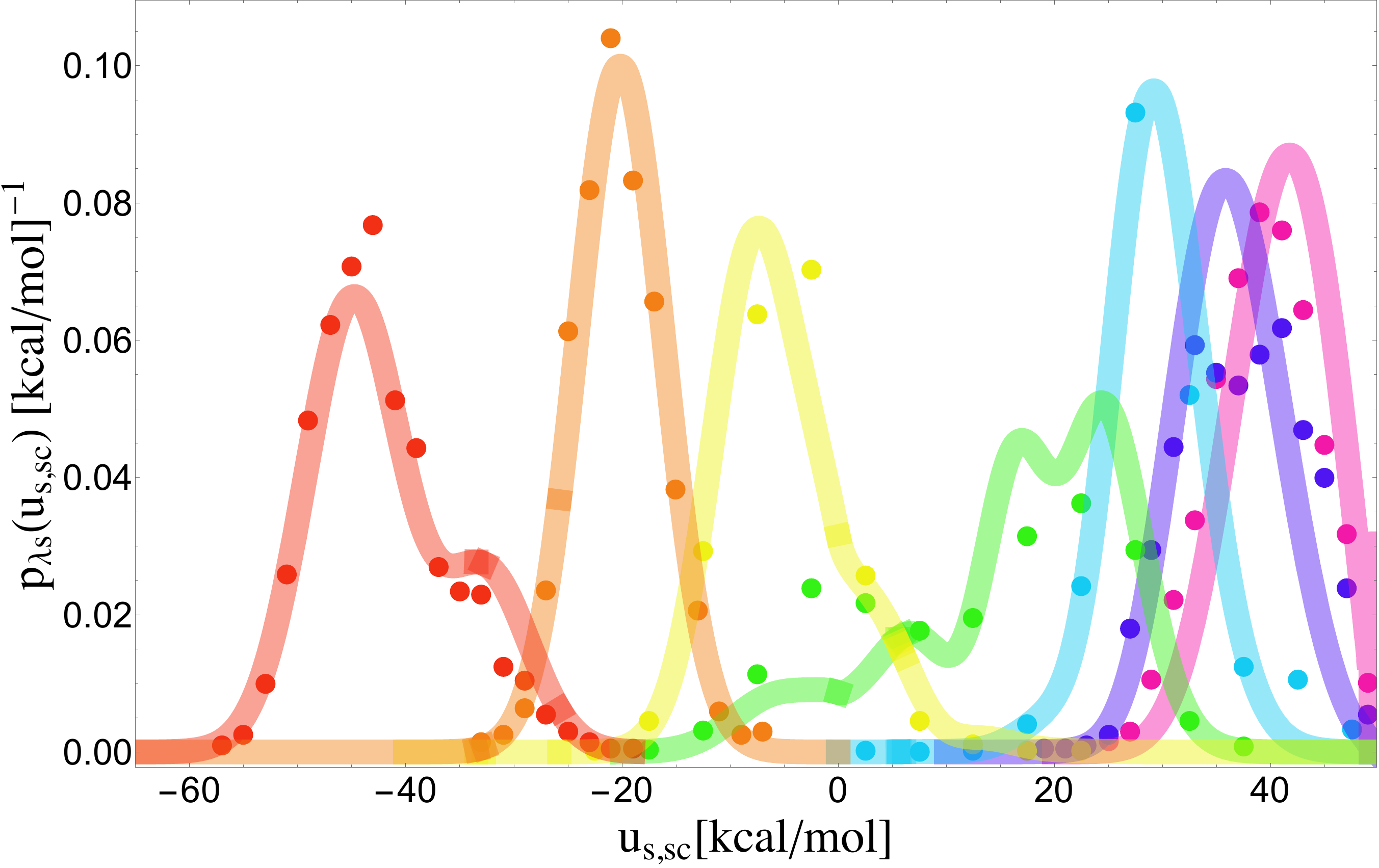} }} 
\captionsetup{justification=Justified}
\caption{Probability densities  $p_{\lambda{\rm h}} (u_{\rm h,sc})$ and   $p_{\lambda{\rm s}} (u_{\rm s,sc})$ of the soft-core interaction energies for the coupling of a guest to TEMOA (left) and the coupling of a guest to water (right) collected from simulations (dots) and predicted from the analytical model's descriptions of the probability densities at the decoupled state  $p_{0{\rm h}} (u_{\rm h})$ and   $p_{0{\rm s}} (u_{\rm s})$ (lines). Each color corresponds to an alchemical state: pink $\lambda = 0$, purple $\lambda = 0.1$, blue $\lambda = 0.25$, green $\lambda = 0.4$, yellow $\lambda = 0.5$, orange $\lambda = 0.7$, and red $\lambda = 1$.}
\label{fig:coupl-solv-upto-g5}
\end{figure}

The coupling models for the guests to TEMOA are more complex, reflecting higher conformational heterogeneity. This is particularly evident for the bound state complexes of TEMOA with G1 and G3 whose perturbation energy probability densities at $\lambda=1$ have two modes (Figures \ref{fig:coupl-solv-upto-g2} and \ref{fig:coupl-solv-upto-g5}) each corresponding to a conformational state with a significant population that contributes to binding. In general, we found that the coupling models of the guests to TEMOA are described by at least three modes: a binding-competent mode characterized by a small statistical weight (the $w$ parameter in Table \ref{tab:coupl_parameters}) and a relatively small probability of clashes (large $b$), another mode extremely unfavorable to binding dominated by clashes, and a moderate mode that is not as unfavorable (Table \ref{tab:coupl_parameters}). Consistently with their more favorable coupling free energies (Table \ref{tab:FE_estimates}) and higher overall binding affinities, the binding-competent modes of the TEMOA-G2 and TEMOA-G4 complexes (mode 1) tend to have larger $b$ values (smaller chance of atomic clashes) and smaller $\bar{u}_0$ values (stronger interactions with the host) than the other complexes.

The modes unfavorable to binding correspond to positions and orientations of the guest that cause frequent and severe atomic clashes with TEMOA, as indicated by the large $n_l$ parameter, which is related to the number of atoms of the ligand that experience very repulsive energies, and a small $b$ value, which reflects the probability of finding a configuration in which the ligand binds to TEMOA without atomic collisions. As expected based on their relative sizes, water has a greater probability of binding without collisions than the larger guests as reflected by the smaller $b$ values of the latter. We found that various combinations of collisional parameters ($\epsilon$, $\tilde u$, and $n_l$) fit the simulation data equally well; the values reported should be considered order-of-magnitude estimates.  Nevertheless, the optimized $\epsilon$, $\tilde u$, and $n_l$ values tend to be larger in magnitude than those of the coupling model of TEMOA-H$_2$O (first column of Table \ref{tab:coupl_parameters} and Fig. \ref{fig:coupl-solv-upto-g2} top), reflecting the larger role of atomic collisions for the larger molecular guests, and generally increase in step with how favorable a mode is towards binding.

\begin{table*}
  \caption{Parameters for coupling models of the various molecules to host TEMOA.}
\begin{ruledtabular}
\begin{tabular}{lccccccc}
       & $w_i$               & $b$ & $\bar{u}_{0h}$\footnote{kcal/mol} & $\sigma_{h}^{\rm a} $ & $\epsilon^{\rm a}$ & $\tilde{u}^{\rm a}$ & $n_l$ \\ \hline 
       \tabularnewline
\multicolumn{8}{c}{TEMOA--H$_2$O}\tabularnewline
mode 1 & $4.46 \times 10^{-1}$ & $1.40 \times 10^{-2}$ & $-0.51$ & $2.66$ & $1.0$ & $1.0$ & $3.0$ \tabularnewline
mode 2 & $5.54 \times 10^{-1}$ & $6.90 \times 10^{-4}$ & $0.00$ & $3.12$ & $1.0$ & $2.5$ & $5.4$ \tabularnewline
\multicolumn{8}{c}{TEMOA--G1}\tabularnewline
mode 1 & $1.23 \times 10^{-5}$ & $2.71 \times 10^{-9}$ & $-19.14$ & $3.72$ & $1.0$ & $1.0$ & $15.9$ \tabularnewline
mode 2 & $7.10 \times 10^{-3}$ & $2.12 \times 10^{-18}$ & $-22.23$ & $4.69$ & $1.0$ & $1.0$ & $49.3$ \tabularnewline
mode 3 & $9.93 \times 10^{-1}$ & $9.57 \times 10^{-13}$ & $-8.91$ & $4.78$ & $20.0$ & $200.0$ & $60.0$ \tabularnewline
\multicolumn{8}{c}{TEMOA--G2}\tabularnewline
mode 1 & $2.27 \times 10^{-2}$ & $1.43 \times 10^{-8}$ & $-23.85$ & $2.58$ & $2.1$ & $2.1$ & $7.4$ \tabularnewline
mode 2 & $1.99 \times 10^{-1}$ & $1.49 \times 10^{-6}$ & $-15.95$ & $3.17$ & $5.2$ & $22.4$ & $17.3$ \tabularnewline
mode 3 & $7.79 \times 10^{-1}$ & $1.35 \times 10^{-6}$ & $-9.48$ & $3.83$ & $9.0$ & $89.8$ & $46.3$ \tabularnewline
\multicolumn{8}{c}{TEMOA--G3}\tabularnewline
mode 1 & $1.37 \times 10^{-4}$ & $8.56 \times 10^{-7}$ & $-15.38$ & $3.72$ & $1.0$ & $1.0$ & $11.7$ \tabularnewline
mode 2 & $3.48 \times 10^{-2}$ & $1.52 \times 10^{-8}$ & $-6.58$ & $4.51$ & $1.0$ & $1.0$ & $48.7$ \tabularnewline
mode 3 & $9.65 \times 10^{-1}$ & $2.62 \times 10^{-14}$ & $-11.87$ & $4.80$ & $18.0$ & $179.8$ & $60.0$ \tabularnewline
\multicolumn{8}{c}{TEMOA--G4}\tabularnewline
mode 1 & $3.90 \times 10^{-5}$ & $1.55 \times 10^{-9}$ & $-27.20$ & $3.00$ & $1.0$ & $1.0$ & $14.3$ \tabularnewline
mode 2 & $3.07 \times 10^{-2}$ & $3.29 \times 10^{-10}$ & $-20.79$ & $3.54$ & $1.0$ & $1.0$ & $50.5$ \tabularnewline
mode 3 & $9.69 \times 10^{-1}$ & $8.32 \times 10^{-11}$ & $-11.57$ & $4.49$ & $18.8$ & $188.0$ & $60.0$ \tabularnewline
\multicolumn{8}{c}{TEMOA--G5}\tabularnewline
mode 1 & $2.82 \times 10^{-5}$ & $7.28 \times 10^{-9}$ & $-24.56$ & $2.85$ & $1.0$ & $1.0$ & $11.4$ \tabularnewline
mode 2 & $3.00 \times 10^{-2}$ & $4.75 \times 10^{-9}$ & $-15.31$ & $3.80$ & $1.0$ & $1.0$ & $44.1$ \tabularnewline
mode 3 & $9.70 \times 10^{-1}$ & $5.04 \times 10^{-28}$ & $1.99$ & $5.23$ & $12.9$ & $123.8$ & $60.0$ \tabularnewline
\end{tabular} 
\end{ruledtabular}
\label{tab:coupl_parameters}
\end{table*}

\subsection{Hydration Models}

Table \ref{tab:solv_parameters} lists the optimized parameters for the hydration models of the guests. The model for coupling a water molecule to the water solvent is represented by one mode (Table \ref{tab:solv_parameters}). With the exception of G3, which is the most flexible guest (and G5, although its modes 2 and 3 are nearly equivalent), the $p_{0s}(u_{s})$ models for the guests are described by two modes. The interplay between the modes is particularly evident for the hydration of G1 and G3 that present bimodal distributions in the $\lambda=1$ hydrated state (Figures \ref{fig:coupl-solv-upto-g2} and \ref{fig:coupl-solv-upto-g5}). Similarly to the models for the coupling to the solvated host, the first mode is more favorable toward hydration with stronger interactions with the solvent (the $\bar{u}_{0s}$ parameter) and a smaller chance of atomic clashes (the $b$ parameter). 

The $b$ parameter, which measures the probability of finding configurations free of clashes in the uncoupled ensemble,\cite{Widom1982} has special significance in the theory of solvation and hydrophobicity,\cite{Chandler:Weeks:Andersen:83,ashbaugh2023cavity,Berne:96,Huang:Chandler:2000,Gallicchio:Kubo:Levy:2000} since it is related to the free energy cost of forming a cavity in the solvent of the size and shape of the solute, $\Delta G_{\rm cav} = -k_B T \log b$. Specifically, the smaller the $b$ parameter, the higher the free energy cost of cavity formation. Indeed, the optimized $b$ parameters we obtained of the SAMPL8 guests track the size of the guests (Table \ref{tab:solv_parameters}), with the smaller guests (G2 and G5) having the largest $b$ for mode 1 while G1, the largest guest, has the smallest $b$ and the highest free energy of cavity formation. The model predicts that the probability of finding a cavity the size of a water molecule is approximately $5.77 \times 10^{-3}$ (Table \ref{tab:solv_parameters}) corresponding to a cavity formation free energy cost of about $3$ kcal/mol. Estimating free energies of cavity formation through this approach is significant because, similar to early information-theory models,\cite{Hummer:Pratt:96} our analytical model does not assume a specific form of the solute-solvent repulsive interaction potential.\cite{Chandler:Weeks:Andersen:83,Gallicchio:Kubo:Levy:2000} Rather, it exploits the distinct statistical signatures of atomic collisions, which, arguably, are the fundamental defining characteristic of the solute cavity.

\begin{table*}[htbp]
  \caption{Parameters for hydration models of the SAMPL8 host-guest complexes.}
\begin{ruledtabular}
\begin{tabular}{lccccccc}
       & $w_i$               & $b$ & $\bar{u}_{0s}$\footnote{kcal/mol} & $\sigma_{s}^{\rm a} $ & $\epsilon^{\rm a}$ & $\tilde{u}^{\rm a}$ & $n_l$\tabularnewline

\multicolumn{8}{c}{H$_2$O}\tabularnewline
mode 1 & $1.00$ & $5.77 \times 10^{-3}$ & $2.41$ & $3.46$ & $3.9$ & $3.9$ & $2.5$ \tabularnewline
\multicolumn{8}{c}{G1}\tabularnewline
mode 1 & $2.79 \times 10^{-5}$ & $2.20 \times 10^{-11}$ & $-7.51$ & $4.27$ & $1.0$ & $1.0$ & $26.2$ \tabularnewline
mode 2 & $9.99 \times 10^{-1}$ & $3.20 \times 10^{-14}$ & $13.42$ & $6.27$ & $2.3$ & $2.3$ & $60.0$ \tabularnewline
\multicolumn{8}{c}{G2}\tabularnewline
mode 1 & $1.27 \times 10^{-3}$ & $4.60 \times 10^{-7}$ & $-6.87$ & $3.37$ & $1.0$ & $1.0$ & $13.4$ \tabularnewline
mode 2 & $9.99 \times 10^{-1}$ & $2.41 \times 10^{-9}$ & $3.90$ & $4.45$ & $1.0$ & $1.0$ & $36.2$ \tabularnewline
\multicolumn{8}{c}{G3}\tabularnewline
mode 1 & $2.98 \times 10^{-6}$ & $1.17 \times 10^{-9}$ & $-11.28$ & $3.12$ & $1.0$ & $1.0$ & $15.1$ \tabularnewline
mode 2 & $5.57 \times 10^{-3}$ & $2.98 \times 10^{-16}$ & $-10.69$ & $4.44$ & $1.0$ & $1.0$ & $33.5$ \tabularnewline
mode 3 & $9.94 \times 10^{-1}$ & $1.57 \times 10^{-13}$ & $-4.38$ & $4.05$ & $1.5$ & $1.5$ & $60.0$ \tabularnewline
\multicolumn{8}{c}{G4}\tabularnewline
mode 1 & $8.64 \times 10^{-5}$ & $1.23 \times 10^{-9}$ & $-6.83$ & $4.11$ & $1.0$ & $1.0$ & $20.2$ \tabularnewline
mode 2 & $9.99 \times 10^{-1}$ & $1.70 \times 10^{-12}$ & $9.74$ & $5.68$ & $1.0$ & $1.0$ & $57.9$ \tabularnewline
\multicolumn{8}{c}{G5}\tabularnewline
mode 1 & $7.94 \times 10^{-8}$ & $3.68 \times 10^{-6}$ & $-5.86$ & $3.99$ & $1.0$ & $1.0$ & $9.2$ \tabularnewline
mode 2 & $1.60 \times 10^{-3}$ & $2.30 \times 10^{-16}$ & $-11.04$ & $4.49$ & $1.0$ & $1.0$ & $26.3$ \tabularnewline
mode 3 & $9.98 \times 10^{-1}$ & $0.00$ & $-10.40$ & $4.47$ & $1.0$ & $1.0$ & $56.9$ \tabularnewline
\end{tabular} 
\label{tab:solv_parameters}
\end{ruledtabular}
\end{table*}

\subsection{Analytical Models of Alchemical Transfer}

\begin{figure}[htpb]
    \subfloat[\centering TEMOA-H$_2$O transfer leg 1]{{\includegraphics[width=3.2in]{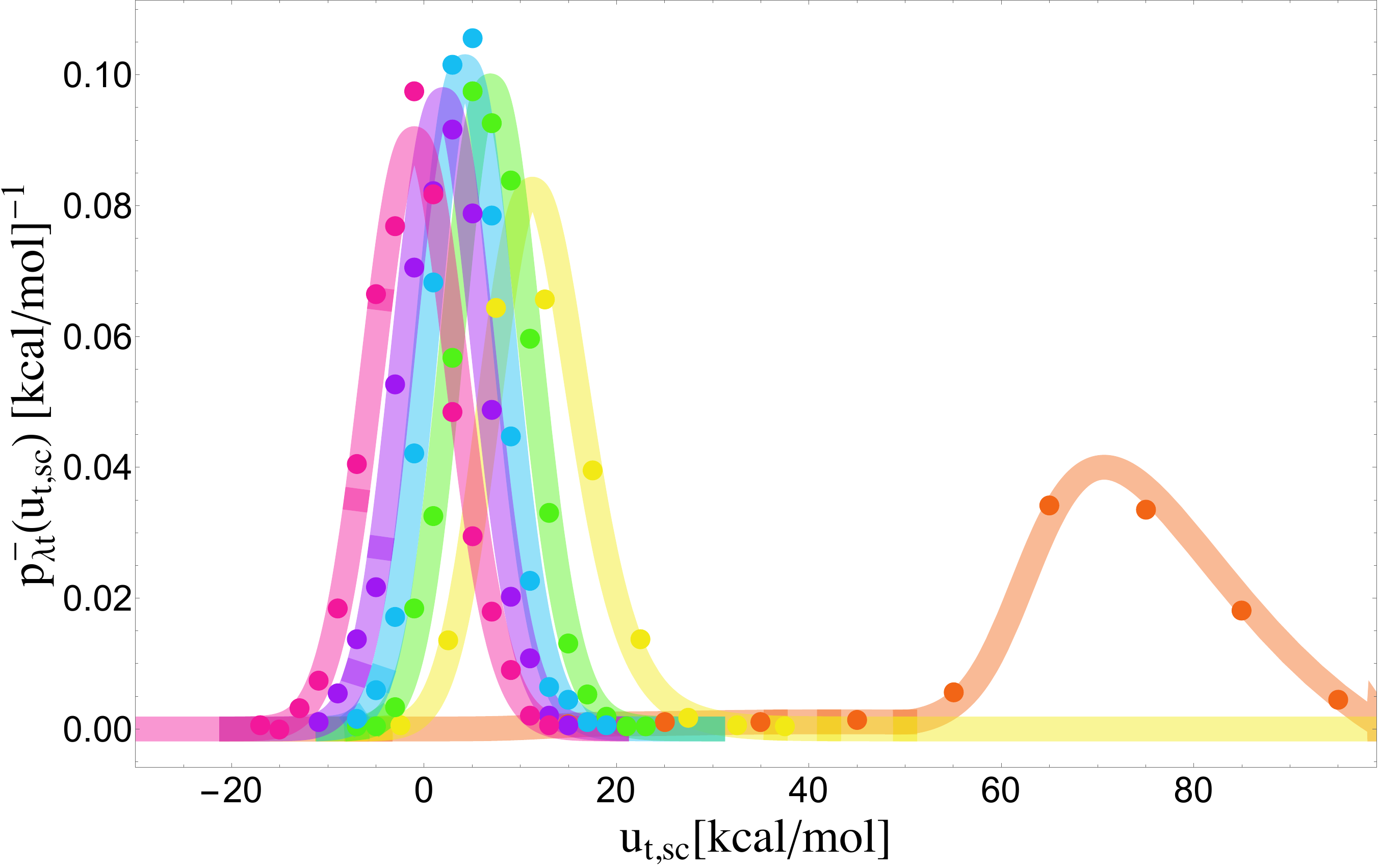} }} 
    \subfloat[\centering TEMOA-H$_2$O transfer leg 2]{{\includegraphics[width=3.2in]{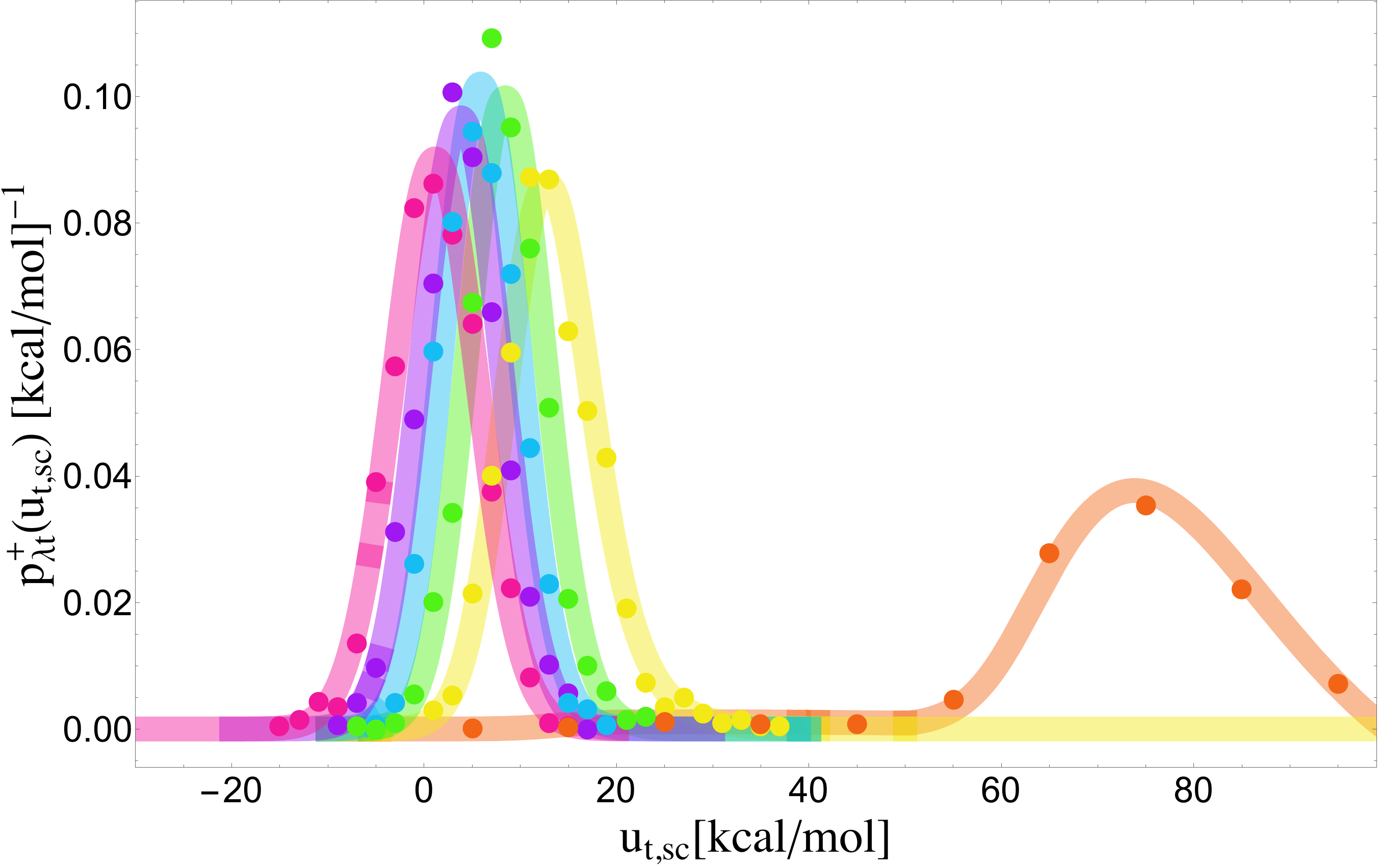} }}   \\
    \subfloat[\centering TEMOA-G1 transfer leg 1]{{\includegraphics[width=3.2in]{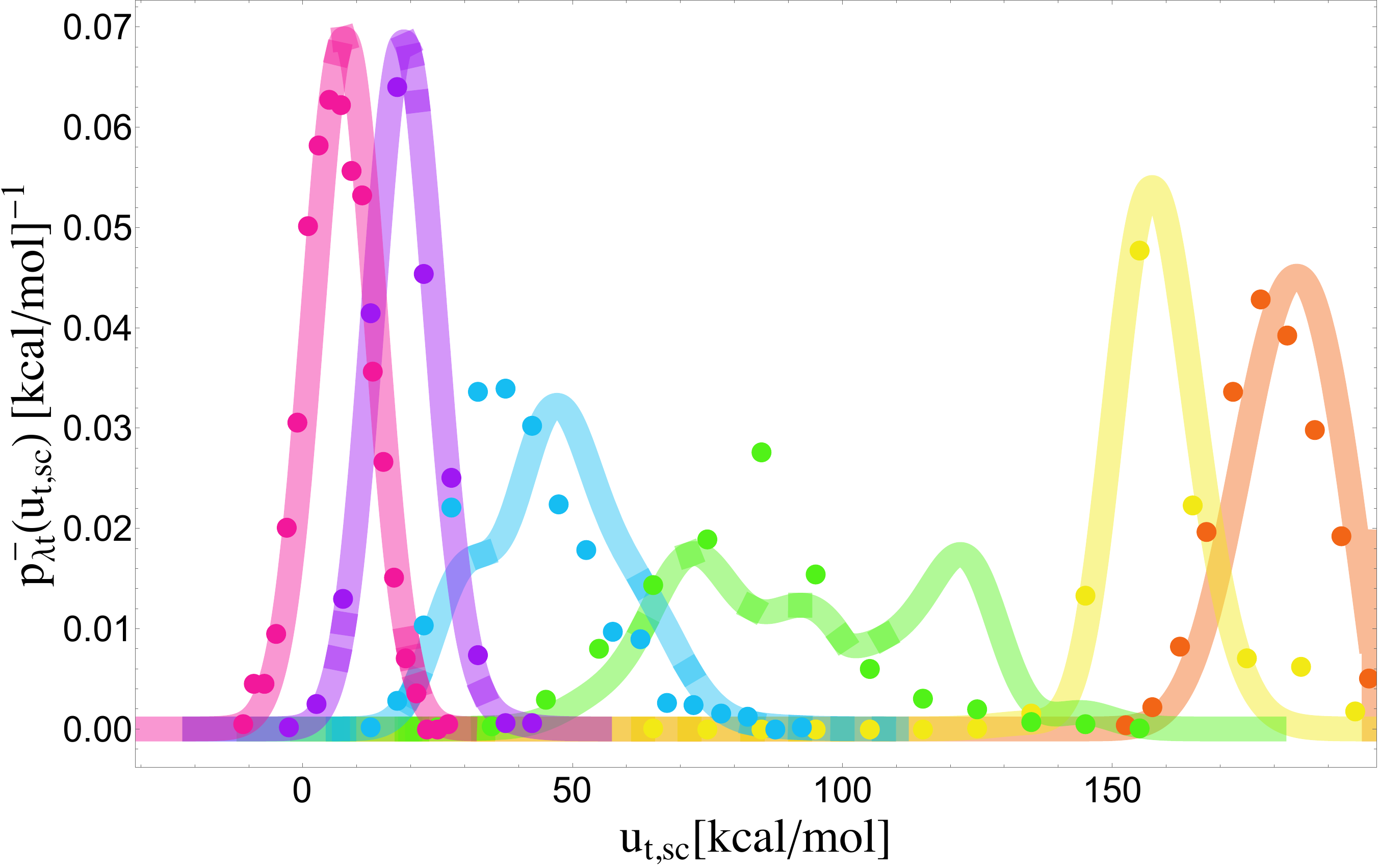} }} 
    \subfloat[\centering TEMOA-G1 transfer leg 2]{{\includegraphics[width=3.2in]{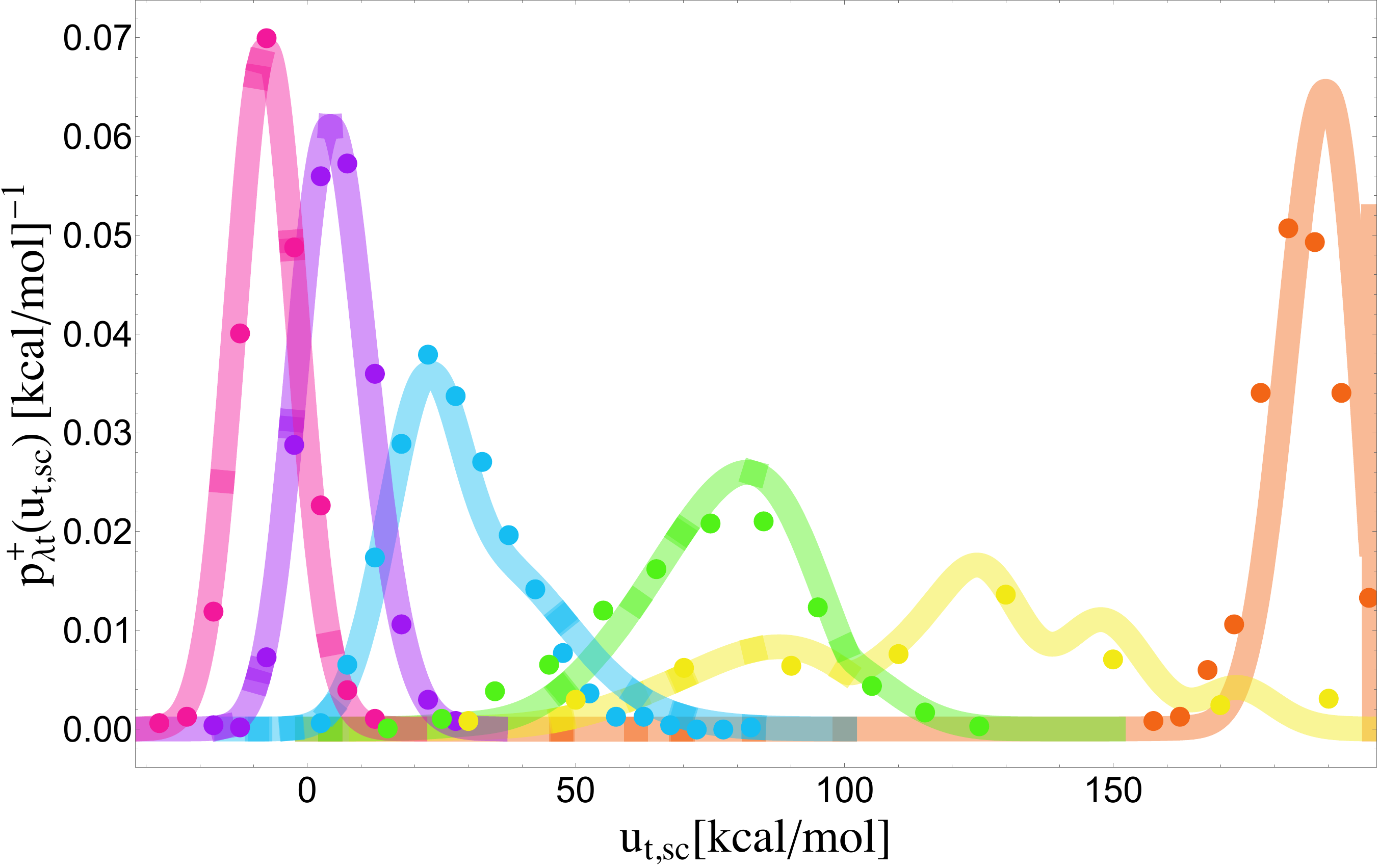} }} \\
    \subfloat[\centering TEMOA-G2 transfer leg 1]{{\includegraphics[width=3.2in]{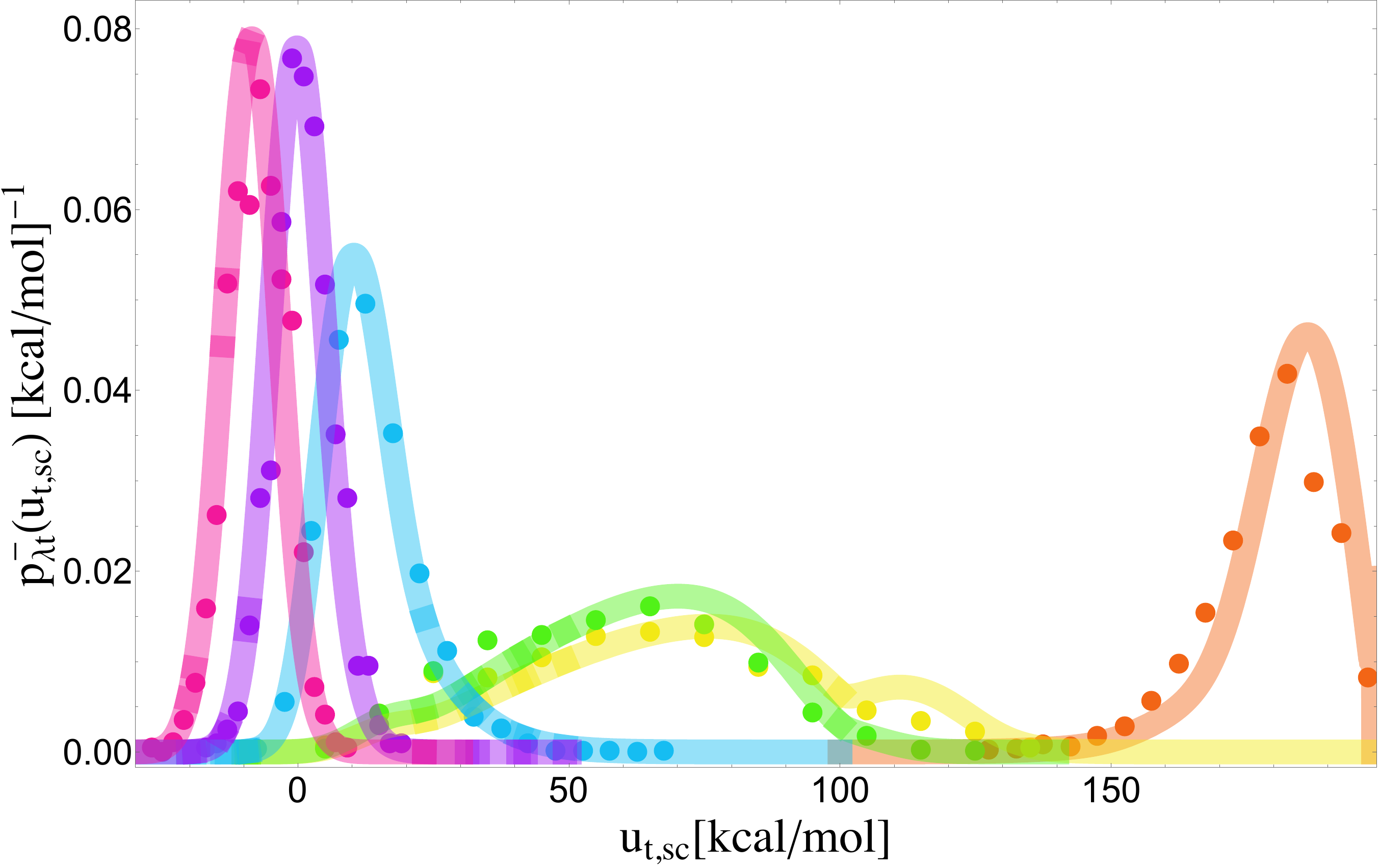} }} 
    \subfloat[\centering TEMOA-G2 transfer leg 2]{{\includegraphics[width=3.2in]{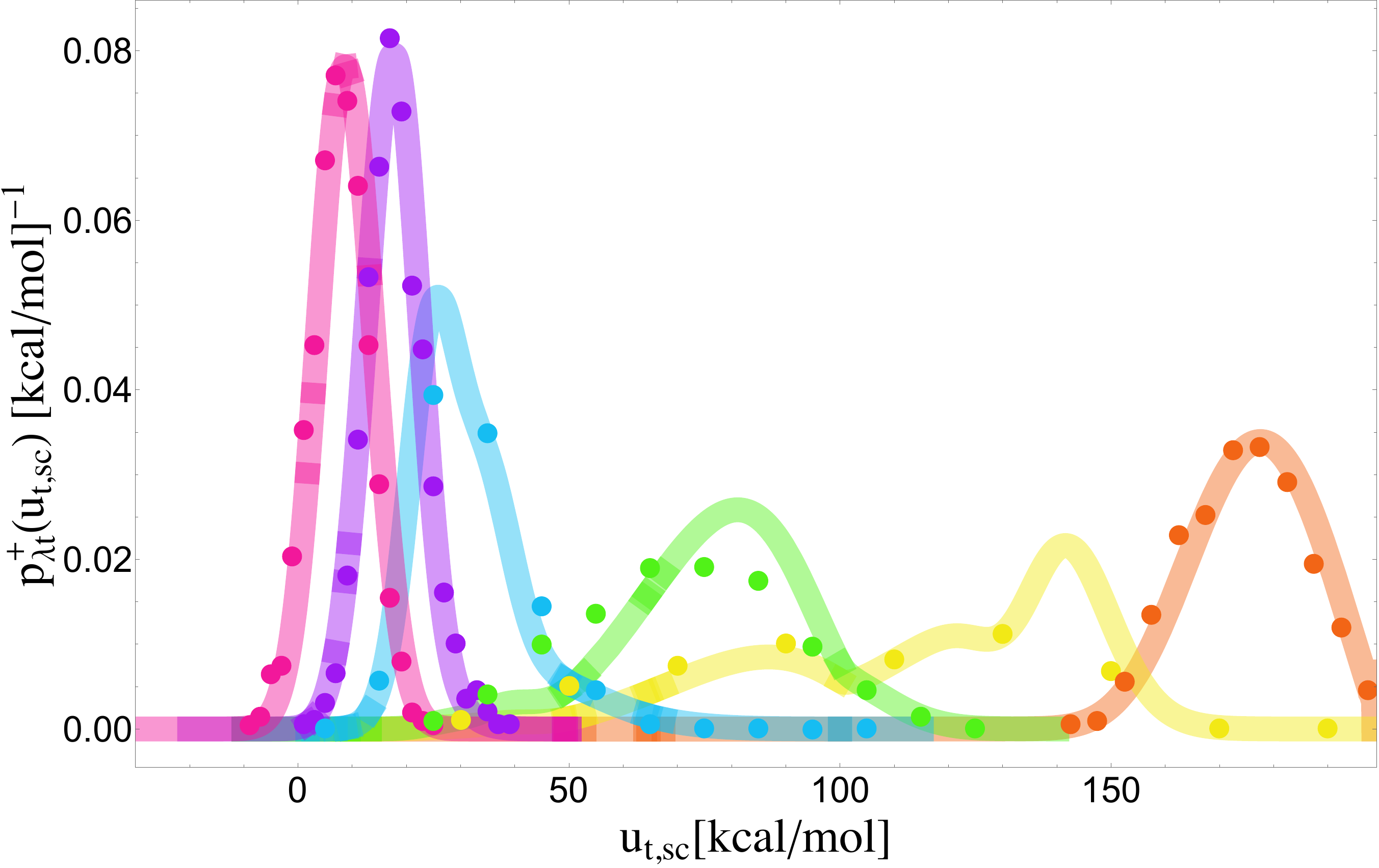} }} 
\captionsetup{justification=Justified}
\caption{
  Probability densities  $p_{\lambda{\rm t}} (u_{\rm t,sc})$ of the soft-core perturbation energy of alchemical transfer collected from simulations (dots) and predicted from the analytical model's description of $p_{0{\rm t}}(u_{\rm t})$ at the initial state (lines). Each color represents a distinct alchemical state: pink $\lambda = 0.5$, purple $\lambda = 0.4$, blue $\lambda = 0.3$, green $\lambda = 0.2$, yellow $\lambda = 0.1$, and orange $\lambda = 0$. Left: leg 2, which describes unbinding. Right: leg 1, which describes binding. The x-axis is in units of kcal/mol.
}
\label{fig:temoa-h2o_ligs_transf}
\end{figure}

\begin{figure}[htbp]
    \subfloat[\centering TEMOA-G3 transfer leg 1]{{\includegraphics[width=3.2in]{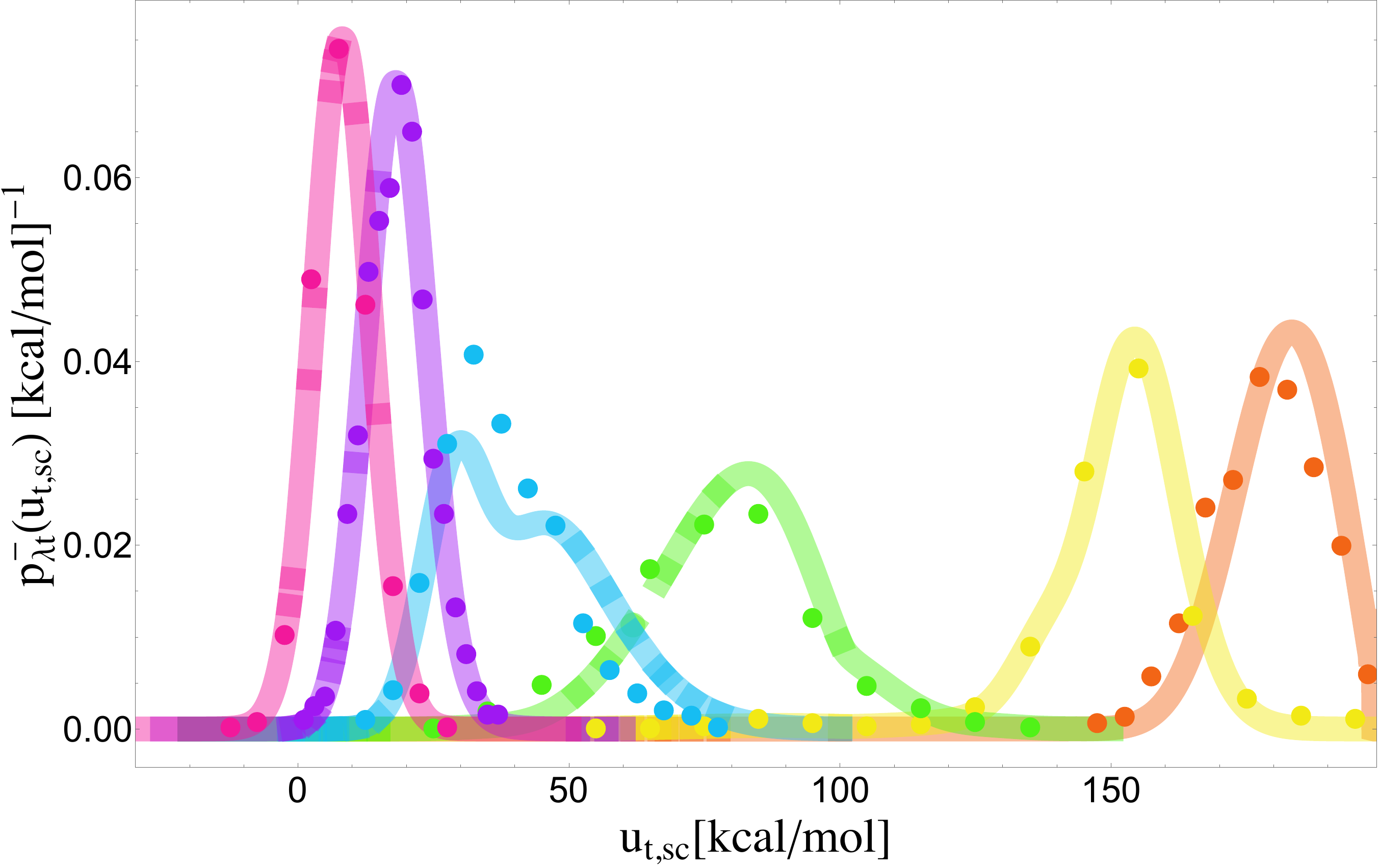} }} 
    \subfloat[\centering TEMOA-G3 transfer leg 2]{{\includegraphics[width=3.2in]{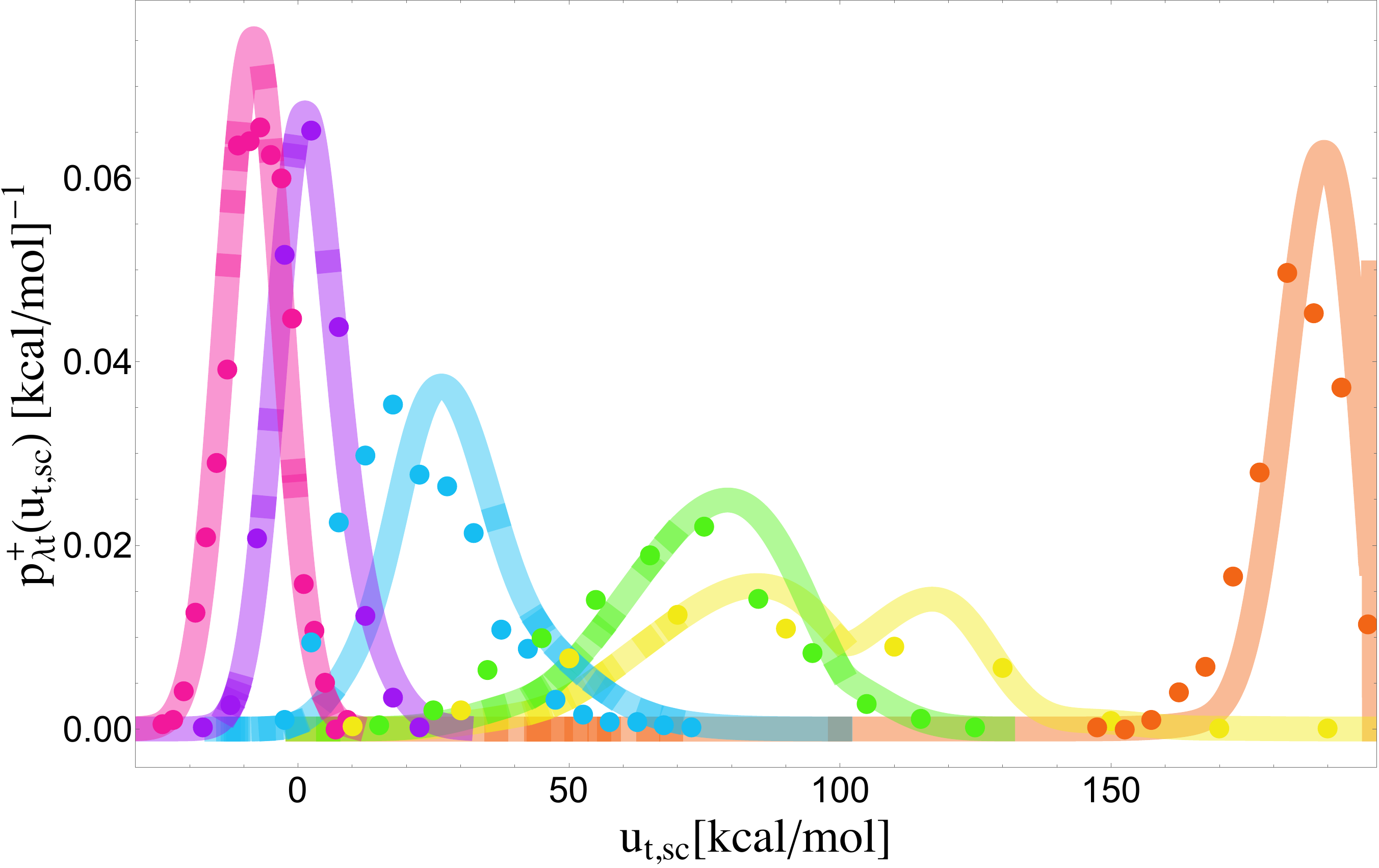} }}   \\
    \subfloat[\centering TEMOA-G4 transfer leg 1]{{\includegraphics[width=3.2in]{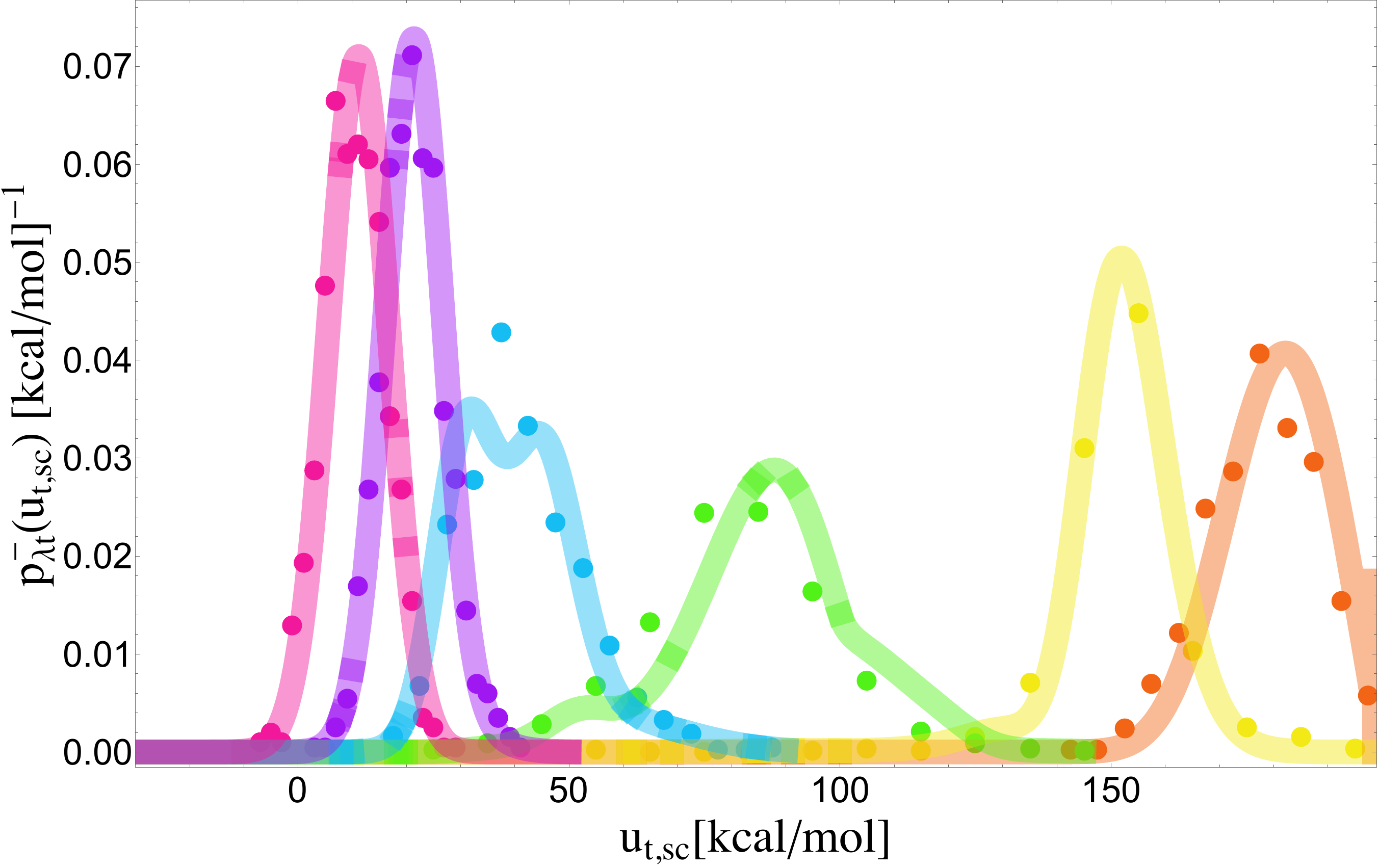} }} 
    \subfloat[\centering TEMOA-G4 transfer leg 2]{{\includegraphics[width=3.2in]{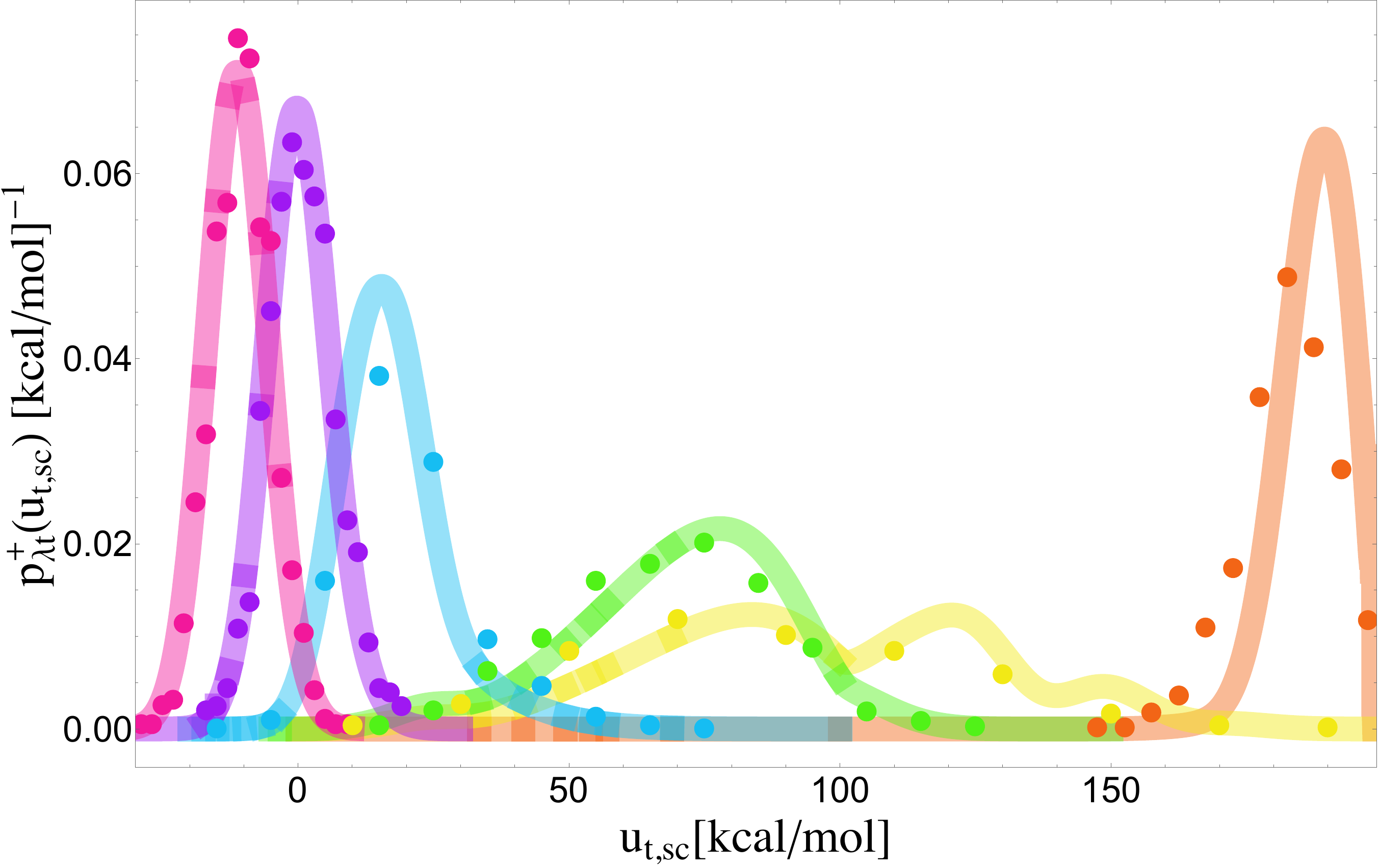} }} \\
    \subfloat[\centering TEMOA-G5 transfer leg 1]{{\includegraphics[width=3.2in]{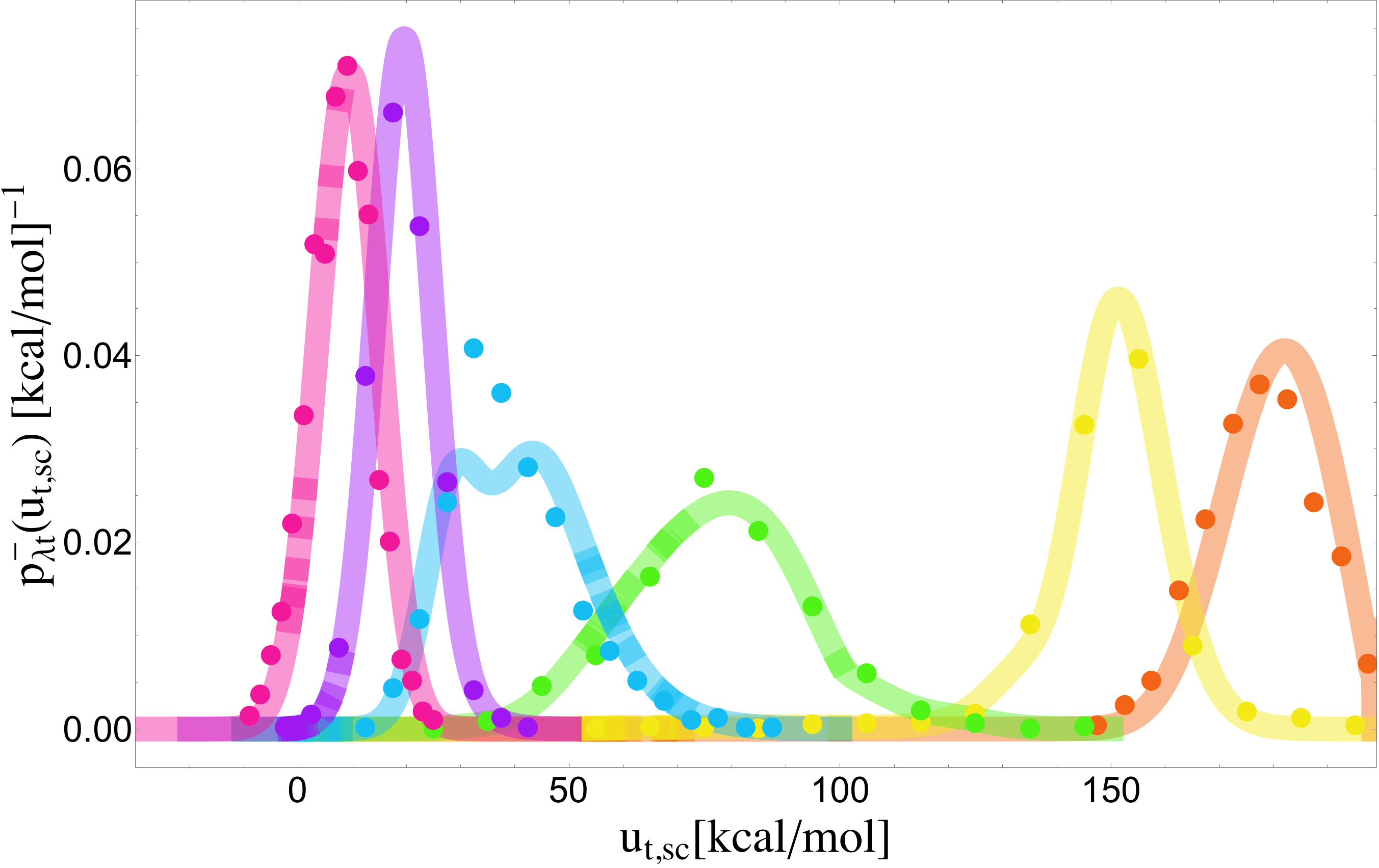} }} 
    \subfloat[\centering TEMOA-G5 transfer leg 2]{{\includegraphics[width=3.2in]{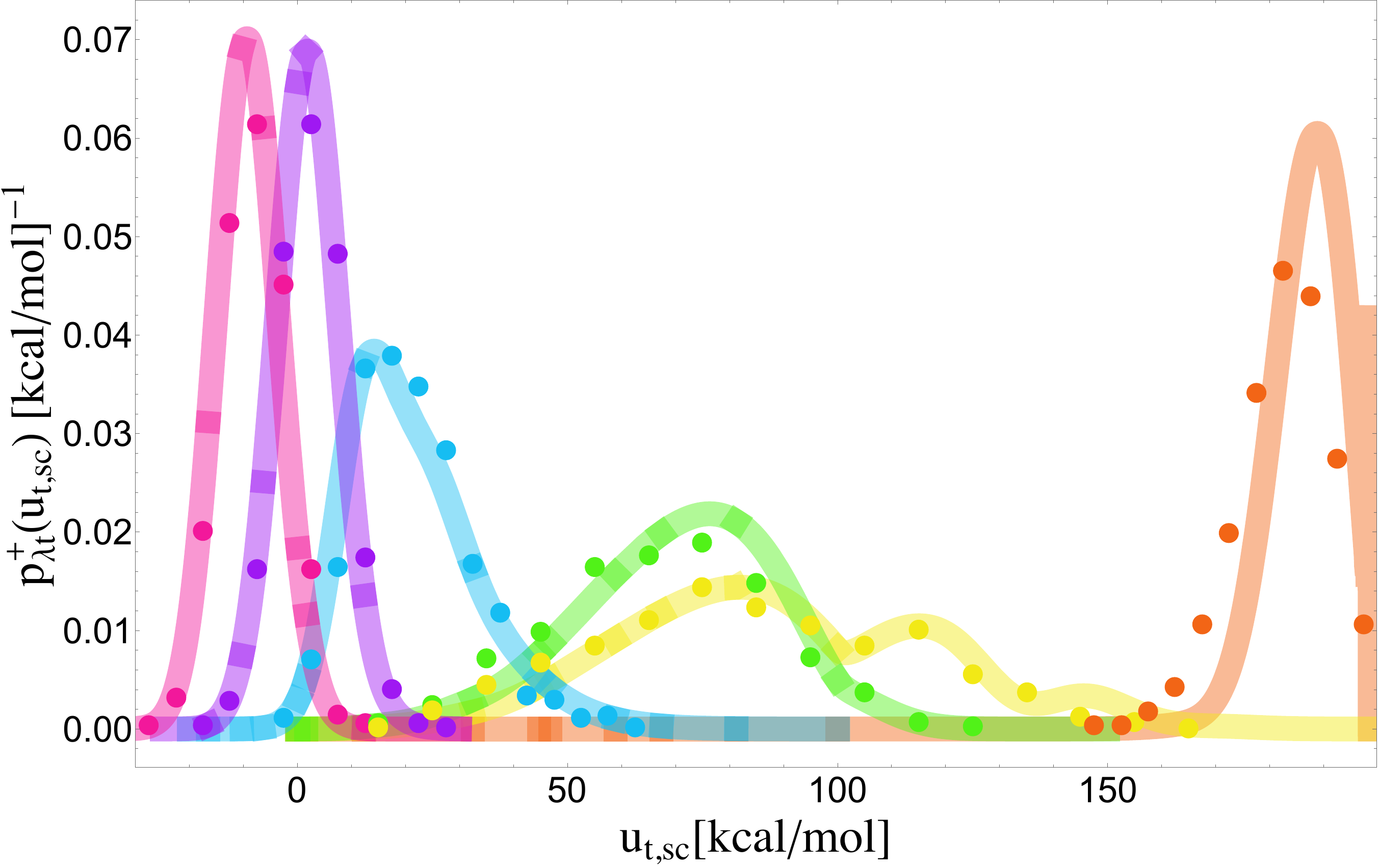} }} 
\captionsetup{justification=Justified}
\caption{
   Probability densities  $p_{\lambda{\rm t}} (u_{\rm t,sc})$ of the soft-core perturbation energy of alchemical transfer collected from simulations (dots) and predicted from the analytical model's description of $p_{0{\rm t}}(u_{\rm t})$ at the initial state (lines). Each color represents a distinct alchemical state: pink $\lambda = 0.5$, purple $\lambda = 0.4$, blue $\lambda = 0.3$, green $\lambda = 0.2$, yellow $\lambda = 0.1$, and orange $\lambda = 0$. Left: leg 2, which describes unbinding. Right: leg 1, which describes binding.
}
\label{fig:temoa_ligs_transf}
\end{figure}

As discussed in the Theory section and illustrated in Figure \ref{fig:conv_scheme}, the models for the perturbation energy distributions of the alchemical transfer of a guest from solution to the host and back are constructed from the convolution of the coupling and hydration models described above [Eqs. (\ref{eq:p0t_model})--(\ref{eq:c0t_model})]. Alchemical transfer is described by two thermodynamic legs, the first starting from the host and the guest dissociated in solution and the other from the guest bound to the host. Both legs terminate at the same alchemical intermediate state at $\lambda=1/2$. We present the models for each leg individually (Figures \ref{fig:temoa-h2o_ligs_transf} and \ref{fig:temoa_ligs_transf}). Unlike the coupling transformations, the initial states at $\lambda=0$ do not represent uncoupled states but rather states where host and guest are coupled to the bulk solvent or to each other.

Each pair of coupling modes, one for coupling to the host and one for coupling to the solvent, combine to produce a transfer mode for the first leg and one for the second leg. Hence, for example, the $p_{0t}^{+}(u_t)$ model for leg 1 of the TEMOA-G1 complex is composed of six modes (SI Table I.) one for each combination of the three coupling modes of G1 to the host (Table \ref{tab:coupl_parameters}) and the two coupling modes to the solvent (Table \ref{tab:solv_parameters}). One of the more intricate systems in the set is the alchemical transfer for TEMOA-G3 which is modeled by nine modes from all possible combinations of the three modes for hydration and host's coupling. Depending on their resulting statistical weights, some of these modes are not apparent in the distributions in Figure \ref{fig:temoa_ligs_transf}, and others contribute significantly only at some $\lambda$-values and in a specific range of perturbation energies.

The parameters for each transfer mode are calculated using straightforward relations described below Eqs. (\ref{eq:p0t_model})--(\ref{eq:c0t_model}): the statistical weight is the product of the statistical weights of the coupling modes, the $\bar{u}_{0t}$ parameter is given by the difference between the corresponding parameter of the host coupling model and the average solute-solvent interaction energy in the fully solvated state, the $\sigma_{t}$ parameter is the geometric average of the corresponding coupling parameters, and the $b$, $\epsilon$, $\tilde u$, and $n_l$ parameters are inherited directly from the coupling mode to the host. The parameters for the $p_{0t}^{-}(u_t)$ model for leg 2 are derived similarly, except that the initial and final states are reversed. For example, the $\bar{u}_{0t}$ parameter for a mode of leg 2 is given by the difference between the corresponding parameter of the solvent coupling model and the average host-guest interaction energy in the bound state.

The parameters of the alchemical transfer models are presented in Tables I through V in Supplementary Information, section B. Figures \ref{fig:temoa-h2o_ligs_transf} and \ref{fig:temoa_ligs_transf} show the corresponding perturbation energy probability densities (continuous curves). Despite their complexities, the analytical models' predictions agree well with the perturbation energy distributions obtained from numerical alchemical transfer (ATM) simulations (dots, in Figures \ref{fig:temoa-h2o_ligs_transf} and \ref{fig:temoa_ligs_transf}). This result confirms the validity of the theory and the assumptions on which it is based, and it provides a physical interpretation of the distributions observed from alchemical transfer calculations.

Similarly to the host coupling and hydration processes (Figures \ref{fig:coupl-solv-upto-g2} and \ref{fig:coupl-solv-upto-g5}), near $\lambda=0$, when the guest interacts only with the solvent, the distributions for transfer towards binding (leg 1 in Figures \ref{fig:temoa-h2o_ligs_transf} and \ref{fig:temoa_ligs_transf}) lie at very unfavorable perturbation energies because they are dominated by clashes between the guest and the atoms of the host (and bound water molecules) when the guest is transferred from the solvent bulk. Analogously, clashes between the guest and the solvent's atoms dominate the probability distributions of the perturbation energy for transfer in leg 2. The shapes of the distributions in this regime are determined by the collisional parameters of the corresponding coupling processes. The distributions progressively shift to lower perturbation energies as $\lambda$ increases and the interactions at the transfer position are turned on. The complex multimodal behavior near $u_{sc} = 100$ kcal/mol is an artifact of the soft-core function that terminates there (see Computational Details). As expected, near $\lambda = 1/2$, the distributions display linear response behavior. The $\lambda=1/2$ states of leg 1 and leg 2 are equivalent and the corresponding distributions differ only in the sign of the perturbation energy [Eqs.\ \ref{eq:ATM-perte-leg1} and \ref{eq:ATM-perte-leg2}]. Hence, as evidenced in Figures \ref{fig:temoa-h2o_ligs_transf} and \ref{fig:temoa_ligs_transf}, they are related by mirror symmetry about zero.

\begin{table*}[htpb]
  \caption{\label{tab:temoa-h2o-transf}Parameters for transfer model of one water molecule binding to host TEMOA.}
\begin{ruledtabular} 
\begin{tabular}{lccccccc}
       & $w_i$               & $b$ & $\bar{u}_{0t}$\footnote{kcal/mol} & $\sigma_{t}^{\rm a} $ & $\epsilon^{\rm a}$ & $\tilde{u}^{\rm a}$ & $n_l$ \\ \hline 
       \tabularnewline
     
\multicolumn{8}{c}{Leg 1}\tabularnewline
mode 1 & $4.62 \times 10^{-1}$ & $5.82\times 10^{-3}$ & $14.8$             & $4.36$              & $3.65$           &   $4.34$          & $2.50$ \tabularnewline
mode 2 & $5.38 \times 10^{-1}$ & $5.82\times 10^{-3}$ & $18.8$             & $4.66$              & $3.65$           &   $4.34$           & $2.50$ \tabularnewline

\multicolumn{8}{c}{Leg 2}\tabularnewline
mode 1 & $4.18 \times 10^{-1}$ & $1.49 \times 10^{-2}$ & $17.1$             & $4.36$              & $1.00$           &   $1.00$          & $2.95$ \tabularnewline
mode 2 & $5.82 \times 10^{-1}$ & $6.50 \times 10^{-4}$ & $17.7$             & $4.66$              & $1.00$           &   $0.00$           & $6.51$ \tabularnewline

\end{tabular} 
\end{ruledtabular}
\end{table*}

\section{Discussion}

The Potential Distribution Theorem (PDT)\cite{PDTbook:2006} offers a useful formalism to describe alchemical transformations.\cite{kilburg2018analytical,pal2019perturbation} It leads naturally to a representation of alchemical coupling in terms of the probability density distribution $p_0(u)$ of the interaction energy $u$ between a ligand and its environment (a solvent or a receptor) collected in the uncoupled ensemble.\cite{Gallicchio2010,Gallicchio2011adv} The PDT formulas [Eqs.\ (\ref{eq:pdt}) and (\ref{eq:Klambda})] relate $p_0(u)$ to the free energy profile and the sequence of perturbation energy distributions along the alchemical pathway.

Because $p_0(u)$ does not depend on the alchemical $\lambda$-dependent potential energy function, it describes any alchemical process that connects the uncoupled and coupled states of the system. The central role of $p_0(u)$ in the statistical mechanics of alchemical transformations is analogous to that of the density of states $\Omega(E)$ in standard statistical mechanics.\cite{Hill86book} Note, for instance, the parallel between the PDT relationship [Eq.\ (\ref{eq:pdt})], which gives the perturbation energy distributions as a function of the alchemical progress parameter $\lambda$, and the well known canonical ensemble relationship $p_\beta(E) \propto \Omega(E) \exp(-\beta E)$ relating the energy distribution of a system's energy as a function of temperature to the density of states.\cite{pal2019perturbation} In both cases, knowledge of a single function (the density of states or $p_0(u)$) uniquely determines the statistical behavior of the system for all values of a system's parameter (the temperature in the case of the canonical ensemble or $\lambda$ in the case of an alchemical process). Hence, $p_0(u)$ can be considered a master function to describe the thermodynamics of alchemical states in the same way that the density of states describes the thermodynamics of physical systems.  

Taking advantage of the fact that perturbation energy for transfer is the sum of the coupling energy of the ligand to the solvated receptor and the uncoupling energy from the solvent, in this work, we developed an analytical PDT description of alchemical transfer for binding by expressing the $p_0(u)$ function for transfer as the convolution of the corresponding functions for uncoupling the ligand to the solvent and to the solvated receptor. The $p_0(u)$ function of each coupling process is expressed in terms of the parameters of the analytical model of alchemical coupling of Kilburg and Gallicchio\cite{kilburg2018analytical,pal2019perturbation} obtained by maximum likelihood analysis of double-decoupling alchemical simulations. We showed that the resulting analytical model for transfer reproduces the perturbation energy distributions observed in alchemical transfer simulations of host-guest complexes in explicit solvent.

Alchemical Transfer (ATM)\cite{wu2021alchemical} is a method developed recently to compute the absolute and relative binding free energies of molecular complexes, including those of protein-ligand complexes relevant to computer-aided drug discovery.\cite{armacost2020novel,azimi2022relative,chen2023performance,sabanes2023validation} Because it is based on a direct coordinate transformation and a simple dual-topology implementation,\cite{rocklin2013separated,konig2020alternative} ATM is easily applied to complex scaffold-hopping and charge-changing transformations.\cite{chen2023performance} ATM is also applicable with any force field, including many-body and machine-learned potentials that are increasingly deployed in drug discovery projects.\cite{eastman2023openmm,sabanes2024enhancing} This work builds a solid theoretical foundation for alchemical transfer and provides physical insights on the origin of the complex perturbation energy distributions that are often observed yet overlooked in alchemical transfer simulations. Furthermore the results of this work verify the theoretical and numerical consistency betweeen alchemical transfer and the more established double-decoupling alchemical descriptions of binding.\cite{Gilson:Given:Bush:McCammon:97,Mey2020Best} 

More generally, the work reinforces the benefits of a view of alchemical processes in terms of the progressive modifications of the statistical distributions of the system. For example, we adopted this approach to develop a graphical scheme to optimize the form and the parameters of alchemical potential energy functions to enhance convergence by avoiding alchemically induced phase transitions.\cite{pal2019perturbation} In this work, we use a similar approach based on the PDT to illustrate how all of the alchemical pathways originating from one state and ending in another are interdependent. Specifically, we showed that distributions, particularly those of intermediate $\lambda$ states that are unphysical, learned from double-decoupling simulations of binding provide information to reproduce those observed in alchemical transfer calculations.

Unfortunately, the insights obtained here do not immediately transfer to some of the popular alchemical models in current use.  The PDT formalism assumes an alchemical potential energy function that depends on one or, at most, a few collective variables with interpretable statistical distributions. Hence, the PDT does not apply to the double-decoupling alchemical models based on parameter interpolation and soft-core pair potentials implemented in some MD engines,\cite{farhi2017novel,abel2017advancing,lee2020alchemical,gapsys2020large} whose perturbation energy depends in complex ways on atomic coordinates directly. However, the results obtained here should be applicable to most dual-topology alchemical models based on energy interpolation.\cite{raman2020automated,konig2021efficient} 

\section{Conclusions}

We presented a Potential Distribution Theory (PDT)\cite{PDTbook:2006,Gallicchio2011adv} description of the Alchemical Transfer Method (ATM)\cite{khuttan2021alchemical,wu2021alchemical,azimi2022application} for molecular binding. The probability density $p_0(u)$ of the perturbation energy for transfer at the dissociated state of the solvated complex, which is the central quantity for PDT, is expressed as the convolution of the probability densities for decoupling the ligand from the solvent and coupling it to the solvated receptor obtained from double-decoupling alchemical calculations. We tested the theory on the alchemical binding of five guests to the TEMOA host from the SAMPL8 benchmark set. In each case, the probability densities of the perturbation energy for transfer along the alchemical transfer pathway obtained from numerical calculations match those predicted from the double-decoupling distributions represented using the analytical model of alchemical coupling of Kilburg and Gallicchio.\cite{kilburg2018analytical,pal2019perturbation} The results of the work provide a solid theoretical foundation for alchemical transfer, provide physical insights on the form of the probability densities observed in alchemical transfer calculations, and confirm the conceptual and numerical equivalence between the alchemical transfer and double-decoupling processes. 

\section{Software and Data Availability}

The software and the input files used in this work are available on public Github repositories {\tt https://github.com\-/Gallicchio-Lab\-/analytical-model-transfer}, {\tt https://github.com\-/Gallicchio-Lab\-/femodel-tf-optimizer}, {\tt https://github.com\-/Gallicchio-Lab\-/AToM-OpenMM} as described in Computational Details. Molecular dynamics trajectories are available from the corresponding author upon request.

\section{Acknowledgements}

We acknowledge support from the National Science Foundation (NSF CAREER 1750511) and the National Institute of General Medical Sciences (NIH 1R15GM151708).

\clearpage

\clearpage

%

\clearpage

\section{Appendix}

\noindent This appendix contains other representations of $p_0(u)$ such as $\log p_0(u)$ and $\lambda_0(u)$ for the systems studied in the main work. It also contains parameters of the transfer models constructed from the coupling and hydration models of the systems. 

\subsection{Other Representations of $p_0(u)$}

The predicted $\log p_0(u_{\rm sc})$ functions and the corresponding $\lambda$-functions $\lambda_0(u_{\rm sc})$ with respect to the soft-core perturbation energies of the molecular complexes studied in this work. The analytical models of alchemical coupling to the solvated host are in pink, and those for alchemical hydration are in blue. The yellow lines represent Gaussian kernel estimates from samples collected from numerical alchemical simulations.

\begin{figure}[h!]
    \subfloat{\includegraphics[width=5.5in]{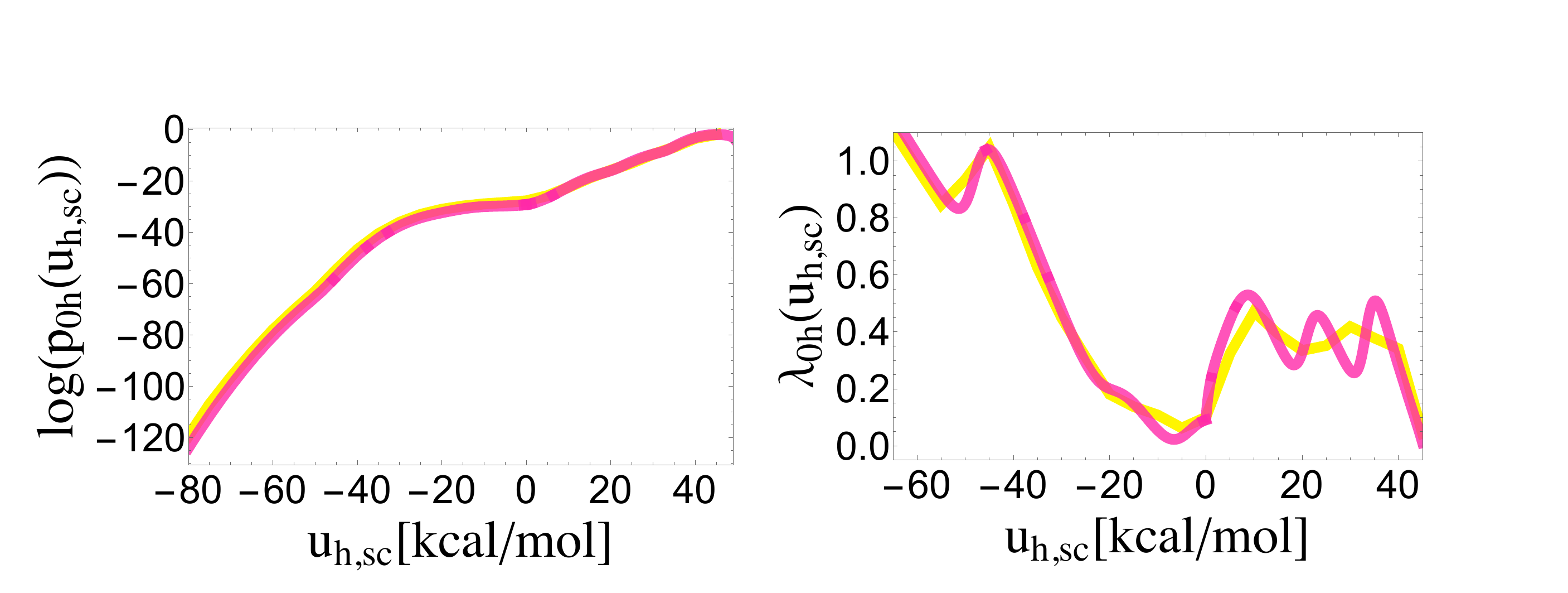}} \\
    \subfloat{\includegraphics[width=5.5in]{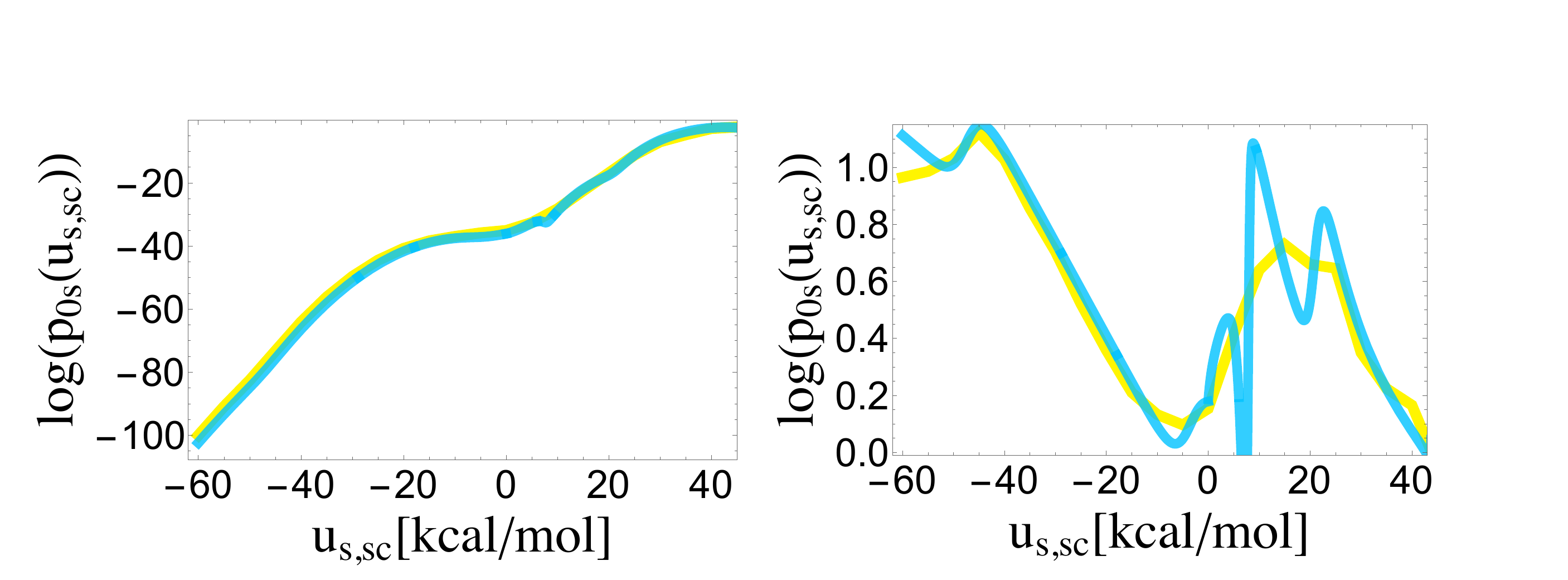}} \\
    \caption{TEMOA-G1.}
\end{figure}

\begin{figure}[h!]
    \subfloat{\includegraphics[width=5.5in]{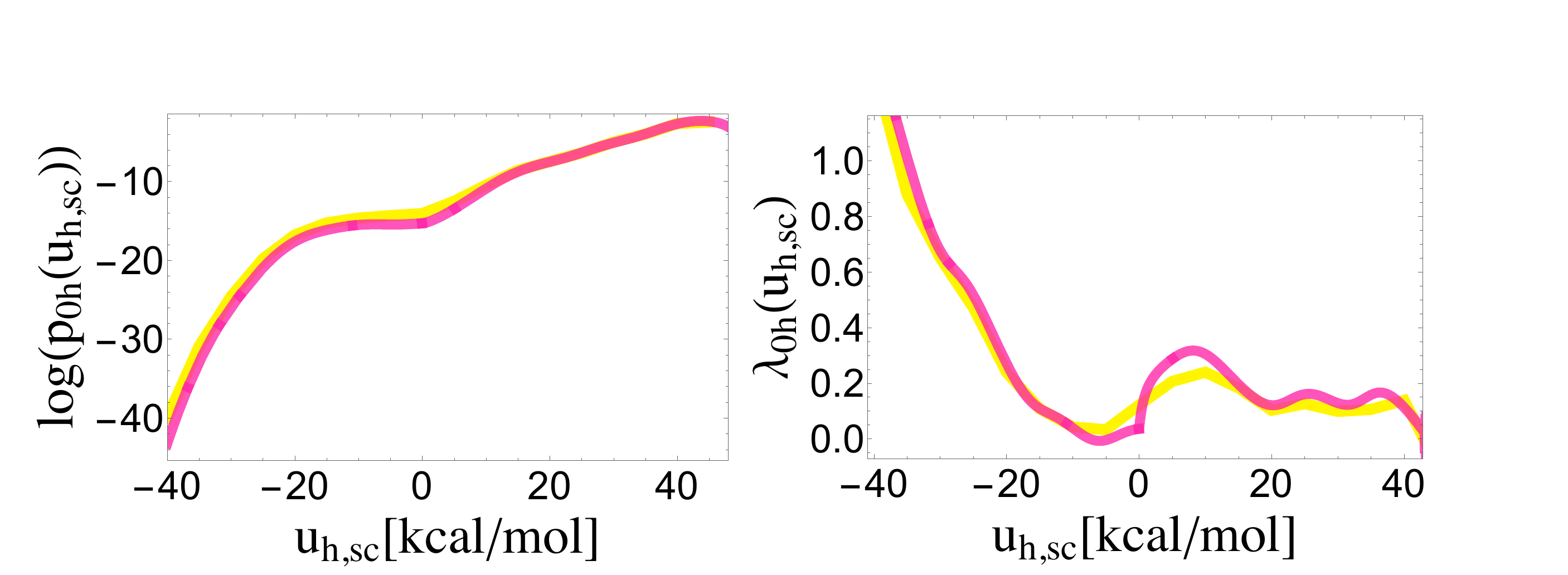}} \\
    \subfloat{\includegraphics[width=5.5in]{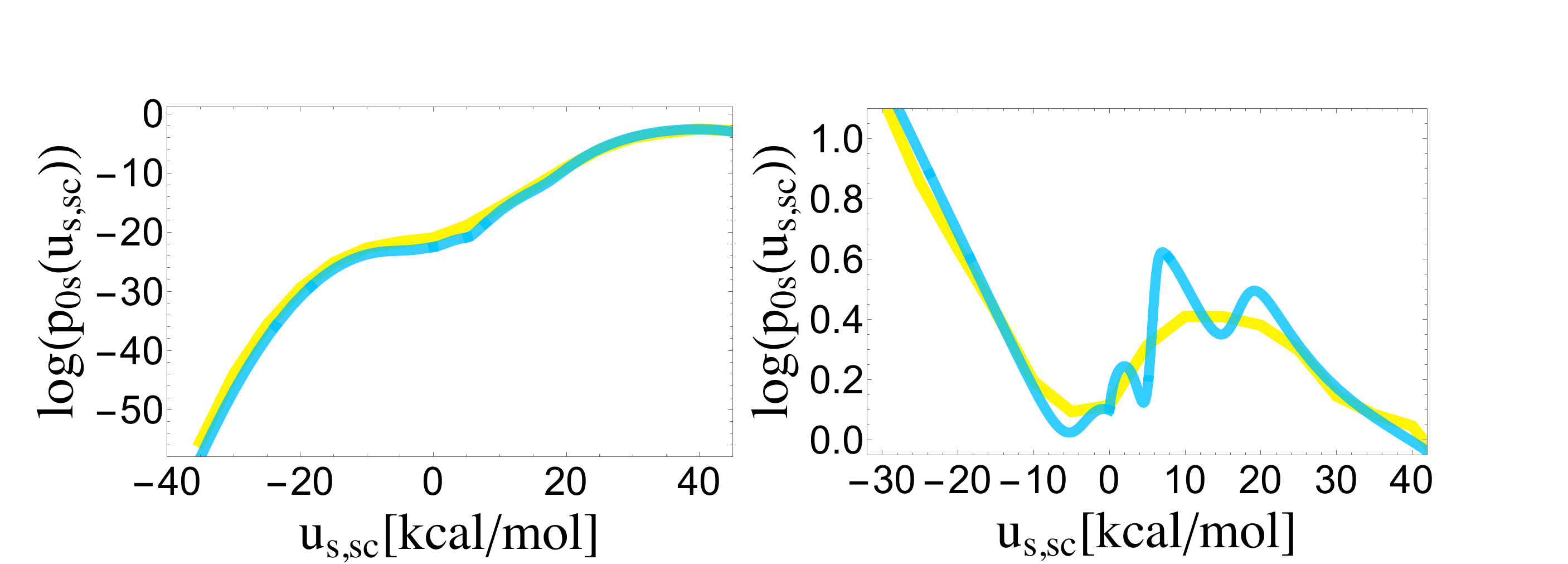}} \\
    \caption{TEMOA-G2.}
\end{figure}

\begin{figure}[h!]
    \subfloat{\includegraphics[width=5.5in]{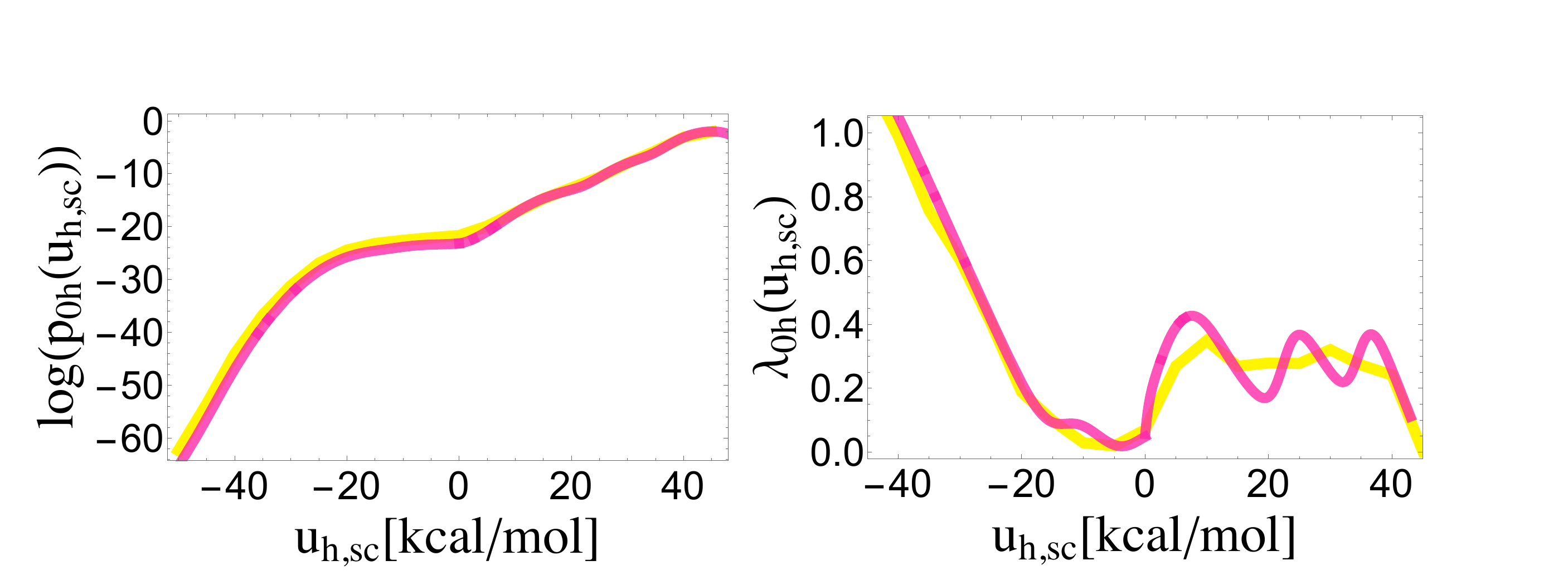}} \\
    \subfloat{\includegraphics[width=5.5in]{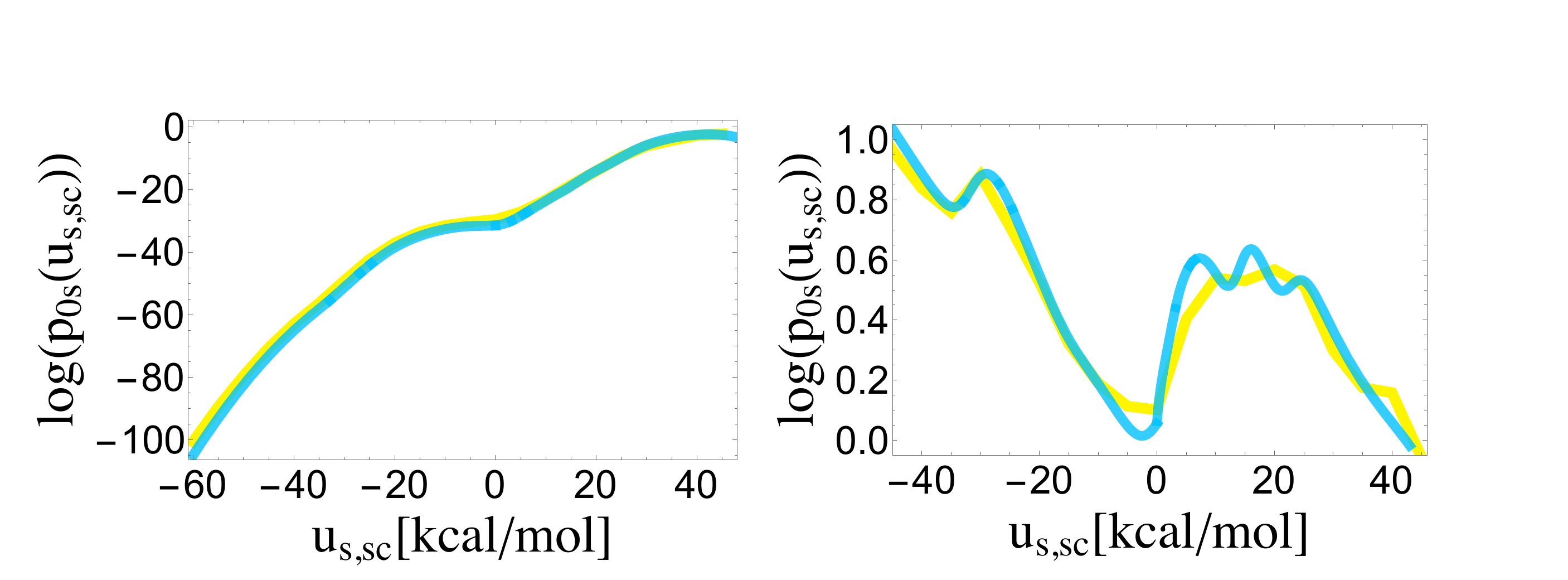}} \\
    \caption{TEMOA-G3.}
\end{figure}

\begin{figure}[h!]
    \subfloat{\includegraphics[width=5.5in]{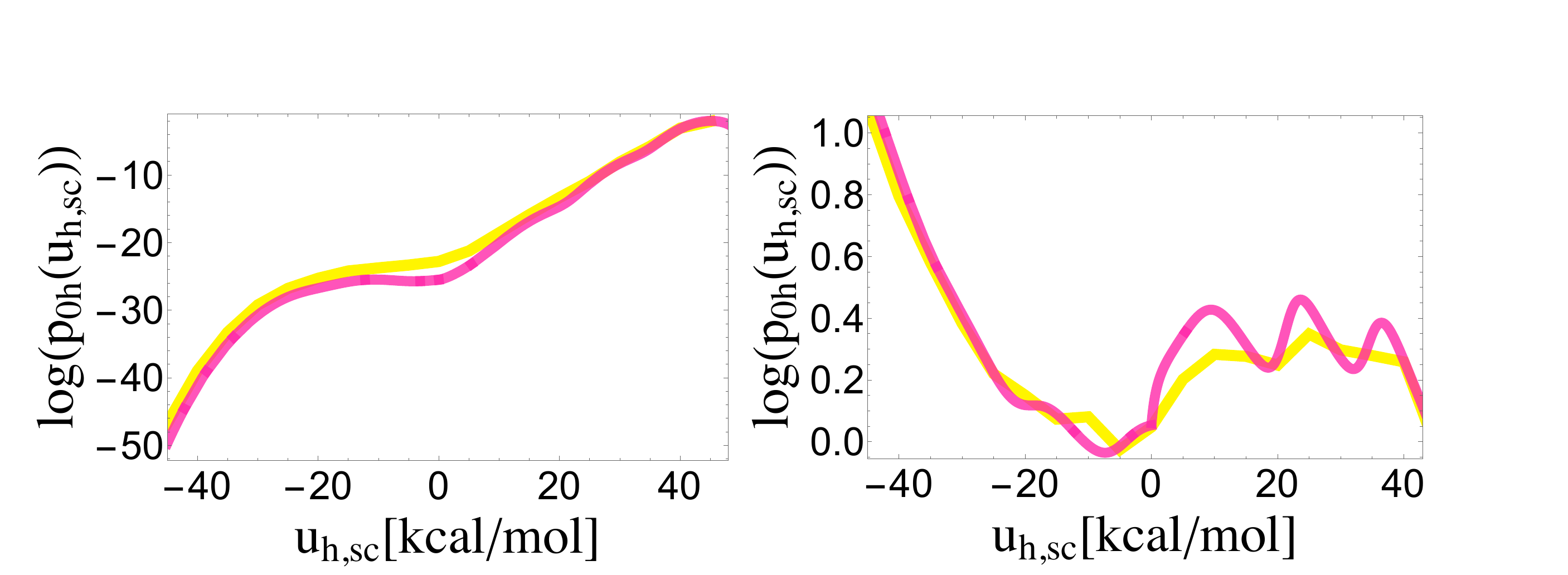}} \\
    \subfloat{\includegraphics[width=5.5in]{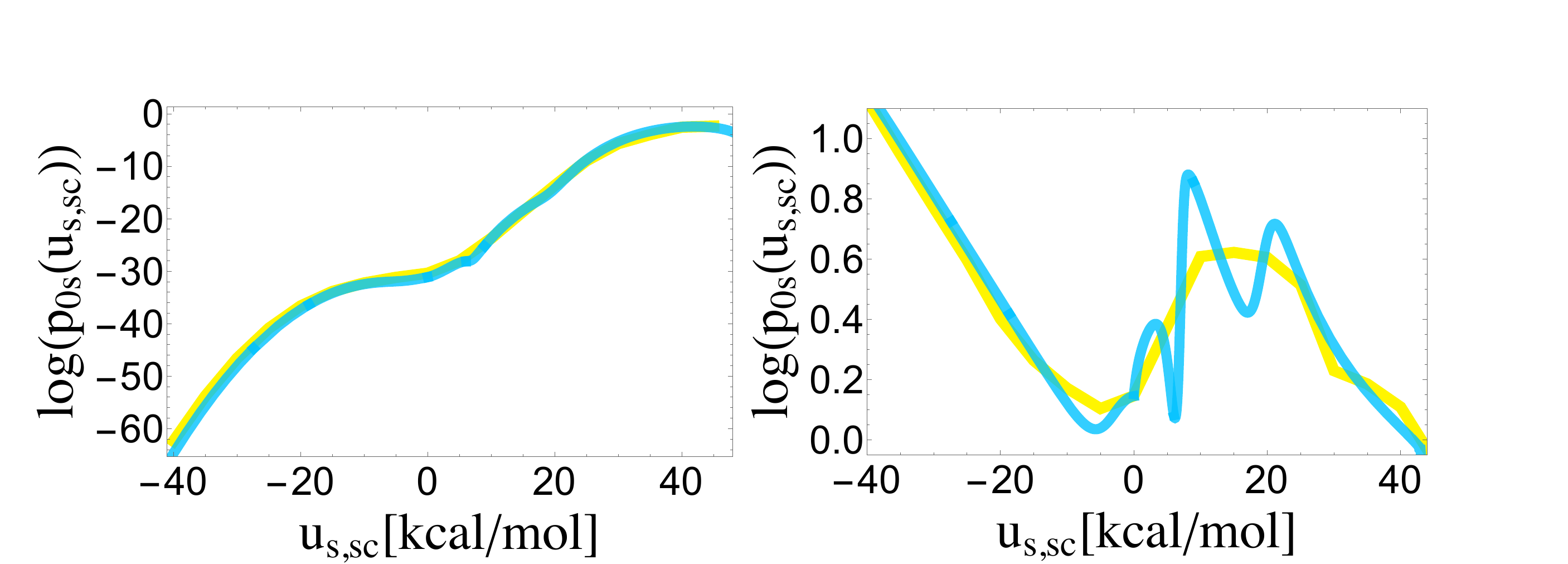}} \\
    \caption{TEMOA-G4.}
\end{figure}

\begin{figure}[h!]
    \subfloat{\includegraphics[width=5.5in]{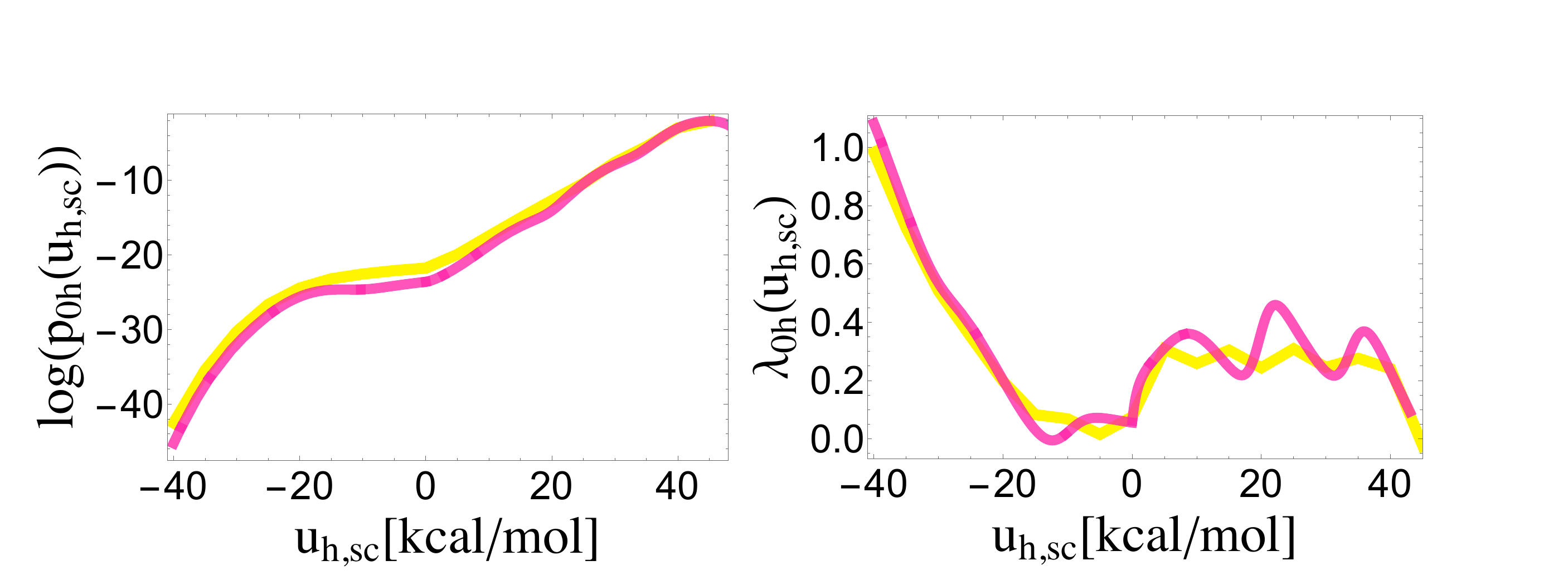}} \\
    \subfloat{\includegraphics[width=5.5in]{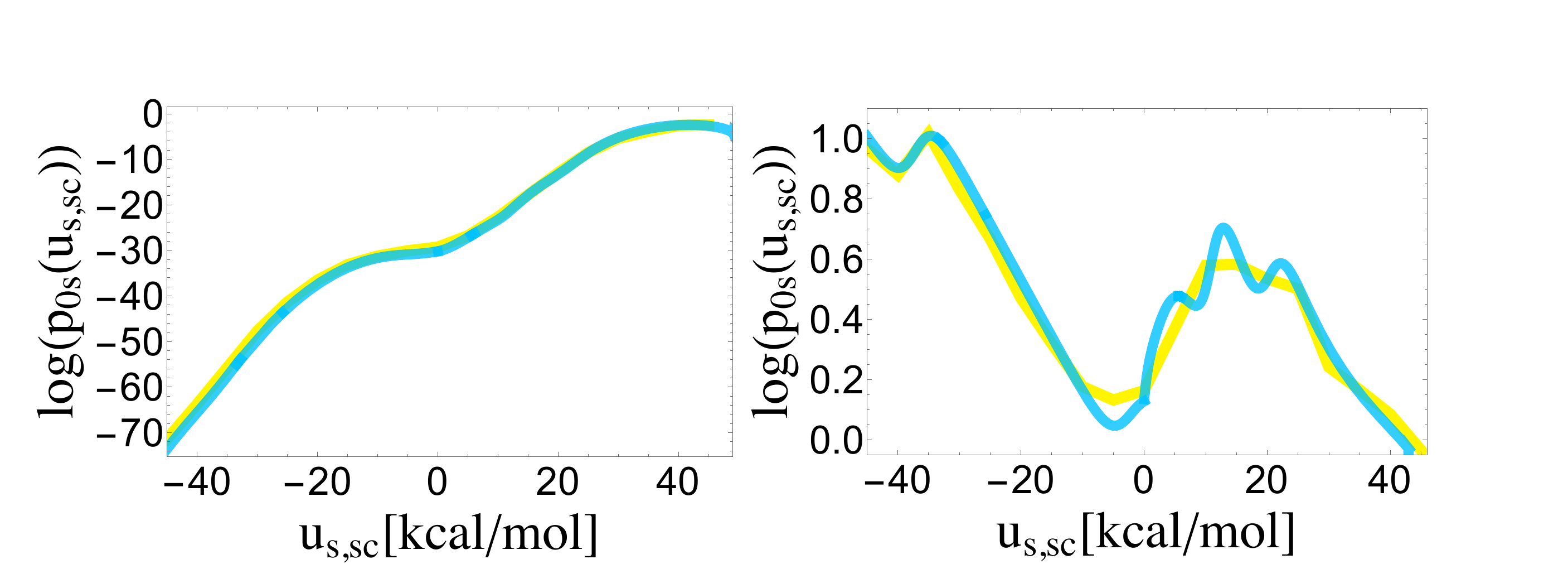}} \\
    \caption{TEMOA-G5}
    \label{fig:logP0_lambdf_temoa-g5}
\end{figure}

\clearpage

\subsection{Optimized Parameters of the Analytical Model of Alchemical Transfer}

\begin{table*}[h!]
  \caption{\label{tab:transf-temoa-g1}Parameters for the transfer model of guest G1 binding to host TEMOA.}
\begin{ruledtabular}
\begin{tabular}{lccccccc}
       & $w_i$              & $b$ & $\bar{u}_{0t}$\footnote{kcal/mol} & $\sigma_{t}^{\rm a} $ & $\epsilon^{\rm a}$ & $\tilde{u}^{\rm a}$ & $n_l$ \\ \hline 
       \tabularnewline

\multicolumn{8}{c}{Leg 1}\tabularnewline
mode 1 & $8.91 \times 10^{-7}$ & $6.40\times 10^{-11}$ & $34.4$             & $5.59$              & $13.8$           &   $386$          & $4.89$ \tabularnewline
mode 2 & $1.07 \times 10^{-1}$ & $1.39\times 10^{-14}$ & $52.6$             & $7.05$              & $13.6$           &   $0.00$           & $41.0$ \tabularnewline
mode 3 & $1.23 \times 10^{-7}$ & $6.40\times 10^{-11}$ & $36.1$             & $6.05$              & $13.8$           &   $386$           & $4.89$ \tabularnewline
mode 4 & $5.38 \times 10^{-1}$ & $1.39\times 10^{-14}$ & $54.3$             & $7.42$              & $13.6$           &   $0.00$           & $41.0$ \tabularnewline
mode 5 & $5.38 \times 10^{-1}$ & $6.40\times 10^{-11}$  & $51.1$             & $6.41$              & $13.8$           &   $386$           & $4.89$ \tabularnewline
mode 6 & $5.38 \times 10^{-1}$ & $1.39\times 10^{-14}$ & $69.2$             & $7.72$              & $13.6$           &   $0.00$           & $41.0$ \tabularnewline

\multicolumn{8}{c}{Leg 2}\tabularnewline
mode 1 & $6.69 \times 10^{-6}$ & $1.63\times 10^{-9}$ & $18.1$             & $5.59$              & $13.5$           &   $237$          & $3.83$ \tabularnewline
mode 2 & $7.71 \times 10^{-1}$ & $2.26\times 10^{-13}$ & $25.5$             & $6.05$              & $24.4$           &   $1220$           & $50.0$ \tabularnewline
mode 3 & $9.73 \times 10^{-3}$ & $7.18\times 10^{-18}$ & $18.0$             & $6.41$              & $18.6$           &   $0.00$           & $30.0$ \tabularnewline
mode 4 & $1.88 \times 10^{-6}$ & $1.63\times 10^{-9}$ & $30.9$             & $7.05$              & $13.5$           &   $237$           & $3.83$ \tabularnewline
mode 5 & $2.16 \times 10^{-1}$ & $2.26\times 10^{-13}$  & $38.2$             & $7.42$              & $24.4$           &   $1220$           & $50.0$ \tabularnewline
mode 6 & $2.73 \times 10^{-3}$ & $7.18\times 10^{-18}$ & $30.8$             & $7.72$              & $18.6$           &   $0.00$           & $30.0$ \tabularnewline

\end{tabular} 
\end{ruledtabular}
\end{table*}

\begin{table*}[h!]
  \caption{\label{tab:transf-temoa-g2}Parameters for the transfer model of guest G2 binding to host TEMOA.}
\begin{ruledtabular}
\begin{tabular}{lccccccc}
       & $w_i$               & $b$ & $\bar{u}_{0t}$\footnote{kcal/mol} & $\sigma_{t}^{\rm a} $ & $\epsilon^{\rm a}$ & $\tilde{u}^{\rm a}$ & $n_l$ \\ \hline 
       \tabularnewline
     
\multicolumn{8}{c}{Leg 1}\tabularnewline
mode 1 & $1.03 \times 10^{-2}$ & $2.60\times 10^{-6}$ & $7.23$             & $4.58$              & $3.50$           &   $3.52$          & $5.29$ \tabularnewline
mode 2 & $9.02 \times 10^{-1}$ & $4.08\times 10^{-8}$ & $18.9$             & $3.91$              & $21.9$           &   $305$           & $31.0$ \tabularnewline
mode 3 & $8.05 \times 10^{-2}$ & $1.44\times 10^{-5}$ & $13.3$             & $4.93$              & $8.75$           &   $63.3$           & $7.32$ \tabularnewline
mode 4 & $7.33 \times 10^{-5}$ & $2.60\times 10^{-6}$ & $9.86$             & $5.33$              & $3.50$           &   $3.52$           & $5.29$ \tabularnewline
mode 5 & $6.42 \times 10^{-3}$ & $4.08\times 10^{-8}$  & $21.6$             & $4.77$              & $21.9$           &   $305$           & $31.0$ \tabularnewline
mode 6 & $5.73 \times 10^{-4}$ & $1.44\times 10^{-5}$ & $15.9$             & $5.64$              & $8.75$           &   $62.3$           & $7.32$ \tabularnewline

\multicolumn{8}{c}{Leg 2}\tabularnewline
mode 1 & $6.57 \times 10^{-4}$ & $7.54\times 10^{-7}$ & $28.0$             & $4.58$              & $13.9$           &   $41.9$          & $4.17$ \tabularnewline
mode 2 & $8.50 \times 10^{-1}$ & $1.85\times 10^{-9}$ & $37.9$             & $5.33$              & $7.77$           &   $0.00$           & $22.8$ \tabularnewline
mode 3 & $8.76 \times 10^{-16}$ & $7.54\times 10^{-7}$ & $6.76$             & $3.91$              & $13.9$           &   $41.9$           & $4.17$ \tabularnewline
mode 4 & $1.13 \times 10^{-12}$ & $1.85\times 10^{-9}$ & $16.7$             & $4.77$              & $7.77$           &   $0.00$           & $22.8$ \tabularnewline
mode 5 & $1.15 \times 10^{-4}$ & $7.54\times 10^{-7}$  & $27.5$             & $4.93$              & $13.9$           &   $41.9$           & $4.17$ \tabularnewline
mode 6 & $1.49 \times 10^{-1}$ & $1.85\times 10^{-9}$ & $37.4$             & $5.64$              & $7.77$           &   $0.00$           & $22.8$ \tabularnewline

\end{tabular} 
\end{ruledtabular}
\end{table*}

\begin{table*}[h!]
  \caption{\label{tab:transf-temoa-g3}Parameters for the transfer model of guest G3 binding to host TEMOA.}
\begin{ruledtabular}
\begin{tabular}{lccccccc}
       & $w_i$               & $b$ & $\bar{u}_{0t}$\footnote{kcal/mol} & $\sigma_{t}^{\rm a} $ & $\epsilon^{\rm a}$ & $\tilde{u}^{\rm a}$ & $n_l$ \\ \hline 
       \tabularnewline
     
\multicolumn{8}{c}{Leg 1}\tabularnewline
mode 1 & $3.04 \times 10^{-6}$ & $8.68\times 10^{-10}$ &       $27.3$             & $4.89$              & $7.62$           &   $58.9$          & $5.32$  \tabularnewline
mode 2 & $7.68 \times 10^{-1}$ & $1.38\times 10^{-13}$ & $51.5$             & $5.39$              & $4.36$           &   $39.7$           & $34.4$ \tabularnewline
mode 3 & $6.42 \times 10^{-3}$ & $3.69\times 10^{-14}$ & $37.1$             & $6.10$              & $17.2$           &   $47.5$           & $13.5$ \tabularnewline
mode 4 & $8.34 \times 10^{-7}$ & $8.68\times 10^{-10}$ & $53.6$             & $6.09$              & $7.62$           &   $58.9$           & $5.32$ \tabularnewline
mode 5 & $2.11 \times 10^{-1}$ & $1.38\times 10^{-13}$  & $35.5$             & $6.50$              & $4.36$           &   $39.8$           & $34.4$ \tabularnewline
mode 6 & $1.76 \times 10^{-3}$ & $3.69\times 10^{-14}$ & $52.0$             & $7.10$              & $17.2$           &   $47.5$           & $13.5$ \tabularnewline
mode 7 & $4.95 \times 10^{-8}$ & $8.68\times 10^{-10}$ & $52.0$             & $5.50$              & $7.62$           &   $58.9$           & $5.32$ \tabularnewline
mode 8 & $1.25 \times 10^{-2}$ & $1.38\times 10^{-13}$ & $52.0$             & $5.95$              & $4.36$           &   $39.7$           & $34.4$ \tabularnewline
mode 9 & $1.05 \times 10^{-4}$ & $3.69\times 10^{-14}$ & $52.0$             & $6.61$              & $17.2$           &   $47.5$           & $13.5$ \tabularnewline

\multicolumn{8}{c}{Leg 2}\tabularnewline
mode 1 & $8.40\times 10^{-7}$ & $9.38\times 10^{-7}$    & $12.8$             & $4.89$              & $3.91$           &   $0.00$          & $8.55$  \tabularnewline
mode 2 & $6.51 \times 10^{-3}$ & $7.91\times 10^{-12}$ & $23.0$             & $6.09$              & $26.8$           &   $396$           & $49.6$  \tabularnewline
mode 3 & $1.82 \times 10^{-4}$ & $1.91\times 10^{-8}$ & $21.5$             & $5.50$              & $7.00$           &   $70.0$           & $19.7$ \tabularnewline
mode 4 & $1.01 \times 10^{-6}$ & $9.38\times 10^{-7}$ & $15.0$             & $5.39$              & $3.91$           &   $0.00$           & $8.55$ \tabularnewline
mode 5 & $7.80 \times 10^{-3}$ & $7.91\times 10^{-12}$  & $25.3$             & $6.50$              & $26.8$           &   $396$           & $49.6$  \tabularnewline
mode 6 & $2.18 \times 10^{-4}$ & $1.91\times 10^{-8}$ & $23.8$             & $5.95$              & $7.00$           &   $70.0$           & $19.7$ \tabularnewline
mode 7 & $1.24 \times 10^{-4}$ & $9.38\times 10^{-7}$ & $28.5$             & $6.10$              & $3.91$           &   $0.00$           & $8.55$ \tabularnewline
mode 8 & $9.58 \times 10^{-1}$ & $7.91\times 10^{-12}$ & $38.7$             & $7.10$              & $26.8$           &   $396$           & $49.6$ \tabularnewline
mode 9 & $2.68 \times 10^{-2}$ & $1.91\times 10^{-8}$ & $37.2$             & $6.61$              & $7.00$           &   $70.0$           & $19.7$ \tabularnewline

\end{tabular} 
\end{ruledtabular}
\end{table*}

\clearpage

\begin{table*}[h]
  \caption{\label{tab:transf-temoa-g4}Parameters for the transfer model of guest G4 binding to host TEMOA.}
\begin{ruledtabular}
\begin{tabular}{lccccccc}
       & $w_i$              & $b$ & $\bar{u}_{0t}$\footnote{kcal/mol} & $\sigma_{t}^{\rm a} $ & $\epsilon^{\rm a}$ & $\tilde{u}^{\rm a}$ & $n_l$ \\ \hline 
       \tabularnewline
     
\multicolumn{8}{c}{Leg 1}\tabularnewline
mode 1 & $4.53 \times 10^{-7}$ & $2.99\times 10^{-9}$ & $35.0$             & $5.63$              & $9.94$           &   $49.8$          & $6.69$ \tabularnewline
mode 2 & $1.05 \times 10^{-2}$ & $1.99\times 10^{-12}$ & $51.5$             & $6.85$              & $12.3$           &   $17.9$           & $27.4$ \tabularnewline
mode 3 & $9.61 \times 10^{-7}$ & $2.99\times 10^{-9}$ & $37.1$             & $5.93$              & $9.94$           &   $49.7$           & $6.69$ \tabularnewline
mode 4 & $2.24 \times 10^{-2}$ & $1.99\times 10^{-12}$ & $53.6$             & $7.10$              & $12.3$           &   $17.9$           & $27.4$ \tabularnewline
mode 5 & $4.15 \times 10^{-5}$ & $2.99\times 10^{-9}$  & $35.5$             & $5.35$              & $9.94$           &   $49.7$           & $6.69$ \tabularnewline
mode 6 & $9.67 \times 10^{-1}$ & $1.99\times 10^{-12}$ & $52.0$             & $6.63$              & $12.3$           &   $17.9$           & $27.4$ \tabularnewline

\multicolumn{8}{c}{Leg 2}\tabularnewline
mode 1 & $5.19\times 10^{-3}$ & $1.65\times 10^{-9}$ & $18.2$             & $5.63$              & $10.8$           &   $70.2$          & $13.1$ \tabularnewline
mode 2 & $9.62 \times 10^{-1}$ & $7.08\times 10^{-11}$ & $21.8$             & $5.93$              & $28.6$           &   $587$           & $38.8$ \tabularnewline
mode 3 & $7.79 \times 10^{-6}$ & $5.62\times 10^{-8}$ & $12.5$             & $5.35$              & $7.82$           &   $56.3$           & $3.92$ \tabularnewline
mode 4 & $1.76 \times 10^{-4}$ & $1.65\times 10^{-9}$ & $27.4$             & $6.86$              & $10.8$           &   $70.2$           & $13.1$ \tabularnewline
mode 5 & $3.26 \times 10^{-2}$ & $7.08\times 10^{-11}$  & $31.0$             & $7.10$              & $28.6$           &   $587$           & $38.8$ \tabularnewline
mode 6 & $2.64 \times 10^{-7}$ & $5.62\times 10^{-7}$ & $21.7$             & $6.63$              & $7.82$           &   $56.3$           & $3.92$ \tabularnewline

\end{tabular} 
\end{ruledtabular}
\end{table*}

\begin{table*}[h]
  \caption{\label{tab:transf-temoa-g5}Parameters for the transfer model of guest G5 binding to host TEMOA.}
\begin{ruledtabular}
\begin{tabular}{lccccccc}
       & $w_i$               & $b$ & $\bar{u}_{0t}$\footnote{kcal/mol} & $\sigma_{t}^{\rm a} $ & $\epsilon^{\rm a}$ & $\tilde{u}^{\rm a}$ & $n_l$ \\ \hline 
       \tabularnewline
     
\multicolumn{8}{c}{Leg 1}\tabularnewline
mode 1 & $5.22 \times 10^{-8}$ & $5.52\times 10^{-6}$ &       $33.1$             & $4.95$              & $1.99$           &   $19.9$          & $4.44$  \tabularnewline
mode 2 & $1.03 \times 10^{-3}$ & $2.65\times 10^{-14}$ & $32.9$             & $5.63$              & $3.79$           &   $22.9$           & $12.6$ \tabularnewline
mode 3 & $9.96 \times 10^{-1}$ & $8.61\times 10^{-18}$ & $39.9$             & $5.99$              & $24.3$           &   $0.00$           & $26.7$ \tabularnewline
mode 4 & $9.96 \times 10^{-11}$ & $5.52\times 10^{-6}$ & $30.7$             & $5.63$              & $1.99$           &   $19.9$           & $4.44$ \tabularnewline
mode 5 & $3.37 \times 10^{-6}$ & $2.65\times 10^{-14}$  & $30.4$             & $6.23$              & $3.79$           &   $22.9$           & $12.6$ \tabularnewline
mode 6 & $3.26 \times 10^{-3}$ & $8.61\times 10^{-18}$ & $37.4$             & $6.560$              & $24.3$           &   $0.00$           & $26.7$ \tabularnewline
mode 7 & $4.10 \times 10^{-34}$ & $5.52\times 10^{-6}$ & $30.8$             & $6.42$              & $1.99$           &   $19.9$           & $4.44$ \tabularnewline
mode 8 & $8.08 \times 10^{-30}$ & $2.65\times 10^{-14}$ & $30.6$             & $6.95$              & $3.79$           &   $22.9$           & $12.6$ \tabularnewline
mode 9 & $7.82 \times 10^{-27}$ & $8.61\times 10^{-18}$ & $37.6$             & $7.25$              & $24.3$           &   $0.00$           & $26.7$ \tabularnewline

\multicolumn{8}{c}{Leg 2}\tabularnewline
mode 1 & $4.86 \times 10^{-6}$ & $2.92 \times 10^{-9}$    & $7.99$             & $4.95$              & $10.9$           &   $152$          & $3.05$   \tabularnewline
mode 2 & $5.01 \times 10^{-2}$ & $1.40\times 10^{-9}$ &     $22.5$             & $5.63$              & $22.0$           &   $217$           & $16.9$  \tabularnewline
mode 3 & $1.73 \times 10^{-1}$ & $5.04\times 10^{-38}$  & $38.3$             & $6.42$              & $50.5$           &   $646$           & $43.1$  \tabularnewline
mode 4 & $1.69 \times 10^{-5}$ & $2.92\times 10^{-9}$ & $20.2$             & $5.63$              & $10.9$           &   $152$           & $3.05$  \tabularnewline
mode 5 & $1.74 \times 10^{-1}$ & $1.40\times 10^{-9}$  & $34.8$             & $6.23$              & $22.0$           &   $217$           & $16.9$  \tabularnewline
mode 6 & $6.02 \times 10^{-1}$ & $5.04\times 10^{-38}$ & $50.6$             & $6.95$              & $50.5$           &   $646$           & $43.1$  \tabularnewline
mode 7 & $1.51 \times 10^{-8}$ & $2.92\times 10^{-9}$ & $20.3$             & $5.99$              & $10.9$           &   $152$           & $3.05$  \tabularnewline
mode 8 & $1.56 \times 10^{-4}$ & $1.40\times 10^{-9}$ & $34.8$             & $6.56$              & $22.0$           &   $217$           & $16.9$  \tabularnewline
mode 9 & $5.39 \times 10^{-4}$ & $5.04\times 10^{-38}$ & $50.6$             & $7.25$              & $50.5$           &   $646$           & $43.1$ \tabularnewline

\end{tabular} 
\end{ruledtabular}
\end{table*}

\end{document}